\documentclass[12pt]{article}
\RequirePackage[authoryear]{natbib}

\usepackage[margin=1in]{geometry}
\usepackage{amsmath}
\usepackage{natbib}
\usepackage{amssymb}
\usepackage{bbm}
\usepackage{tikz}
\usetikzlibrary{shapes,arrows}
\usetikzlibrary{decorations.pathreplacing}
\usetikzlibrary{positioning}
\usepackage{graphicx}
\usepackage{multirow}
\usepackage{enumitem}
\usepackage{booktabs,caption,subcaption}
\usepackage{url}
\usepackage{float}
\usepackage{longtable}

\newtheorem{theorem}{Theorem}

\newcommand{\R}{\mathbb{R}}
\newcommand{\Rc}{\mathcal{R}}

\newcommand{\Y}{\mathbf{Y}}

\newcommand{\1}{\mathbf{1}}
\newcommand{\0}{\mathbf{0}}

\newcommand{\bbeta}{\boldsymbol\beta}
\newcommand{\btheta}{\boldsymbol\theta}
\newcommand{\bmu}{\boldsymbol\mu}
\newcommand{\bpsi}{\boldsymbol\psi}

\newcommand{\bSigma}{\boldsymbol\Sigma}
\newcommand{\bx}{\mathbf{x}}
\newcommand{\bX}{\mathbf{X}}
\newcommand{\bH}{\mathbf{H}}
\newcommand{\bh}{\mathbf{h}}
\newcommand{\bA}{\mathbf{A}}
\newcommand{\ba}{\mathbf{a}}
\newcommand{\calC}{\mathcal{C}}
\newcommand{\bd}{\mathbf{d}}

\newcommand{\bZ}{\mathbf{Z}}
\newcommand{\bC}{\mathbf{C}}
\newcommand{\pr}{P}
\newcommand{\E}{\mathbb{E}}

\allowdisplaybreaks

\begin{document}

\title
{\bf Interim Monitoring of Sequential Multiple Assignment Randomized Trials Using Partial Information}

\author{Cole Manschot$^{1}$, Eric Laber$^{2}$, and Marie Davidian$^{1}$ \\  \\
$^{1}$Department of Statistics, North Carolina State University \\
$^{2}$Department of Statistical Science and Department of Biostatistics \\ 
\& Bioinformatics,
Duke University \\
}
\date{}

\maketitle
 
\bigskip

\begin{abstract}

  The sequential multiple assignment randomized trial (SMART) is the
  gold standard trial design to generate data for the evaluation of multi-stage
  treatment regimes.  
  As with conventional (single-stage) randomized clinical trials,
  interim monitoring allows early stopping;
  however, there are few methods for principled
  interim analysis in SMARTs.  
  Because SMARTs involve multiple stages of treatment, a key challenge 
  is that not all enrolled participants will have progressed
  through all treatment stages at the time of an interim analysis. 
  \citet*{wu2021} propose basing interim analyses on an estimator for the mean
  outcome under a given regime that uses data only from participants who have
  completed all treatment stages.
  We propose an estimator for the mean outcome
  under a given regime
  that gains efficiency by using partial information from enrolled
  participants regardless of their progression through treatment stages.
  Using the asymptotic distribution of this estimator, 
  we derive associated Pocock and O'Brien-Fleming testing
  procedures for early stopping.  
  In simulation experiments, the
  estimator controls type I error and achieves nominal power while
  reducing expected sample size relative to the method of \citet[]{wu2021}. 
  We present an illustrative application of the proposed estimator based
  on a recent SMART evaluating behavioral pain interventions for breast
  cancer patients.
  
\end{abstract}

\noindent%
\vfill

\maketitle

\section{Introduction}
\label{s:intro}

Treatment of chronic diseases and disorders involves a series of
treatment decisions made at critical points in the 
progression of a patient's health status.
To optimize long-term health outcomes, these decisions
must adapt to evolving patient information, including response to
previous treatments.  Strategies for adapting treatment decisions over
time are formalized as treatment regimes, which comprise a sequence of
decision rules, one per stage of intervention, that map accrued
patient information to a recommended treatment
\citep{chakraborty2013statistical, tsiatis2019dynamic}.  
The value of a regime is the expected utility if the regime is used to select treatments in the population of interest. 
A regime is optimal if it has maximal value. 
Much of the
statistical literature on treatment regimes has focused on estimation
and inference for optimal regimes
\citep{kosorok2019precision}.  
However, scientific interest often
focuses on comparison of a small number of pre-specified treatment
regimes, either with each other or against a control,
on the basis of mean outcome.

The gold standard for data collection for the evaluation of treatment
regimes is the sequential multiple assignment randomized trial
\citep[SMART;][]{lavori2004, murphy2005}.  
A SMART contains multiple stages of randomization, with each stage corresponding to a key decision point. 
In a SMART, if, when, and to whom a treatment might be randomly assigned is allowed to depend on a patient's treatment and outcome history, leading to a rich and flexible class of designs. 
In the past decade, the use of SMARTs has increased dramatically;
SMARTs have been conducted in a range of disease and disorder areas,
including cancer
\citep{wang2012evaluation,thall2015chapter,kelleher2017optimizing},
behavioral sciences
\citep{almirall2014introduction,kidwell2016adaptive},
and mental health
\citep*{manschreck2007catie,sinyor2010sequenced}.
For a comprehensive list of SMARTs, see \citet*{bigirumurame2022}.

\begin{figure}[ht!]
\begin{center}
\includegraphics[scale=0.5]{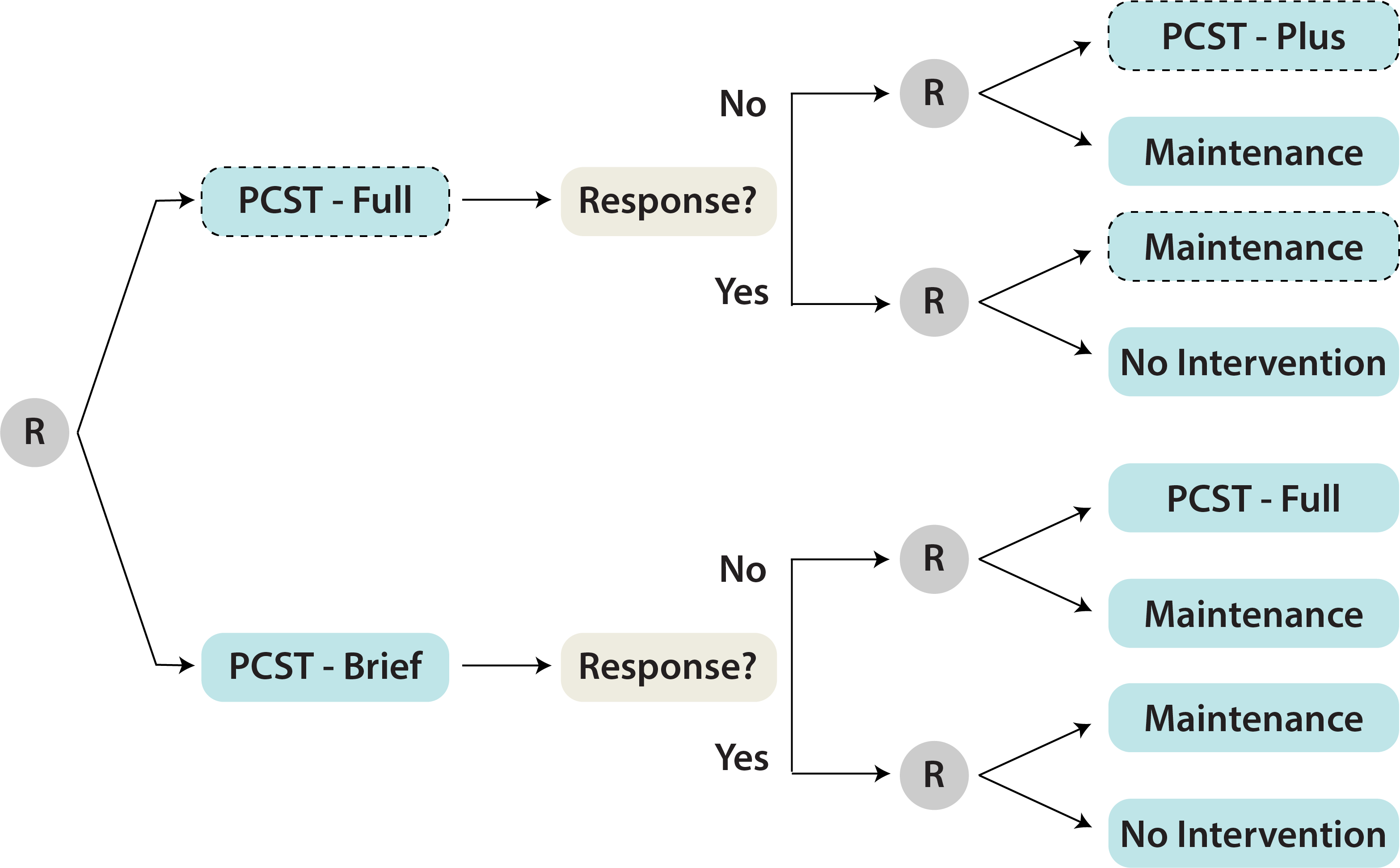}
\caption{
\label{fig:PCST-schema}
    Schema for the SMART evaluating regimes involving behavioral
    interventions for pain management in breast cancer patients embedding
    eight regimes of the form ``Give intervention $a$; if non-response, give
    $b$; otherwise, if response give $c$.''  The embedded regime
    determined by $a$ = PCST-Full, $b$= PCST-Plus, $c$ = Maintenance is
    shown with dashed lines around the treatments.
    Regimes $l =1,\ldots,8$ take $(a,b,c)$ to be 
    (Full, Plus, Maintenance), 
    (Full, Plus, No Intervention), 
    (Full, Maintenance, Maintenance), 
    (Full, Maintenance, No Intervention), 
    (Brief, Full, Maintenance), 
    (Brief, Full, No Intervention), 
    (Brief, Maintenance, Maintenance), and
    (Brief, Maintenance, No Intervention), 
    respectively. 
    This figure appears in color in the electronic version of this article.
}
\end{center}
\end{figure}

Every SMART can be equivalently represented as randomizing subjects at baseline among a set of fixed regimes known as the trial's ``embedded regimes.''
Primary analyses in a SMART often focus on comparisons of the embedded regimes against each other or a control \citep{lavori2004, murphy2005}.
These comparisons are often used for sizing a SMART 
\citep{seewald2020, artman2020}.  
For example, Figure~\ref{fig:PCST-schema} shows a two-stage SMART schema for behavioral interventions for pain management in cancer patients with eight embedded regimes \citep{kelleher2017optimizing, nct02791646}. 
Each embedded regime takes the form 
``give intervention $a$; if response, give
$b$; if non-response, give $c$;'' e.g., 
give pain coping skills training (PCST) Full initially; if response, give maintenance; otherwise, give PCST-Plus. 
We discuss this study further in Section~\ref{s:app}.

Interim monitoring allows early stopping for efficacy or futility, which can reduce cost and accelerate evaluation of candidate treatments.  
Group sequential methods allowing early stopping are well established for conventional clinical trials
 \citep{jennison1999}. 
However, analogous methodology for SMARTs is limited. 
\cite*{wu2021} propose an interim test for a difference in mean outcome among embedded regimes in two-stage SMARTs.
However, their approach is based on the inverse probability weighted estimator (IPWE),
which does not incorporate baseline and accruing patient information that could be used to enhance efficiency
\citep[][]{zhang2013}.
\citet{chao2020} consider interim analysis for
a small-$n$, two-stage SMART restricted to the specific situation
in which the same treatments are available at each stage and the goal is to remove futile treatments.

We develop a class of interim analysis methods for SMARTs based on an augmented inverse probability weighted estimator (AIPWE) for the value of a regime that
increases statistical efficiency by using partial information from individuals with incomplete regime trajectories.
Our method applies to SMARTs with an arbitrary number of stages and treatments,
as well as those in which the set of allowable treatments depends on a patient's history. 
We present the statistical framework in Section~\ref{s:background}.
In Section~\ref{s:aipw_est}, we review the AIPWE
for the value of a regime when all participants have progressed
through all stages.  We introduce the proposed Interim AIPWE in Section~\ref{s:interim-est}.
In Section~\ref{s:interim}, we discuss testing
procedures, stopping boundaries, and sample size formul{\ae} for interim analysis. In
Section~\ref{s:sim}, we evaluate the empirical performance of the
proposed procedure in a series of simulation experiments, and we
present a case study based on the cancer pain management SMART in
Section~\ref{s:app}.

\section{Statistical framework} \label{s:background}

Consider a SMART with $K$ stages and a planned total sample size of
$N$. Each subject completing the trial generates a trajectory of the
form $(\bX_1, A_1, \bX_2, A_2, \dots, \bX_K, A_K, Y)$, where 
$A_k \in \mathcal{A}_k$, $k=1,\ldots,K$, is the treatment assigned at
stage $k$; $\mathcal{A}_k$ is a finite set of treatment options at decision $k$;
$\bX_1\in\mathbb{R}^{p_1}$ comprises baseline subject variables; 
$\bX_k\in\mathbb{R}^{p_k}$, $k=2,\ldots,K$, comprises subject
variables collected between stages $k-1$ and $k$; and 
$Y\in\mathbb{R}$ is an outcome measured at the end of follow up, coded
so that higher values are better.  
Let
$\overline{\bX}_k = (\bX_1,\ldots, \bX_k)$ and
$\overline{\bA}_k = (A_1,\ldots, A_k)$, and define $\bH_1 = \bX_1$ and
$\bH_k = (\overline{\bX}_k, \overline{\bA}_{k-1})$, $k\ge 2$, so
that $\bH_k$ is the information available at the time $A_k$ is
assigned.  
Let $\mathcal{H}_k = \mathrm{dom}\ \bH_k$ and let $2^{\mathcal{A}_k}$ denote the power set of $\mathcal{A}_k$. 
We assume there exists a set-valued function 
$\Psi_k: \mathcal{H}_k \rightarrow 2^{\mathcal{A}_k}$  
so that the set of allowable treatments for a
subject with $\bH_k = \bh_k$ at stage $k$ is 
$\Psi_k(\bh_k) \subseteq \mathcal{A}_k$
\citep{van2007causal,tsiatis2019dynamic}.  

In this setting, a treatment regime is a sequence of decision rules,
$\bd= (d_1,\ldots,d_K)$, where
$d_k : \mathcal{H}_k \to \mathcal{A}_k$ and
$d_k(\bh_k)\in\Psi_k(\bh_k)$ for all $\bh_k \in \mathcal{H}_k$,
$k=1,\ldots, K$.  
Let $Y^*(\overline{\ba}_K)$ denote the potential
outcome under treatment sequence
$\overline{\ba}_{K} = (a_1,\ldots,a_K)$, 
and let
$\bX_k^*(\overline{\ba}_{k-1})$ denote the potential intermediate variables under sequence $\overline{\ba}_{k-1}$ at stage $k \ge 2$.
Define
$\overline{\bX}^*_k(\overline{\ba}_{k-1}) = \{\bX_1, \bX^*_2(a_1),
  \ldots, \bX^*_k(\overline{\ba}_{k-1})\}$,
  $\bH_k^*(\overline{\ba}_{k-1}) =
  \{\overline{\bX}^*_k(\overline{\ba}_{k-1}), \overline{\ba}_{k-1}\}$, and
  $\bH_1^*(a_0) = \bH_1$. 
The potential covariates and outcome for an individual
receiving treatment according to regime $\bd$ are 
\allowdisplaybreaks
\begin{align*}
\bX^*_k(\bd) &= \sum_{\overline{\ba}_{k-1} \in \mathcal{A}_1 \times \cdots  \times \mathcal{A}_{k-1}}\bX^*(\overline{\ba}_{k-1})
\prod_{k=1}^{k-1}
I\left[
a_k = d_k\left\lbrace
\bH_k^*(\overline{\ba}_{k-1})
\right\rbrace
\right], k=2,\ldots,K, 
\\
Y^*(\bd) &= \sum_{\overline{\ba}_K \in \mathcal{A}_1 \times \cdots  \times \mathcal{A}_K}Y^*(\overline{\ba}_K)
\prod_{k=1}^KI\left[
a_k = d_k\left\lbrace
\bH_k^*(\overline{\ba}_{k-1})
\right\rbrace
\right].
\end{align*}
The mean outcome, or value, for a regime $\bd$ is $V(\bd) = \mathbb{E} \{ Y^*(\bd) \}$. 

In a SMART, primary analyses often focus on inference on $V(\bd)$ for regimes
$\bd$ that are embedded in the trial.  
Let
$\pi_k(a_k, \bh_k) = \pr(A_k = a_k|\bH_k=\bh_k)$ be the probability
(propensity) of being randomized to treatment $a_k \in \Psi_k(\bh_k)$
at stage $k$ for a subject with history $\bh_k$.
It is well
known that $V(\bd)$ is identifiable under the following conditions:
positivity ($\pi_k(a_k, \bh_k) > 0$ for all
$\bh_k\in \mathcal{H}_k$ and $a_k\in\Psi_k(\bh_k)$);
sequential
randomization 
\allowbreak($\{\bX_1, \bX^*_2(a_1), \ldots, \bX^*_k(\overline{\ba}_{k-1}), Y^*(\overline{\ba}_K)\}_{\overline{\ba}_K \in \mathcal{A}_1 \times \cdots \times \mathcal{A}_K} \perp\!\!\!\perp A_k \vert \bH_k$ 
at each stage $k$ for all
$\overline{\ba}_K$, where $\perp\!\!\!\perp$ denotes independence), which holds by design in a SMART;
consistency,
$Y = Y^*(\overline{\bA}_K)$ and
$\bH_k = \bH_k^*(\overline{\bA}_{k-1})$; 
and no interference among
subjects \citep{tsiatis2019dynamic}. 
Hereafter, we assume that these conditions hold. 

Take $\bd^\ell$,
$\ell = 1,\ldots,L$, to be the regimes embedded in the SMART and
$\bd^0$ a possible control, e.g., a treatment or regime representing the standard of care.
For definiteness, we consider two null hypotheses that address the efficacy of the embedded regimes:
\begin{align} 
    &\text{Homogeneity} && H_{0H}: V(\bd^1)= \cdots = V(\bd^L)  \label{h:h0H}  \\
    &\text{Superiority} && H_{0D}: V(\bd^\ell) - V(\bd^0) \leq \delta \text{ for all } \ell= 1,\ldots,L.  \label{h:h0D} 
\end{align}
These hypotheses are analogous to those used in multi-arm, multi-stage and platform trials (\citealp[][Chapter 16]{jennison1999}; \citealp{wason2019}). 
The control value $V(\bd^0)$ may be fixed or estimated from an additional control arm. 
The methods presented here apply to futility testing with minor modification. 
Hypotheses that stop the trial for either a single regime or all regimes falling below an efficacy boundary are possible in this construction.

\section{AIPWE for complete data} \label{s:aipw_est}

We briefly review the AIPWE of the value in the setting where one observes $N$ complete 
independent and identically distributed trajectories
$\left\lbrace \bX_{1,i}, A_{1,i},\ldots, \bX_{K,i}, A_{K,i}, Y_i
\right\rbrace_{i=1}^N$.
For any regime $\bd$, define
$\calC_{\bd,k} = \prod_{j=1}^k I\left\lbrace A_j = d_j(\bH_j)
\right\rbrace$
to be an indicator that treatment
is consistent with $\bd$ through the first $k$ decisions, and let
$\calC_{\bd,0} = 1$.  
For each $k=1,\ldots,K$, let 
$\pi_k(a_k, \bh_k;\btheta_k)$ be a
posited model for $\pi_k(a_k, \bh_k)$ indexed by $\btheta_k\in\pmb{\Theta}_k$.
Although the
propensities are known in a SMART, estimating them based on correctly
specified models can increase efficiency \citep{tsiatis2006book}.
Let $\widehat{\btheta}_{k,N}$ be an estimator of $\btheta_k$.
The form of the AIPWE for $V(\bd)$ is (\citealp{zhang2013}; \citealp[][Section 6.4.4]{tsiatis2019dynamic})
\begin{multline} \label{eq:AIPW} 
    \widehat{V}_{\mathrm{A}}(\bd) = 
    N^{-1} 
    \sum_{i=1}^N \Bigg[ \frac{ Y_i \ 
    \mathcal{C}_{\bd, K, i} }{ 
     \prod_{k=1}^K \pi_k(A_{k,i}, \bH_{k,i}; 
    \widehat{\btheta}_{k,N}) 
     } \\
    + \sum_{k=1}^K \left\lbrace 
    \frac{ \mathcal{C}_{\bd,k-1, i} }{ 
    \prod_{v=1}^{k-1}{\pi}_{v}
    (A_{v,i}, \bH_{v,i}; \widehat{\btheta}_{v,N})} 
    -
    \frac{ \mathcal{C}_{\bd,k, i} }{ 
    \prod_{v=1}^k{\pi}_{v}(
    A_{v,i}, \bH_{v,i}; \widehat{\btheta}_{v,N})} 
    \right\rbrace L_k(\overline{\bX}_{ki}) \Bigg],
\end{multline}
where $L_k(\overline{\bx}_{k})$ is an arbitrary function of $\overline{\bx}_{k}$ and we define
$\prod_{v=1}^{0}{\pi}_{v} (A_{v,i}, \bH_{v,i};
\widehat{\btheta}_{v,N}) = 1$.  
Setting
$L_k(\overline{\bx}_k) \equiv 0$ yields the IPWE 
that forms the basis for the approach of \citet{wu2021}; the IPWE uses only the observed outcomes and no covariate information.
It can be an inefficient estimator for $V(\bd)$ when there are covariates that are correlated with the outcome. 
The efficient choice for $L_k$ is
$L_k^{\bd}( \overline{\bx}_k) = E\{Y^*(\bd) | \overline{\bX}_k =
\overline{\bx}_k, \overline{\bA}_{k-1}=
\overline{\bd}_{k-1}(\overline{\bx}_{k-1}) \}$, 
where
$\overline{\bd}_k(\overline{\bx}_k)$ is defined as follows: 
\allowbreak$\overline{d}_1(\bx_1) = d_1(\bx_1)$,
\allowbreak$\overline{\bd}_2(\overline{\bx}_2) = [ d_1(\bx_1), d_2\{
\overline{\bx}_2, d_1(\bx_1)\}], \ldots,
\overline{\bd}_k(\overline{\bx}_k) = [ d_1(\bx_1), d_2\{
\overline{\bx}_2, d_1(\bx_1)\}, \ldots, d_k\{ \overline{\bx}_k,
\overline{\bd}_{k-1}(\overline{\bx}_{k-1})\} ]$, 
$k=1,\ldots, K$;
and
$\overline{\bA}_k = \overline{\bd}_k(\overline{\bX}_k)$ is the event
that 
all treatments received are consistent with $\bd$ through decision $k$. 

In practice, the functions 
$L_k^{\bd}(\overline{\bx}_k),\,k=1,\ldots,K$ are unknown, but
they can be estimated using 
$Q$-learning as follows \citep[][Section 6.4.2]{tsiatis2019dynamic}.  
Posit 
a model $Q_K(\overline{\bx}_K, \overline{\ba}_K; 
\bbeta_K)$ for
$Q_K(\overline{\bx}_{K}, \overline{\ba}_K)
= \mathbb{E}(Y|\overline{\bX}_K=\overline{\bx},
\overline{\bA}_K=\overline{\ba}_K)
$ indexed by $\bbeta_K \in \mathcal{B}_K\subseteq
\mathbb{R}^{P_{Q_K}}$.  
Obtain an estimator $\widehat{\bbeta}_{K,N}$
for $\bbeta_K$ by an appropriate regression method, e.g., least squares,
and take 
$\widehat{L}_K^{\bd}(\overline{\bx}_K) = 
Q_K\{
\overline{\bx}_{K}, \overline{\bd}_K(\overline{\bx}_K);
\widehat{\bbeta}_{K,N} \}$.  
Define the pseudo-outcomes at stage $k = K-1, \ldots, 1$ as
$Q_{k+1}^{\bd}[ \overline{\bX}_{k+1,i}, 
\left\lbrace 
\overline{\bA}_{k,i}, 
d_{k+1}(\overline{\bX}_{k+1,i}, \overline{\bA}_{k,i})
\right\rbrace; \widehat{\bbeta}_{k+1,N}]$,
the predicted outcome using the fitted model when individuals receive consistent treatments at stage $k+1$.
Then, obtain $\widehat{\bbeta}_{k,N}^\bd$ by a suitable regression method using the pseudo-outcomes as the response, e.g., for least squares, 
\begin{equation*} \label{eq:ols}
\widehat{\bbeta}_{k,N}^{\bd} = \arg\min_{\bbeta_k}
\sum_{i=1}^n\left\lbrace 
Q_{k+1}^{\bd}\left[
\overline{\bX}_{k+1,i}, 
\left\lbrace 
\overline{\bA}_{k,i}, 
d_{k+1}(\overline{\bX}_{k+1,i}, \overline{\bA}_{k,i})
\right\rbrace; \widehat{\bbeta}_{k+1,N}
\right] - 
Q_k^{\bd}\left(
\overline{\bX}_k, \overline{\bA}_k; \bbeta_k
\right)
\right\rbrace^2,
\end{equation*}
and $\widehat{L}^\bd_k(\overline{\bx}_k) = Q^\bd_{k} \{ \overline{\bx}_{k}, \overline{\bd}_{k}(\overline{\bx}_{k}); \widehat{\bbeta}_{k,N} \}$.
For individuals with only one treatment available at stages $k$ to $k'$, we use 
pseudo-outcome $Q_{k' + 1}^{\bd}[ \overline{\bX}_{k' + 1,i}, 
\left\lbrace 
\overline{\bA}_{k',i}, 
d_{k'+1}(\overline{\bX}_{k'+1,i}, \overline{\bA}_{k',i})
\right\rbrace; \widehat{\bbeta}_{k'+1,N}]$
for $k' < K - 1$ and $Y$ for $k'=K-1$ \citep[][Section 6.4.2]{tsiatis2019dynamic}. 

The estimator
(\ref{eq:AIPW}) is doubly robust, i.e., it consistently estimates
$V(\bd)$ if either of the sets of models
$\pi_k(a_k, \bh_k;\btheta_k)$, $k=1,\ldots,K$, or
$Q_K(\overline{\bx}_K, \overline{\ba}_K; \bbeta_K)$,
$Q_k^{\bd}(\overline{\bx}_k, \overline{\ba}_k;\bbeta_k)$,
$k=1,\ldots,K$, is correctly specified 
\citep{han2014, vermeulen2015, leudtke2018, tsiatis2019dynamic}.
Consistency is guaranteed in SMARTs because propensities are known.

\section{Interim AIPW estimator} \label{s:interim-est}

The interim AIPW estimator (IAIPWE) uses partial information
from individuals who have yet to complete follow up at the times interim 
analyses are conducted;
the IAIPWE includes the
IPWE and AIPWE for complete data as special cases.
Assume that the enrollment process is
independent of all subject information
and that the time between stages is fixed, as is the case for many SMARTs.
Let $S$ be the number of planned analyses. 
Let $\Gamma(t) \in \lbrace 0, 1\rbrace$ be an indicator that a
participant has enrolled in the SMART at study time $t$, where $t=0$ denotes the
start of the study (in calendar time). 
In addition, let
$\kappa(t) \in \left\lbrace 0, 1,\ldots ,K \right\rbrace$ 
be the furthest stage reached by a participant at time $t$ with
$\Gamma(t) = 0 \Rightarrow \kappa(t) = 0$; and let
$\Delta(t)\in\lbrace 0, 1\rbrace$ be an indicator that a participant
has completed follow up, 
i.e., they have completed all $K$ stages and have had their outcome ascertained.   
Thus, the number of participants enrolled at time $t$ is
$n(t) = \sum_{i=1}^{N}\Gamma_i(t)$.  
We evaluate either the fixed set of $L$ embedded regimes 
$\left\lbrace \bd^{\ell}\right\rbrace_{\ell=1}^L$ for null hypothesis \eqref{h:h0H}, 
or the embedded regimes along with a control regime, $\bd^0$, for null hypothesis \eqref{h:h0D}.  
The control regime may be estimated from a separate trial arm or may have a predetermined fixed value. 
We use superscript $\ell$ to indicate
that a quantity is being computed for regime $\bd^{\ell}$, e.g.,
$\widehat{\bbeta}_{k,N}^{\ell}$ is shorthand for
$\widehat{\bbeta}_{k,N}^{\bd^{\ell}}$.

We define the ``full data'' under regime $\bd^\ell$ as 
$W^*_{\bd^\ell} = \{ Y^*(\bd^\ell), \overline{\bX}_K^*(\bd^\ell) \}$,
which comprises the potential outcome $Y^*(\bd^\ell)$ and 
associated potential covariates $\overline{\bX}_K^*(\bd^\ell) = \{\bX_1, \bX^*_2(\bd^\ell), \ldots, \\ \bX^*_k(\bd^\ell) \}$.
The observed data for an individual at time $t$ are therefore
\allowbreak
$
W(t)= \Gamma(t)\big[ 1, \ 
\kappa(t),\ \bX_{1},\ \\
A_{1},\  I\{\kappa(t) >1\} \bX_{2}, \ 
I\{\kappa(t) >1\} A_{2},\ \ldots \ ,\    
I\{\kappa(t) >K-1\} \bX_{K},\  
I\{\kappa(t) >K-1\} A_{K},\   \Delta(t),\ 
\\ 
\Delta(t)Y  \big]. 
$
For a given time $t$ and regime $\bd^{\ell}$, let
$\Rc^\ell(t) \in \left\lbrace 1,\ldots, 2K, \infty \right\rbrace$ be a
discrete coarsening variable, which is defined as follows:
\begin{align*}
    & \Rc^{\ell}(t) = 1 &\text{ if }&\quad A_1 \ne 
        d_1^{\ell}(\bH_1) \\
    & \Rc^{\ell}(t) = 2 &\text{ if }&\quad \mathcal{C}_{\bd^{\ell}, {1}} = 1, 
    \kappa(t) = 1 \\
    & \Rc^{\ell}(t) = 3 &\text{ if }&\quad \mathcal{C}_{\bd^{\ell}, {1}} = 1,  
    \kappa(t) = 2, A_2 \ne d_2^{\ell}(\bH_2) \\
    & \Rc^{\ell}(t) = 4 &\text{ if }&\quad \mathcal{C}_{\bd^{\ell}, {2}} = 1, 
    \kappa(t) = 2 \\
    & & \quad \vdots \\
    & \Rc^{\ell}(t) = 2k - 1 &\text{ if }&\quad 
    \mathcal{C}_{\bd^{\ell}, {k-1}} 
    = 1, 
    \kappa(t) = k, A_k \ne d_k^{\ell}(\bH_k) \\
    & \Rc^{\ell}(t) = 
    2k  &\text{ if }&\quad \mathcal{C}_{\bd^{\ell}, k} = 1,
     \kappa(t) = k \\
    & & \quad \vdots \\
    & \Rc^{\ell}(t) = 2K 
    &\text{ if }&\quad \mathcal{C}_{\bd^{\ell}, {K}} = 1, 
    \kappa(t) = K, \Delta(t) \ne 1 \\
    & \Rc^{\ell}(t) = \infty &\text{ if }&\quad 
    \mathcal{C}_{\bd^{\ell}, {K}} = 1, 
    \kappa(t) = K,  \Delta(t) = 1. 
\end{align*}
Thus, $\Rc^{\ell}(t) = \infty$ corresponds to a participant having completed follow up and being consistent with $\bd^{\ell}$ for all treatment
decisions at time $t$.  For $\Rc^{\ell}(t) < \infty$, 
$\lfloor \Rc^{\ell}(t)/2\rfloor$ is the number of stages at which
a participant is consistent with $\bd^{\ell}$ at time $t$, and
$\Rc^{\ell}(t) \mod 2$ encodes whether the number of consistent stages
is due to time-related censoring, i.e., not having yet completed the current stage,
or having been assigned a treatment that is inconsistent with
$\bd^{\ell}$.
See Appendix A for an example of how $\Rc^\ell(t)$ is determined.

The observed data $W(t)$ are a coarsened version of the full data  $W^*_{\bd^\ell}$. 
The coarsening is monotone in that the full data coarsened to level $\Rc^\ell(t)=r$ are a many-to-one function of the full data coarsened to level $\Rc^\ell(t)=r+1$ at time $t$. 
Moreover, the data are coarsened at random, as
$P \{ \Rc^\ell(t) = r \vert W^*_{\bd^\ell} \} = P \{ \Rc^\ell(t) = r \vert W(t) \} $ 
(\citealp[][Chapter 7]{tsiatis2006book}; \citealp{zhang2013}), 
which follows from the consistency and sequential randomization assumptions in Section~\ref{s:background}.
Define
the coarsening hazard function
$\lambda_r^\ell(t) = P\left\lbrace \Rc^\ell(t) = r \big| \Rc^\ell(t) \geq
  r, W(t)\right\rbrace$ to be the conditional probability that an
individual is coarsened to level $r$ given that they are at risk of being coarsened.  
Because the data are coarsened at random, $\lambda_r^\ell(t)$ is a function of the observed data. 
Let the
probability that an individual is coarsened after $r$ be
$K^\ell_r(t) = P\left\lbrace \Rc^\ell(t) > r \big| W(t) \right\rbrace$, which is also a function of the observed data.
Let $\widehat{K}_{r,n(t)}^{\ell}(t)$ be an estimator of $K_r^{\ell}(t)$.
Let $\nu_{k}(t) = P\{\kappa(t) \geq k \vert \Gamma(t) =1 \}$, $k=1,\ldots,K$; 
$\nu_{K+1}(t) = P\{\Delta(t)=1 \vert \Gamma(t)=1 \}$; 
and $k(r)$ map coarsening level $r$ to corresponding decision $k$. 
We can express both $\lambda^\ell_r(t)$ and $K^\ell_r(t)$ in terms of propensities $\pi_k(A_k, \bH_k)$ and $\nu_k(t)$ for $k=1,\ldots,K+1$. 
For $\Rc^\ell(t) = r$, $r$ odd, 
$\lambda^\ell_r(t) = \pi_{k(r)}(A_{k(r)}, \bH_{k(r)})^{1 - C_{d^\ell, k(r)}} \{1-\pi_k(A_{k(r)}, \bH_{k(r)}) \}^{ C_{d^\ell, k(r)}}$ and
$K^\ell_r(t) = \nu_{k(r)}(t) \pi_{k(r)}(A_{k(r)}, \bH_{k(r)})^{C_{d^\ell, k(r)}} \{1-\pi_k(A_{k(r)}, \bH_{k(r)}) \}^{1-C_{d^\ell, k(r)}} 
\prod_{v=1}^{k(r)-1} \pi_{v}(A_{v}, \bH_{v})$.
For $\Rc^\ell(t) = r$, $r$ even, 
$\lambda^\ell_r(t) = \{ \nu_{k(r)}(t) - \nu_{k(r)+1}(t) \} / \nu_{k(r)}(t)$ and
$K^\ell_r(t) = \nu_{k(r)+1}(t) \prod_{v=1}^{k(r)} \pi_{v}(A_{v}, \bH_{v})$.
It is straightforward to posit models for $\pi_{k}(A_k, \bH_k)$ or $\nu_{k}(t)$ using logistic regression or simple averages and estimate $\lambda_r^\ell(t)$ and $K_r^\ell(t)$.
The form of the IAIPWE for regime $\bd^{\ell}$ at time
$t$ is
\begin{align} 
\widehat{V}_{\mathrm{IA}}^{\ell}(t)  &= 
n(t)^{-1} \sum_{i=1}^N \Gamma_i(t) 
\bigg[
\frac{
I\left\lbrace 
\Rc_i^{\ell}(t) = \infty
\right\rbrace 
}{
\widehat{K}^\ell_{2K, i, n(t)}(t) 
}Y_i \label{eq:IAIPWk}
\\ &\quad + 
\sum_{r=1}^{2K} \frac{
    I\left\lbrace
        \Rc^{\ell}_i(t) = r
     \right\rbrace - 
    \widehat{\lambda}^{\ell}_{r,i}(t) 
    I\left\lbrace 
        \Rc^{\ell}_{i}(t) \geq r
    \right\rbrace
    }{
    \widehat{K}^\ell_{r,i,n(t)}(t)
    }
L^{\ell}_{k(r)} \left(
\overline{\bX}_{k(r), i} \right)
\bigg], \nonumber
\end{align}
where $L_{k(r)}^{\ell}(\overline{\bx}_{k(r)})$ is an arbitrary function of $\overline{\bx}_{k(r)}$.
The estimator is doubly robust and thus guaranteed to be consistent in a SMART with a specified enrollment process.
We include a proof in Appendix B. 
Similar to the AIPWE, we estimate the efficient choice of unknown functions $L_{k(r)}^{\ell}(\overline{\bx}_k) = \mathbb{E} \{Y^*(\bd) \vert  \overline{\bX}_k = \overline{\bx}_k, \overline{\bA}_{k-1} = \overline{\bd}_{k-1}(\overline{\bx}_{k-1}) \}$ 
using $Q$-learning; however,
because the IAIPWE uses individuals with incomplete treatment trajectories, 
the $Q$-learning procedure for $L_{k(r)}(\overline{\bx}_{k(r)})$ is more complicated.
Posit a model 
$Q^\ell_K\{\overline{\bx}_K, \overline{\ba}_K; \bbeta_K(t) \}$ for $Q_K(\overline{\bx}_K, \overline{\ba}_K) 
= \mathbb{E} (Y \vert \overline{\bX}_K = \overline{\bx}_K, \overline{\bA}_K = \overline{\ba}_K)$ indexed by 
$\bbeta_K(t) \in \mathcal{B}_K\subseteq
\mathbb{R}^{P_{Q^\ell_K}}$.  
Construct an estimator $\widehat{\bbeta}_{K}$
for $\bbeta_K$ by an appropriate regression method, e.g., least squares,
using only individuals who have completed all treatment stages, i.e., $\Delta(t) = 1$,
and subsequently take $\widehat{L}^\ell_K(\overline{\bx}_K) = 
Q^\ell_K\{
\overline{\bx}_{K}, \overline{\bd}_K(\overline{\bx}_K);
\widehat{\bbeta}_{K} \}$.
Posit models 
$Q^\ell_{k(r)}(\overline{\bx}_{k(r)}, \overline{\ba}_{k(r)}; \bbeta_{k(r)})$
for
\allowbreak$
Q^\ell_{k(r)} (\overline{\bx}_{k(r)}, \overline{\ba}_{k(r)})
= 
\mathbb{E}\bigl(
Q^\ell_{k(r)+1}
\bigl[
\overline{\bX}_{k(r)+1}, 
\bigl\lbrace 
\overline{\bA}_{k(r)}, 
d_{k(r)+1}(\overline{\bX}_{k(r)+1}, \overline{\bA}_{k(r)})
\bigr\rbrace
\bigr]  \big| 
\overline{\bX}_{k(r)} = \overline{\bx}_{k(r)},
\overline{\bA}_{k(r)} = \overline{\ba}_{k(r)}
\bigr)
$
for $k(r)=K-1,\ldots,1$.
Estimating $\widehat{\bbeta}_{k(r)}$ requires pseudo-outcomes, which may be missing when using individuals who have been observed through stage $k(r) + 1$, i.e., $\kappa(t) > k(r)$, but have no observed outcome $Y$ or estimable pseudo-outcome from stages $k(r)+2$ or later.
In such cases, we define the pseudo-outcomes for estimating $\widehat{\bbeta}_{k(r)}$ as
\begin{align*}
    \Tilde{Q}&^\ell_{k(r)+1}(\overline{\bx}_{k(r)+1}, \overline{\ba}_{k(r)+1}; \widehat{\bbeta}_{k(r)+1}, \ldots, \widehat{\bbeta}_{K})= 
    \bigl[ I \{ \vert \Psi_{k(r)+1}(\bh_{k(r)+1}) \vert \ne 1 \}  \\
    &  + I \{\kappa(t) = k(r) +1 , \Delta(t) = 0, \vert \Psi_{k(r)+1}(\bh_{k(r)+1}) \vert = 1   \} 
    \bigr] Q^\ell_{k(r)+1}(\overline{\bx}_{k(r)+1}, \overline{\ba}_{k(r)+1}; \widehat{\bbeta}_{k(r)+1}) \\
    &  + I \{ \kappa(t) = k(r) +2, \Delta(t) = 0, \vert \Psi_{k(r)+1}(\bh_{k(r)+1}) \vert = 1 \} Q^\ell_{k(r)+2}(\overline{\bx}_{k(r)+2}, \overline{\ba}_{k(r)+2}; \widehat{\bbeta}_{k(r)+2}) \\
    & + \ldots + I \{\Delta(t) = 1, \vert \Psi_{k(r)+1}(\bh_{k(r)+1}) \vert = 1 \}  Y.
\end{align*}
This approach uses individuals with incomplete information to fit the $Q$-functions for greater efficiency.
When all observed individuals have completed their regimes, this strategy is equivalent to the pseudo-outcome method outlined in Section~\ref{s:aipw_est}. 
We obtain $\widehat{\bbeta}^\ell_{k(r)}$ by a suitable regression method, 
using
$Q^\ell_{k(r)+1}(\overline{\bx}_{k(r)+1}, \overline{\ba}_{k(r)+1}; \widehat{\bbeta}_{k(r)+1})$ with 
\allowbreak$\Tilde{Q}^\ell_{k(r) +1}(\overline{\bx}_{k(r)+1}, \overline{\ba}_{k(r)+1}; \widehat{\bbeta}_{k(r)+1}, \\ \ldots, \widehat{\bbeta}_{K})$ 
when necessary,
and 
\allowbreak$L^\ell_{k(r)}(\overline{\bx}_{k(r)}) 
= Q^\ell_{k(r)}(\overline{\bx}_{k(r)}, \overline{\ba}_{k(r)}; \widehat{\bbeta}_{k(r)})$. 

To make clear the connection between the IAIPWE and the (A)IPWE, we express
$\widehat{V}_{\mathrm{IA}}^{\ell}(t) $ in (\ref{eq:IAIPWk}) in an alternate form.  
For definiteness,
consider $K=2$ decisions at fixed times, and 
let $\nu_2(t)$ and $\nu_3(t)$ be estimated by
$\widehat{\nu}_{2, n(t)}(t) = \sum_{i=1}^N I\{ \kappa_i(t) = 2 \} / \sum_{i=1}^N \Gamma_i(t)$, and
$\widehat{\nu}_{3, n(t)}(t) = \sum_{i=1}^N \Delta_i(t) / \sum_{i=1}^N \Gamma_i(t)$.  
It is shown in Appendix A that in this case 
(\ref{eq:IAIPWk}) is equivalent to 
\begin{align}
\widehat{V}_{\mathrm{IA}}^{\ell}(t) &=
n(t)^{-1} \sum_{i=1}^N \Gamma_i(t) \bigg( 
\frac{ \Delta_i(t) \mathcal{C}^\ell_{2,i} Y_i}{\pi_1(A_{1,i},
  \bH_{1,i}; \widehat{\btheta}_{1,n(t)}) \pi_2(A_{2,i}, \bH_{2,i};
  \widehat{\btheta}_{2,n(t)}) \widehat{\nu}_{3,n(t)}(t)} \nonumber\\
&- \bigg[ \frac{ I\{ A_{1i}=d^\ell_1(\bH_{1i})\} I\{\kappa_i(t)=2\} }{\pi_1(A_{1,i},
  \bH_{1,i}; \widehat{\btheta}_{1,n(t)}) \widehat{\nu}_{2,n(t)}(t)} -1\bigg]
L^{\ell}_1(\overline{\bX}_{1i}) \label{eq: vhat}\\
&- \frac{ I\{ A_{1i}=d^\ell_1(\bH_{1i})\} I\{\kappa_i(t)=2\} }{\pi_1(A_{1,i},
  \bH_{1,i}; \widehat{\btheta}_{1,n(t)}) \widehat{\nu}_{2, n(t)}(t)} 
\bigg[ \frac{I\{ A_{2i}=d^\ell_2(\bH_{2i})\} \Delta_i(t)\widehat{\nu}_{2,n(t)}(t) }{\pi_2(A_{2,i}, \bH_{2,i};
  \widehat{\btheta}_{2,n(t)}) \widehat{\nu}_{3,n(t)}(t)} -1\bigg]
L^{\ell}_2(\overline{\bX}_{2i}) 
\bigg).  \nonumber
\end{align}
If $\Gamma_i(t) = \Delta_i(t) = 1$ for all $i$, so that $n(t) = N$, as
at the time of the final analysis, (\ref{eq: vhat}) reduces to the
AIPWE (\ref{eq:AIPW}) with $K=2$.  
    The augmentation terms in (\ref{eq:IAIPWk}) use partial
      information from participants who are enrolled at the time of an
      interim analysis but who do not yet have complete follow up.  In
      contrast, the IPWE (obtained by setting
      $L_k^\ell(\overline{\bX}_{ki})\equiv 0$, $k=1,2,\ldots,K$) uses data only
      from those subjects who are consistent with the regime under
      consideration at all stages of the study and who have completed
      the trial.  The AIPWE (\ref{eq:AIPW}) also uses information only from
      subjects who have completed the trial, but it additionally uses a series
      of regression models, one for each stage, to impute information
      for subjects who are not consistent with the regime under
      consideration starting from the stage at which their treatment
      first deviates from the regime.  The IAIPWE (\ref{eq:IAIPWk}) furthermore uses data from
      all subjects in fitting the regression models in the AIPWE and
      thereby uses more information and further improves efficiency.

As our goal is to use the IAIPWE for interim monitoring and analyses, we need to characterize its sampling distribution. 
The following result shows that the IAIPWE for the embedded regimes is asymptotically normal; 
we use this result to construct tests and decision boundaries in subsequent sections. 
A proof is given in the Appendix C. 
\begin{theorem} \label{thm:asy_norm}
Let $\widehat{\mathcal{V}}(t) = \{\widehat{V}_{\mathrm{IA}}^0(t),
\widehat{V}_{\mathrm{IA}}^1(t),
\ldots, \widehat{V}_{\mathrm{IA}}^L(t)\}^{\top}$ be the stacked value estimators at time $t$ across all regimes, and let $n(t) / N \overset{p}{\to} c$, a constant. 
Under standard regularity conditions stated in the Appendix C, 
$N^{1/2} \{ \widehat{\mathcal{V}}(t) - {\mathcal{V}}(t) \} \overset{d}{\to} \mathcal{N}(\0, \bSigma)$ as $N \rightarrow \infty$.
\end{theorem}
A consistent estimator $\widehat{\bSigma}$ of $\bSigma$ can be obtained using the sandwich estimator or the bootstrap. 
Comparisons among the $L+1$ regimes can be constructed using a contrast vector and are asymptotically normal via a simple Taylor series argument (see Appendix C). 
When there is no control regime, $\mathcal{V}(t)$ is indexed only by $\ell=1,\ldots,L$.

\section{Interim analysis for SMARTs} \label{s:interim}

\subsection{Hypothesis testing} \label{s:hypothesis_testing}

For simplicity, consider  
$S=2$ planned analyses at study times $t_1$ (interim
analysis) and $t_2$ (final analysis).
We present the extension to an arbitrary $S$ in Appendix D.
We discuss the interim analysis procedure in the
context of superiority; the procedure for homogeneity follows under
minor modifications.

Define the test statistics at analysis time $t_s$, $$Z^\ell(t_s) = \{ \widehat{V}_{\mathrm{IA}}^{\ell}(t_s) -
\widehat{V}_{\mathrm{IA}}^0(t_s)\}/\mathrm{SE}\{ \widehat{V}_{\mathrm{IA}}^{\ell}(t_s)
-\widehat{V}_{\mathrm{IA}}^0(t_s)\}, \hspace*{0.1in} \ell = 1, \ldots, L,$$ where
$\widehat{V}_{\mathrm{IA}}^0(t_s)$ can be estimated as the sample average of response $Y_i$ for individuals receiving $\bd^0$ and
the denominator is obtained from the approximate normal sampling
distribution for $\widehat{\mathcal{V}}(t_s)$ in Theorem~\ref{thm:asy_norm}.
If regime means are compared to a fixed control value
$V^0$, replace $\widehat{V}_{\mathrm{IA}}^0(t_s)$ by $V^0$.  
At each analysis $s$, we propose to stop the trial if any test statistic exceeds a stopping boundary $c_s(\alpha)$, which will be discussed in the next section. 
Heuristically, the testing
procedure at significance level $\alpha$ across all $t_s$ is as
follows.
\begin{itemize}
\item[(1)] At time $t_1$, compute $Z^\ell(t_1)$, $\ell = 1,\ldots,L$.
  If $Z^\ell(t_1) > c_{\alpha}(1)$, for any $\ell$, reject $H_{0}$ and terminate the trial; else,
  continue the trial.
\item[(2)] At time $t_2$, compute $Z^\ell(t_2)$, $\ell = 1,\ldots,L$.
  If $Z^\ell(t_2) > c_{\alpha}(2)$ for any $\ell$. 
  reject $H_0$; otherwise, fail to reject $H_0$. Terminate the trial.
\end{itemize}
A trial with more than two planned analysis repeats step (1) for
all interim analyses, terminating when a test statistic is greater
than the corresponding stopping boundary. 

This formulation can be adapted to any set of hypotheses involving
functions of the values of regimes of interest. 
For example, testing
the homogeneity hypothesis \eqref{h:h0H} 
would involve calculation of
chi-square test statistics based on the distributions of
$\widehat{\mathcal{V}}(t_s)$, $s=1,2$, analogous to \citet{wu2021},
which would be compared to corresponding stopping boundaries.

\subsection{Stopping boundaries} \label{s:stop_bound}

We discuss boundary selection and sample size
calculations for superiority null hypothesis \eqref{h:h0D}, 
which involves multiple comparisons
of embedded regimes against a control regime.  
We seek to determine
stopping boundaries $c_\alpha(s)$, $s=1, 2$, that control the
family-wise error rate across all planned analyses at level $\alpha$; i.e.,
\begin{equation} \label{eq:bound_cup}
    \pr \lbrace \text{Reject}\  H_{0D} \vert H_{0D} \ \text{is true} \rbrace
    = \pr \left[ \left.\bigcup_{\ell=1}^L \bigcup_{s=1}^2 \lbrace Z^\ell(t_s)
      \geq c_{\alpha}(s) \rbrace  \right| H_{0D} \right] \leq \alpha.
\end{equation}
Common approaches to calculating boundaries that satisfy (\ref{eq:bound_cup}) include the Pocock boundary,
which takes $c_{\alpha}(s) = c_{\alpha}$ for some $c_\alpha$ for
$s=1,2$ \citep{pocock1977}; the O'Brien-Fleming (OBF) boundary
$\left\lbrace 
c_{\alpha}(1), c_{\alpha}(2)\right\rbrace = \{ \iota c_{\alpha}, c_{\alpha}\}$
\citep{of1979}, where $\iota$ is the reciprocal of the square root of the statistical information (e.g., inverse of the variance of the numerator of the associated $Z$-score) available at analysis $s$ divided by the statistical information available at final analysis $S$; or the
broader $\alpha$-spending approach \citep{demets1994}.  
If the information proportion between the interim and final analysis varies by regimes, practitioners may elect to use a regime-dependent $\iota^\ell$ in the spirit of OBF. 
For a detailed discussion about if and when each boundary type might be preferable, see \citet[]{jennison1999}.

Define the stacked vector of sequential test statistics
\begin{equation} \label{eq:z_stats}
    \bZ = \{  Z^1(t_1),\ldots, Z^L(t_1), Z^1(t_2),\ldots,
    Z^L(t_2) \}^{\top}.
\end{equation}
Boundaries that satisfy \eqref{eq:bound_cup} can
be obtained via the joint cumulative distribution function of $\bZ$
under null hypothesis \eqref{h:h0D}.   
\begin{theorem} \label{thm:zdist_joint}
Under the null hypothesis  \eqref{h:h0D} 
and for $n(t)/N \overset{p}{\to} c$, a constant,  
the test statistics $\bZ$ satisfy
$\bZ \overset{d}{\to} \mathcal{N}(\0, \bSigma_{H_0})$
where $\bSigma_{H_0}$ is a block diagonal matrix 
with diagonal entries 
$\mathrm{Corr}\{ Z^1(t_s),\ldots, Z^L(t_s)\}$
and off-diagonal entries 
$\iota^{-1} \mathrm{Corr}\{ Z^1(t_s),\ldots, Z^L(t_s)\} $,
$\iota$ is the reciprocal of the information proportion between interim analysis $s$ and final analysis $S$. 
\end{theorem}
A proof of Theorem~\ref{thm:zdist_joint} and discussion on calculating $\iota$ and the correlation between the $Z$-statistics are provided in the Appendix E. 
In practice, computation of $c_{\alpha}$ can be done numerically.  
Either the correlation of the test
statistics or the variance of all components of the estimator must be
specified to compute the stopping boundaries. 
We approximate $c_{\alpha}$ through integration of the
corresponding multivariate normal distribution of $\bZ$.
Under the information monitoring approach \citep{tsiatis2006inf}, the
correlation between sequential test statistics for the same regime simplifies
to the square root of the ratio of the information available between the two
time points.  
Because of incomplete information for participants enrolled but who have not yet completed the trial, this quantity does not simplify to the square root of the ratio of the interim sample size to the final planned sample size.
The off-diagonal elements of the covariance matrix, $\bSigma_{H_0}$, may be non-zero for overlapping embedded regimes. 
For these reasons, it may be difficult to specify $\bSigma_{H_0}$. 
An alternative is to specify generative models for
the observed data, i.e., a mean model and distributions for associated covariates, propensities, and enrollment at time of interim analyses, and estimate the correlation structure empirically via simulation.

The choice of the models and estimators for
$\lambda_r^\ell(t)$, $K_r^\ell(t)$, and $L_{k(r)}^\ell(\overline{\bx}_{k(r)})$ impact the correlation structure of $\bSigma_{H_0}$ and can
result in correlated value estimators across non-overlapping embedded
regimes; i.e., regimes that involve different stage 1 treatment
options.  If cohorts enroll sequentially and interim analyses are
planned such that all enrollment occurs within each cohort (i.e.,
$\Delta_i(t_s) = \Gamma_i(t_s)$ for all $i$ for $s=1,2$), then the
test statistics at each analysis use the standard AIPWE
(\ref{eq:AIPW}) computed using data from all participants who have entered the trial.
Therefore, stopping boundaries for trials with such enrollment
procedures are subsumed by this method.

\subsection{Power and sample size}

With stopping boundaries 
$\mathbf{c}_{\alpha} = \{c_{\alpha}(1) \boldsymbol{1}_{L}, c_{\alpha}(2) \boldsymbol{1}_{L} \} \in \R^{2L}$ for $\boldsymbol{1}_{L}$ an L-vector of ones,
and
specified alternative $H_A$, the power of the testing procedure is
\begin{equation*}
\pr \lbrace \text{Reject}\  H_{0D} \vert H_A \ \text{is true} \rbrace
= \pr \left[ \left.\bigcup_{\ell=1}^L \bigcup_{s=1}^2 \lbrace Z^\ell(t_s)
  \geq c_{\alpha}^{\ell}(s) \rbrace  \right| H_A \right] = 1-\beta. 
\end{equation*}
The power of the test under $H_A$, where $\bZ$ has expected value
$\bmu_A = \bmu_A\left\lbrace n(t_1), n(t_2)\right\rbrace$ and covariance $\bSigma_{H_A}$, is approximately
\begin{equation} \label{eq:power}
    1- \int \dots \int_D \frac{1}{(2\pi)^{L} 
    \mathrm{det}(\bSigma_{H_A})^{-1/2}} \mathrm{exp} \lbrace -\frac{1}{2} (\bZ-\boldsymbol{\mu}_A)^{\top} \bSigma_{H_A}^{-1} (\bZ-\boldsymbol{\mu}_A) \rbrace 
    dz^1(1)dz^2(1) \cdots dz^L(2)
\end{equation}
for domain
$D=(-\infty, c_{\alpha}(1)] \times (-\infty, c_{\alpha}(1)]
\times \dots \times (-\infty, c_{\alpha}(2)]$.
As the mean under the alternative, $\boldsymbol{\mu}_A$, is a function of the sample size, so too is
\eqref{eq:power}. Thus, to achieve nominal power $100(1-\beta)$\%, one can
set \eqref{eq:power} equal to $1-\beta$ and solve for the sample size. 
Although our results hold for a general alternative hypothesis $H_A$, we proceed under the simplifying assumption that $\bSigma_{H_0} = \bSigma_{H_A}$, i.e., that the covariance is the same under $H_{0D}$ and $H_{A}$. 
In our implementation, we use a grid search for a fixed enrollment
process and ratio between interim sample sizes to find the total planned sample size $N$ that
attains the desired power.
When the augmentation terms are zero, the analyst must specify the correlation among estimators of the regimes, the information proportion for analyses, the alternative mean outcomes, and the variance of the mean outcomes. 
When augmentation terms are non-zero, all generative models must be specified to determine the sample size and corresponding power.

Specification of all generative models required for the IAIPWE at the design stage may be challenging.  
Accordingly, a practical strategy would be to power the trial and thus determine $N$ conservatively based on the IPWE but base interim analyses on the more efficient IAIPWE, which can lead to increased power and smaller expected sample size.  

As previously stated, the covariance structure, $\bSigma_{H_0}$, depends on the enrollment process through the information proportion at the time of analysis.
Thus, one can compute the
maximum power for a fixed
sample size under differing enrollment processes using
\eqref{eq:power} adjusted for the differences in the information proportion at the time of the analysis. 
One can also consider other objectives such as minimizing the time to
decision or the cost of the trial using these same procedures.

\subsection{Test for homogeneity} \label{s:testhomog}

Exploiting the previous developments, we formulate
a sequential testing procedure using $\bZ(t_s) = \{Z^1(t_s),\ldots,Z^L(t_s)\}$ for the global null hypothesis \eqref{h:h0H}, i.e., that all regimes are equal.
We derive a $\chi^2$ statistics using Theorem \ref{thm:zdist_joint}. 
Let $\bC  = [\mathbf{I}_{L-1} \vert -\boldsymbol{1}_{L-1} ]$ where $\mathbf{I}_q$ is the $(q \times q)$ 
identity matrix and $\boldsymbol{1}_q$ a $q$-vector of ones.  
Let $\bSigma_{H_0}(t_s)$ be the $(L \times L)$ submatrix of $\bSigma_{H_0}$ corresponding to the covariance of $\bZ(t_s)$,
and let $\boldsymbol{\mu}_A(t_s)$ be the $(L \times 1)$ vector corresponding to the alternative mean at time $t_s$. 
The sequential Wald-type test statistic at time $t_s$ is
\begin{equation} \label{eq:chi_sq_statistic}
    T_{\chi^2, \upsilon}(t_s) = \bZ^{\top}(t_s) \bC^{\top} \lbrace \bC \bSigma_{H_0}(t_s) \bC^{\top} \rbrace ^{-} \bC \bZ(t_s) ,
\end{equation}
which follows a $\chi^2$ distribution with degrees of freedom $\upsilon = \mathrm{rank}\lbrace \bC \bSigma_{H_0}(t_s) \bC^{\top} \rbrace$ and non-centrality parameter $\phi_A = \boldsymbol{\mu}_A(t_s) ^\top \lbrace \bC \bSigma_{H_0}(t_s) \bC^{\top} \rbrace ^{-}\boldsymbol{\mu}_A(t_s)$. 
Following the methods in previous sections, the stopping boundaries now come from a $\chi^2$ distribution.
Using simulation, we estimate the stopping boundaries using the correlation structure of $\bZ$ such that $\lbrace c_{\alpha}(1), c_\alpha(2) \rbrace$ satisfy the type I error rate. 
The Pocock boundaries still satisfy
$c_{\alpha}(s) = c_{\alpha}$;
however, the OBF type boundaries satisfy 
$\lbrace c_{\alpha}(1), c_{\alpha}(2) \rbrace = \lbrace \iota^2 c_{\alpha}, c_{\alpha} \rbrace$ with $\iota$ as defined in Section \ref{s:stop_bound}.

After calculating the stopping boundaries, we use the distribution of $\bZ$ for relevant power and sample size calculations. 
We estimate the total planned sample size required to attain power $1-\beta$ numerically; see Appendix F for details on implementation.

\section{Simulation experiments} \label{s:sim}

We report on extensive simulations to evaluate the performance of the IAIPWE.
In our simulation settings, IPWE corresponds to the proposed method of \citet{wu2021}. 
We present results 
here based on $1000$ Monte Carlo replications 
for the schema shown in Figure~\ref{fig:PCST-schema}. 
We evaluate the type I error rates, power, and expected sample sizes for fixed interim analysis times for
the null hypothesis
$H_{0D}$ in
(\ref{h:h0D})
and
alternative hypothesis $H_{AD}: V(\bd^\ell) - V(\bd^0) > \delta$ for at least one $\ell$. 
We also investigate the benefit of leveraging partial information through the IAIPWE over an IPWE in trials with sample size determined by the IPWE.
Finally, we consider how the proportion of enrolled individuals having reached different stages of the trial at an interim analysis affects performance.
We consider both Pocock and OBF boundaries.  
We use correctly specified $Q$-functions for augmented estimators. 
Appendix G includes results for additional schema and settings; the results are qualitatively similar.

We generate data with a dependence between history and outcomes and explore the impact of the enrollment process on interim analyses. 
We generate two baseline covariates
$X_{1,1} \sim \mathrm{Normal}(47.5, 64)$ and 
$X_{1,2} \sim \mathrm{Bernoulli}(0.5)$ as well as 
an interim outcome 
$X_{2,1} \sim \mathrm{Normal}(1.25 X_{1,1}, 9)$.
We simulate the response status 
$R_2 \sim \mathrm{Bernoulli} 
\lbrace \mathrm{expit}(0.01 X_{1,1} + 0.02 X_{1,2} -0.008 X_{2,1}) \rbrace$
where $\mathrm{expit}(u) = e^u / (1 + e^u)$.
Individuals at stage one and responders at stage two are randomized with equal probability to feasible treatments.   The outcome is normally distributed 
with variance 100 and mean 
\begin{align*}
\mu_{S2}&(\overline{\bX}_2, \overline{\bA}_2) = 
I\left\lbrace
A_1 = 0
\right\rbrace \big\lbrace 
\beta_0 + \beta_1X_{1,1} + \beta_2X_{1,2} + \beta_3R_{2}X_{2,1}
+ \beta_4(1-R_2)X_{2,1} \\  
&+ R_2A_2\left(
\beta_5 + \beta_6X_{1,1} + \beta_{7}X_{1,2} + \beta_{8}X_{2,1}
\right) \\
&+ (1-R_2)A_2\left(
\beta_{9} + \beta_{10}X_{1,1} + \beta_{11}X_{1,2} + \beta_{12}X_{2,1}
\right)
\big\rbrace \\ 
&+ I\left\lbrace
A_1 = 0
\right\rbrace \big\lbrace 
\beta_{13} + \beta_{14}X_{1,1} + \beta_{15}X_{1,2} + \beta_{16}R_{2}X_{2,1}
+ \beta_{17}(1-R_2)X_{2,1}  \\
 &+R_2A_2\left(
\beta_{18} + \beta_{19}X_{1,1} + \beta_{20}X_{1,2} + \beta_{21}X_{2,1}
\right)  \\
&+ (1-R_2)A_2\left(
\beta_{22} + \beta_{23}X_{1,1} + \beta_{24}X_{1,2} + \beta_{25}X_{2,1}
\right)
\big\rbrace.
\end{align*}
In the first scenario, we perform an interim analysis at day $500$, and
enrollment times are drawn uniformly between 0 and 1000 days with follow-up
times every 100 days.
We define three value patterns (VPs): 
(VP1), all embedded regimes have value $47.5$;
(VP2), regimes $\ell = 1,\ldots,8$ have values $(49.5, 49.5, 49.5, 49.5, 47.5, 47.5, 47.5, 47.5)$;
and (VP3), regimes $\ell = 1,\ldots,8$ have values $(50.5, 49.0, 49.0, 47.5, 47.5, 47.5, 47.5, 47.5)$.
In each case, $\bbeta$
is chosen to achieve these VPs.
We use the sample size determined to achieve power $80\%$ under a specified VP and estimator. 
This allows us to investigate the performance of the estimators for different alternatives. 
In $H_{0D}$ and $H_{AD}$, $V(\bd^0)$ is a fixed control value equal to $47.5$ and $\delta=0$.

Table \ref{tab:schema2_main3_adj} summarizes the total planned sample size to achieve power $80\%$ under a specified alternative (VP2 for VP1 and VP2, and VP3 for VP3), the proportion of early rejections of null \eqref{h:h0D},
the proportion of total rejections of null \eqref{h:h0D},
the expected sample size, and the expected stopping time.
Results are given for both the total sample size to achieve the desired power for each individual estimator (a) and for the total sample size for the IPWE to achieve the desired power (b). 
The slight differences among the total planned sample size in (a) and (b) are due to Monte Carlo error. 
All estimators achieve nominal power and type I error rate. 
The IAIPWE requires a smaller total planned sample size to achieve nominal power. 
The IAIPWE also exhibits the highest early rejection rate under true alternatives demonstrating the efficiency gain from the augmentation terms and therefore lower expected sample sizes and earlier expected stopping times. 
The AIPWE slightly underperforms relative to the IPWE due to the overestimation of variance using the sandwich matrix for small $n(t_1)$.
It is well known that the performance of the sandwich matrix can deteriorate for small samples. 
As such, alternative estimation of the covariance matrix, such as using the empirical bootstrap, can be used.  
The IAIPWE is less affected by overestimation of the variance than the AIPWE. 
When the total sample size is selected based on the IPWE and an augmented estimator is used, the type I error rate is controlled and the study achieves a higher power. 

\begin{table}[ht!]
\setlength\tabcolsep{5pt}  

\centering
\caption{\label{tab:schema2_main3_adj}
    For the schema in Figure~\ref{fig:PCST-schema}, interim analysis performance results for testing hypothesis \eqref{h:h0D} against $H_{AD}$ with a fixed control value using Pocock and OBF boundaries. 
    VP indicates the true value pattern.
    Method indicates the estimator used. 
    The total planned sample size $N$ is determined by either each method (a) or by the IPWE (b). 
    Total planned sample sizes are determined to maintain a nominal type I error rate of $\alpha=0.05$ and achieve a power of $80\%$ under the respective value patterns, using alternative (VP2) to determine the sample size for the null (VP1).  
    Early Reject and Total Reject are the rejection rates at the first analysis and for the overall procedure, respectively. 
    $\mathbb{E}$(SS) is the expected sample size, i.e, the average number of individuals enrolled in the trial when the trial is completed. 
    $\mathbb{E}$(Stop) is the expected stopping time, i.e., the average number of days that the trial ran. 
    Monte Carlo standard deviations are given in parentheses. 
}

\resizebox{.98\textwidth}{!}{%
 \begin{tabular}{p{1em}l lp{3em}p{3em}ll l lp{3em}p{3em}ll}
    \toprule
    \toprule   
    & & \multicolumn{5}{c}{(a) $N$ Based on Method} & & \multicolumn{5}{c}{(b) $N$ Based on IPWE} \\

    VP & Method & $N$ & Early Reject & Total Reject & $\mathbb{E}$(SS) & $\mathbb{E}$(Stop) & & $N$ & Early Reject & Total Reject & $\mathbb{E}$(SS) & $\mathbb{E}$(Stop)   \\
    \hline
    1 & IPWE & 1049 & & 0.076 & 1049 (0) & 1199 (1) & & 1051 & & 0.059 & 1051 (0) & 1199 (1) \\
    1 & AIPWE & 758 & & 0.042 & 758 (0) & 1199 (1) & & 1051 & & 0.049 & 1051 (0) & 1199 (1) \\
    2 & IPWE & 1049 & & 0.795 & 1049 (0) & 1199 (1) & & 1051 & & 0.781 & 1051 (0) & 1199 (1) \\
    2 & AIPWE & 758 & & 0.814 & 758 (0) & 1199 (1) & & 1051 & & 0.908 & 1051 (0) & 1199 (1) \\
    3 & IPWE & 873 & & 0.833 & 873 (0) & 1199 (1) & & 873 & & 0.833 & 873 (0) & 1199 (1) \\
    3 & AIPWE & 586 & & 0.801 & 586 (0) & 1198 (2) & & 873 & & 0.953 & 873 (0) & 1199 (1) \\
    \multicolumn{3}{l}{Pocock}&  \\
    1 & IPWE & 1212 & 0.040 & 0.072 & 1188 (119) & 1171 (137) & & 1213 &  0.041 & 0.071 & 1188 (120) & 1171 (138) \\
    1 & AIPWE & 872 & 0.024 & 0.042 & 861 (67) & 1182 (107) & & 1213 &  0.030 & 0.043 & 1195 (103) & 1178 (119) \\
    1 & IAIPWE & 869 & 0.032 & 0.049 & 855 (77) & 1177 (123) & & 1213 &  0.037 & 0.059 & 1191 (112) & 1174 (130) \\
    2 & IPWE & 1212 & 0.321 & 0.799 & 1017 (283) &  975 (327) & & 1213 &  0.299 & 0.797 & 1032 (277) & 990 (320) \\
    2 & AIPWE & 872 & 0.256 & 0.800 & 760 (191) & 1020 (305) & & 1213 &  0.339 & 0.912 & 1008 (286) & 962 (331) \\
    2 & IAIPWE & 869 & 0.322 & 0.801 & 729 (203) & 974 (327) & & 1213 &  0.399 & 0.915 & 972 (296) & 920 (343) \\
    3 & IPWE & 987 & 0.297 & 0.835 & 841 (225) & 984 (323) & & 987 &  0.297 & 0.835 & 841 (225) & 984 (323) \\
    3 & AIPWE & 663 & 0.236 & 0.825 & 585 (140) & 1034 (297) & & 987 &  0.364 & 0.950 & 808 (237) & 945 (337) \\
    3 & IAIPWE & 660 & 0.290 & 0.822 & 565 (149) & 996 (317) & & 987 &  0.434 & 0.953 & 774 (244) & 896 (347) \\
    \multicolumn{3}{l}{O'Brien Fleming} & \\
    1 & IPWE & 1052 & 0.000 & 0.071 & 1052 (0) & 1199 (1) & & 1051 &  0.000 & 0.059 & 1051 (0) & 1199 (1) \\
    1 & AIPWE & 758 & 0.000 & 0.042 & 758 (0) & 1199 (1) & & 1051 &  0.000 & 0.050 & 1051 (0) & 1199 (1) \\
    1 & IAIPWE & 756 & 0.000 & 0.043 & 756 (0) & 1199 (1) & & 1051 &  0.000 & 0.050 & 1051 (0) & 1199 (1) \\
    2 & IPWE & 1052 & 0.008 & 0.795 & 1048 (47) & 1194 (62) & & 1051 &  0.006 & 0.781 & 1048 (40) & 1195 (54) \\
    2 & AIPWE & 758 & 0.002 & 0.813 & 757 (17) & 1197 (31) & & 1051 &  0.005 & 0.908 & 1048 (37) & 1196 (49) \\
    2 & IAIPWE & 756 & 0.014 & 0.811 & 751 (44) & 1189 (82) & & 1051 &  0.013 & 0.908 & 1044 (59) & 1190 (79) \\
    3 & IPWE & 873 & 0.012 & 0.833 & 868 (48) & 1191 (76) & & 873 &  0.012 & 0.833 & 868 (48) & 1191 (76) \\
    3 & AIPWE & 585 & 0.003 & 0.801 & 584 (16) & 1196 (38) & & 873 &  0.004 & 0.954 & 871 (28) & 1196 (44) \\
    3 & IAIPWE & 586 & 0.013 & 0.802 & 582 (34) & 1189 (79) & & 873 &  0.021 & 0.954 & 864 (28) & 1184 (100) \\
    \bottomrule
    \end{tabular}
}%

\end{table}

Table \ref{tab:schema2_pocock_value_ests} summarizes 
estimation performance at the interim and final analyses, where a mean square error (MSE) greater than one implies that the indicated estimator is more efficient than the IPWE.
The estimators are all consistent as expected.
Both the AIPWE and IAIPWE are more efficient than the IPWE at both analyses, and the IAIPWE is more efficient than the AIPWE. 
At the interim analysis, the standard errors for the IPWE underestimate the sampling variation in most cases, whereas the standard errors for the AIPWE overestimate the sampling variation. 
The IAIPWE consistently estimates the sampling variation with the exception of regime 6 at the interim analysis.

\begin{table}[ht!]
\setlength\tabcolsep{5pt} 
    \centering
        \caption{\label{tab:schema2_pocock_value_ests}
     For the schema in Figure~\ref{fig:PCST-schema}, interim analysis performance results for testing hypothesis \eqref{h:h0D} against $H_{AD}$ with a fixed control value under Pocock Boundaries under (VP2) and sample size $N$ based on the method.
     MC Mean is the Monte Carlo mean of the estimates, 
     MC SD is the Monte Carlo standard deviation of estimates,
     ASE is the Monte Carlo mean of the standard errors, and 
     MSE Ratio is the ratio of the Monte Carlo mean square error for the IPWE divided by that of the indicated estimator for the tree estimates at the interim analysis (a) and final analysis (b) for $B=1000$ simulations. 
     The true values under (VP2) for regimes $(1,\ldots,8)$ are $(49.5, 49.5, 49.5, 49.5, 47.5, 47.5, 47.5, 47.5)$.
    }
\resizebox{.98\textwidth}{!}{%
    \begin{tabular}{ll llll l llll}
    \toprule
    \toprule
     & & \multicolumn{4}{c}{(a) Interim Analysis} & & \multicolumn{4}{c}{(b) Final Analysis} \\ 
    Method & Regime & MC Mean & MC SD & ASE & MSE Ratio & & MC Mean & MC SD & ASE & MSE Ratio \\
    \hline 
IPWE & 1 & 49.47 & 1.52 & 1.44 & 1.00 & & 49.50 & 0.80 & 0.79 & 1.00 \\ 
IPWE & 2 & 49.52 & 1.51 & 1.44 & 1.00 & & 49.52 & 0.84 & 0.79 & 1.00 \\ 
IPWE & 3 & 49.48 & 1.47 & 1.44 & 1.00 & & 49.48 & 0.78 & 0.79 & 1.00 \\ 
IPWE & 4 & 49.53 & 1.48 & 1.44 & 1.00 & & 49.51 & 0.82 & 0.79 & 1.00 \\ 
IPWE & 5 & 47.51 & 1.47 & 1.43 & 1.00 & & 47.51 & 0.82 & 0.79 & 1.00 \\ 
IPWE & 6 & 47.55 & 1.43 & 1.44 & 1.00 & & 47.55 & 0.78 & 0.79 & 1.00 \\ 
IPWE & 7 & 47.48 & 1.48 & 1.44 & 1.00 & & 47.49 & 0.80 & 0.79 & 1.00 \\ 
IPWE & 8 & 47.52 & 1.47 & 1.44 & 1.00 & & 47.53 & 0.76 & 0.79 & 1.00 \\ 
AIPWE & 1 & 49.47 & 1.43 & 1.48 & 1.14 & & 49.48 & 0.76 & 0.77 & 1.11 \\ 
AIPWE & 2 & 49.50 & 1.42 & 1.47 & 1.13 & & 49.51 & 0.76 & 0.76 & 1.22 \\ 
AIPWE & 3 & 49.49 & 1.39 & 1.46 & 1.13 & & 49.47 & 0.75 & 0.77 & 1.06 \\ 
AIPWE & 4 & 49.52 & 1.40 & 1.48 & 1.12 & & 49.50 & 0.75 & 0.77 & 1.21 \\ 
AIPWE & 5 & 47.50 & 1.39 & 1.46 & 1.12 & & 47.49 & 0.77 & 0.76 & 1.12 \\ 
AIPWE & 6 & 47.53 & 1.33 & 1.45 & 1.16 & & 47.53 & 0.75 & 0.76 & 1.08 \\ 
AIPWE & 7 & 47.42 & 1.45 & 1.45 & 1.04 & & 47.46 & 0.78 & 0.76 & 1.05 \\ 
AIPWE & 8 & 47.44 & 1.41 & 1.46 & 1.07 & & 47.51 & 0.72 & 0.77 & 1.13 \\ 
IAIPWE & 1 & 49.47 & 1.37 & 1.38 & 1.24 & & 49.48 & 0.76 & 0.77 & 1.10 \\ 
IAIPWE & 2 & 49.50 & 1.37 & 1.37 & 1.21 & & 49.51 & 0.76 & 0.77 & 1.21 \\ 
IAIPWE & 3 & 49.50 & 1.33 & 1.37 & 1.23 & & 49.47 & 0.76 & 0.77 & 1.05 \\ 
IAIPWE & 4 & 49.53 & 1.35 & 1.38 & 1.20 & & 49.50 & 0.75 & 0.77 & 1.21 \\ 
IAIPWE & 5 & 47.51 & 1.34 & 1.37 & 1.20 & & 47.49 & 0.77 & 0.77 & 1.13 \\ 
IAIPWE & 6 & 47.54 & 1.27 & 1.35 & 1.26 & & 47.53 & 0.75 & 0.76 & 1.08 \\ 
IAIPWE & 7 & 47.43 & 1.40 & 1.36 & 1.12 & & 47.46 & 0.79 & 0.76 & 1.05 \\ 
IAIPWE & 8 & 47.46 & 1.36 & 1.37 & 1.16 & & 47.51 & 0.72 & 0.77 & 1.12 \\ 
    \bottomrule
    \end{tabular}
}%

\end{table}

In the second scenario, we investigate how different enrollment 
processes affect the proportion of early rejections for hypothesis 
\eqref{h:h0D} with $S=2$ analyses.
To vary the rate of enrollment, we select in which of four time periods 
($[0,500]$, $[501,600]$, $[601,700]$, and $[701,1000]$)
an individual enrolls using a multinomial distribution.
Within each, 
individuals enroll uniformly.
Results for the Pocock stopping boundaries under (VP2) are given in Table~\ref{tab:schema2_enroll}.
The sample sizes are determined to achieve $80\%$ power under (VP2), and
the interim analysis is conducted on day $700$. 
Both the total planned and expected sample sizes are lower for the IAIPWE than the IPWE or AIPWE.
The proportion of early rejections is higher when more individuals have progressed further through the study
due to the increased information available at the time of analysis.
All methods attain the desired power, and the IAIPWE achieves earlier expected stopping times and lower expected sample sizes than the IPWE and AIPWE.

\begin{table}[ht!]
    \centering
    \caption{\label{tab:schema2_enroll}
    For the schema in Figure~\ref{fig:PCST-schema}, interim analysis performance results for testing hypothesis \eqref{h:h0D} against $H_{AD}$ with a fixed control value using Pocock boundaries under varying enrollments. 
    The interim analysis is conducted on day $700$. 
    The percentages $p_1, p_2,$ and $p_3$ are the expected percentage of individuals to have completed the trial, made it to only stage two, and to have made it to only stage one, respectively. 
    Method indicates the estimator used. 
    The total planned sample size $N$ is determined by either each method.
    Total planned sample sizes are determined to maintain a nominal type I error rate of $\alpha=0.05$ and achieve a power of $80\%$ under (VP2).
    Early Reject and Total Reject are the rejection rates at the first analysis and for the overall procedure, respectively. 
    $\mathbb{E}$(SS) is the expected sample size, i.e, the average number of individuals enrolled in the trial when the trial is completed. 
    $\mathbb{E}$(Stop) is the expected stopping time, i.e., the average number of days that the trial ran. 
    Monte Carlo standard deviations are given in parentheses. 
}
\begin{footnotesize}
\begin{tabular}{lll ll p{10mm} p{10mm} ll}
    \toprule
    \toprule
    $p_1$ & $p_2$ & $p_3$ & Method  & $N$ & Early Reject & Final Reject & $\mathbb{E}$(SS) & $\mathbb{E}$(Stop) \\
    \midrule
    50 & 10 & 10 & IPWE & 1179 & 0.472 & 0.802 & 1012 (177) & 964 (249) \\
    50 & 10 & 10 & AIPWE & 844 & 0.423 & 0.787 & 737 (125) & 988 (247) \\
    50 & 10 & 10 & IAIPWE & 839 & 0.472 & 0.792 & 720 (126) & 963 (249) \\
    \midrule
    40 & 20 & 10 & IPWE & 1190 & 0.392 & 0.826 & 1050 (174) & 1003 (243) \\
    40 & 20 & 10 & AIPWE & 856 & 0.367 & 0.811 & 762 (124) & 1016 (241) \\
    40 & 20 & 10 & IAIPWE & 851 & 0.436 & 0.814 & 740 (127) & 981 (247) \\
    \midrule
    30 & 30 & 10 & IPWE & 1216 & 0.312 & 0.811 & 1102 (170) & 1043 (231) \\
    30 & 30 & 10 & AIPWE & 874 & 0.247 & 0.799 & 809 (113) &  1076 (215) \\
    30 & 30 & 10 & IAIPWE & 868 & 0.353 & 0.802 & 776 (125) & 1023 (239) \\
    \midrule
    40 & 10 & 20 & IPWE & 1194 & 0.382 & 0.800 & 1057 (174) & 1009 (243) \\
    40 & 10 & 20 & AIPWE & 859 & 0.340 & 0.782 & 772 (122) & 1029 (236) \\
    40 & 10 & 20 & IAIPWE & 855 & 0.408 & 0.793 & 751 (126) & 995 (245) \\
    \midrule
    30 & 20 & 20 & IPWE & 1215 & 0.329 & 0.812 & 1095 (171) & 1035 (235) \\
    30 & 20 & 20 & AIPWE & 874 & 0.257 & 0.806 & 807 (115) & 1071 (218) \\
    30 & 20 & 20 & IAIPWE & 867 & 0.340 & 0.811 & 779 (123) & 1029 (236) \\
    \midrule
    30 & 10 & 30 & IPWE & 1213 & 0.302 & 0.798 & 1103 (167) & 1048 (229) \\
    30 & 10 & 30 & AIPWE & 873 & 0.271 & 0.800 & 802 (116) & 1064 (222) \\
    30 & 10 & 30 & IAIPWE & 870 & 0.332 & 0.812 & 784 (123) & 1033 (235) \\
    \bottomrule
\end{tabular}
\end{footnotesize}

\end{table}

In Appendix G, we present results for two additional, common designs:
the schema in Figure~\ref{fig:PCST-schema} with a control arm and a schema in which responders are not re-randomized. 
The additional simulations demonstrate that the IAIPWE performs well even under misspecification of the $Q$-functions. 
In small samples, the IAIPWE variance may be overestimated, resulting in the estimated proportion of information at interim analyses being inflated. 
The OBF boundaries may be conservative in these cases. 
The IAIPWE performs well with multiple interim analyses and for the $\chi^2$ testing procedure for $H_{0H}$.

\section{Case study: cancer pain management SMART}\label{s:app}

We present a case study based on a recently completed trial evaluating
behavioral interventions for pain management in breast cancer patients
\citep{kelleher2017optimizing, nct02791646}. A schematic for the trial is
shown in Figure \ref{fig:PCST-schema}.  Initially, patients are
randomized with equal probability to one of two pain coping skills
training interventions: five sessions with a licensed therapist (PCST-Full) or one
60-minute session (PCST-Brief) with a licensed therapist.  After eight weeks (end of stage one),
participants who achieve a $30\%$ reduction in pain from baseline are
deemed responders and randomized with equal probability to maintenance
therapy or no further intervention. Non-responders who received
PCST-Full are randomized with equal probability to either two full
sessions (PCST-Plus) or maintenance.  Non-responders who received
PCST-Brief are randomized with equal probability to PCST-Full or
maintenance.  
The eight embedded regimes are given in Figure~\ref{fig:PCST-schema}.
Follow up occurs eight weeks after administration of stage
two intervention and again six months later. 
Here, we take the outcome of interest to be
percent reduction in pain from baseline at the final six month
assessment
and the primary analysis to be the evalution of the eight embedded regimes via the 
null hypothesis $H_{0D}$ in (\ref{h:h0D}) as described below.

Because the data from the trial are not yet published, we simulate the
trial based on the protocol.  We consider five baseline covariates:
height $X_{1,1}$, weight $X_{1,2}$, presence/absence of comorbidities
$X_{1,3}$, use of pain medication $X_{1,4}$, and whether or not the
participant is receiving chemotherapy $X_{1,5}$.  We observe the
response status $R_{2}$, percent reduction in pain $X_{2,0}$, and
degree of adherence $X_{2,1}$ at the first follow up at the end of
stage one.  Participants enroll uniformly over 1000 days, the end of
stage one occurs eight weeks after enrollment, 
and 
the outcome $Y$ is ascertained eighteen weeks after the end of stage one and thus six months after enrollment.
The distributions of covariates and outcomes are given in Appendix H.
We take $N=284$ to match the sample size of \citet{kelleher2017optimizing}.

An interim analysis is planned for day $500$ and a final analysis at the trial conclusion, a maximum of $1182$ days. 
We test the null hypothesis \eqref{h:h0D} against the alternative that any regime achieves greater than a $22.5\%$ reduction in pain (fixed control value); see Appendix H. 
We consider both Pocock and OBF boundaries, for which, to achieve a
type I error of $\alpha = 0.05$ using our IAIPWE procedure,
 $c_{\alpha=0.05}$ 
$=(2.66, 2.66)$ and $(4.20, 2.43)$ respectively.
For the AIPWE and IPWE, the Pocock and OBF boundaries are 
$=(2.66, 2.66)$ and $(4.30, 2.44)$ respectively.
In this setting, the correlation structure for $\bZ$ is similar for all estimators. 
Therefore the Pocock boundaries are the same even with the difference of available information at the interim analysis. 
As a result, the Pocock boundaries illustrate in part why we expect more early rejections under a true alternative for the IAIPWE than the other estimators. 
By construction, the different OBF boundaries demonstrate the impact of the increased information available using the IAIPWE at the interim analysis.

\begin{figure}
    \centering
    \includegraphics{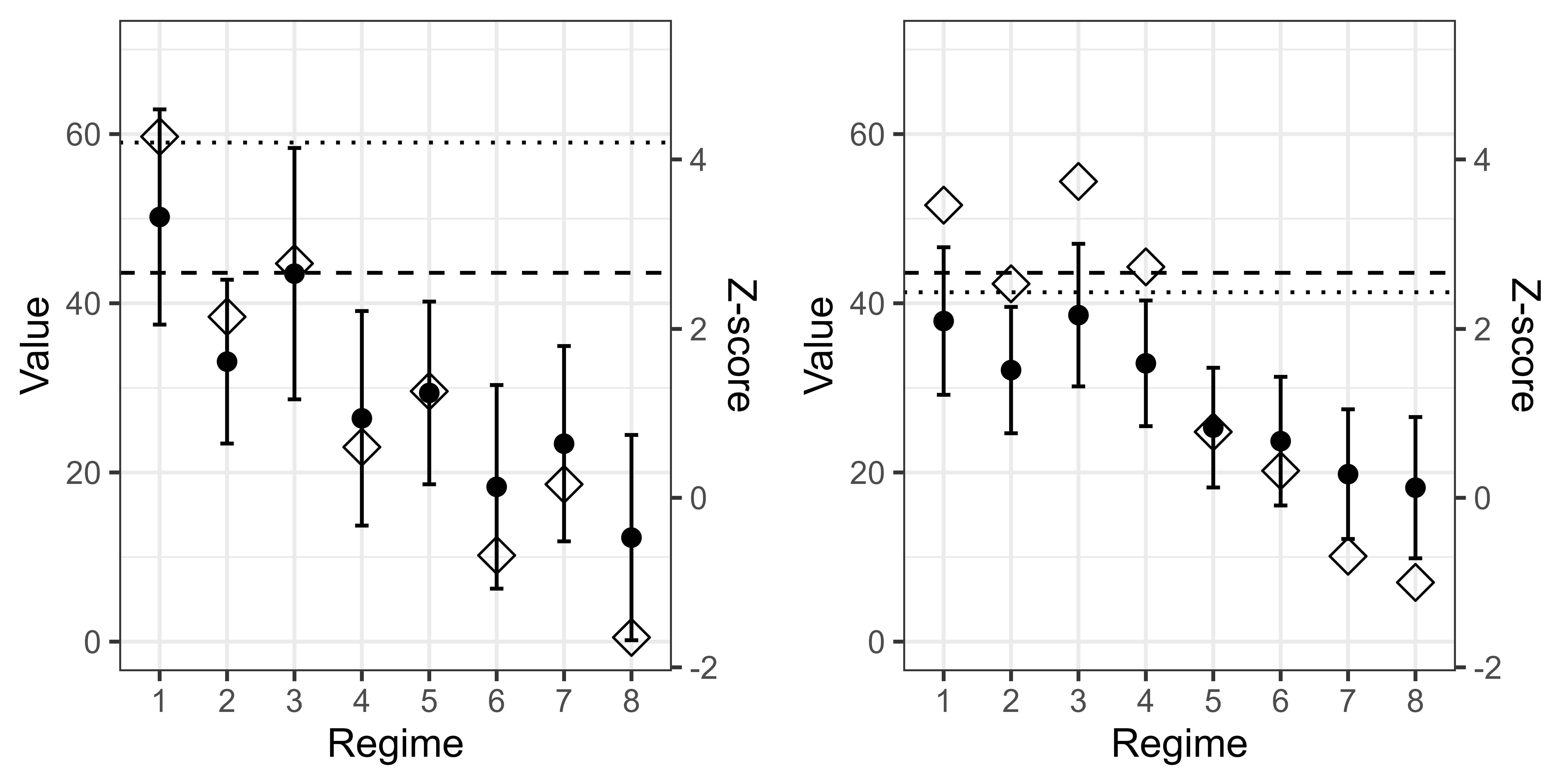}
   
\caption{\label{fig:pcst-regimes}
For the schema in Figure~\ref{fig:PCST-schema}, interim analysis performance results for testing null hypothesis (2) against $H_{AD}$ with a fixed control. Results include the Pocock boundaries (dashed), OBF boundaries (dotted), value estimates (circles) and 95\% confidence bounds, and test statistics (rhombus) at the interim (left) and final (right) analysis time for the behavioral pain management case study data set using the IAIPWE.
}
\end{figure}

The interim analysis occurs at 500 days after the trial enrollment begins, at which point $51.4\%$
of the total planned sample size $N$ has been enrolled, $46.8\%$ of the $N$
planned participants have progressed to the second decision, and
$34.9\%$ have completed the trial.  
Figure \ref{fig:pcst-regimes} summarizes
the estimated values for each regime at the time of analysis, corresponding
$Z$-statistic.
Exact numbers are recorded in a tabular format in Appendix I. 
Regime 1, which starts with PCST-Full, triggers early stopping based
on the test statistic exceeding the OBF boundary. 
Regimes 1 and 3 trigger early stopping based on test statistics exceeding the Pocock boundary. 
The standard errors are smaller than those obtained using the IPWE or AIPWE, which are included in the Appendix I. 
The IPWE and AIPWE trigger early stopping with regimes exceeding the Pocock boundary, but fail to trigger early stopping under the OBF boundary. 
The decision to stop the trial early reduces the sample size from the total possible $284$ subjects to $146$ and the length of
the study by $96$ weeks. 
Early stopping means implementation of behavioral interventions for pain management in breast cancer patients, potentially helping more individuals and avoiding less efficacious regimes for those who otherwise would have enrolled in the trial.

\section{Discussion}
\label{s:discuss}

We proposed interim analysis methods for SMARTs that gain efficiency
by using partial information from participants who have not yet completed all stages of
the study. The approach yields a smaller expected sample size than competing methods while preserving type I error and power.
Simulations demonstrate a potential for substantial resource savings.

We have demonstrated the methodology in the
case of two-stage SMARTs with an interim analysis focused on evaluation of efficacy. 
However, the methods extend readily to studies with $K \geq 2$ decision points, multiple interim looks, and general hypotheses including futility.  
We have consider Pocock and OBF boundaries, though the approach can be
adapted to any monitoring method, such as information-based monitoring
\citep{tsiatis2006inf} and the use of $\alpha$ spending functions
\citep{demets1994}.

    We have made the simplifying assumptions throughout that: (i)
      the time between stages is fixed, which is the case for many SMARTs; and (ii) the final
      outcome is observed on all individuals by the end of
      the trial (so excluding the possibility of drop out).  The
      extension to random times between stages is non-trivial.
      Simulations included in Appendix G suggest that the IAIPWE
      (incorrectly assuming fixed transition times) performs well when
      time per stage varies with subject outcomes.  Due to variability
      in enrollment, an analysis at a predetermined time may have a
      realized power slightly different from the nominal power based
      on the number of individuals enrolled and their realized
      trajectories at the time of analysis.  In such cases, planning
      the interim analysis based on available sample size rather than
      a pre-determined time may be preferred.  Extensions for
      additional levels of coarsening, such as those due to drop out,
      attrition, or time-to-event outcomes requires additional
      augmentation terms or changes to the functions
      $\lambda_r^\ell(t), K_r^\ell(t)$ and
      $L_{k(r)}^\ell(\overline{\bx}_{k(r)}) $.  For a comprehensive
      review of the considerations involved, see Chapter 8 of
      \citet{tsiatis2019dynamic}.  A modified multiple imputation
      strategy may also be used for missing data following that of
      \citet{shortreed2014}.

As demonstrated in our simulation experiments, the sandwich covariance estimator can overestimate the variance of the values and lead to conservative stopping boundaries when the number of parameters is close to the sample size. 
Interim analyses typically have larger sample sizes, so this issue is unlikely to occur in practice.
The information proportion can be checked at each interim analysis to verify the planned proportions against the realized values. 
The IAIPWE stopping boundary and sample size calculations also require the challenge of positing models. 
Although we have studied the performance of the IAIPWE under these conditions to evaluate fully its properties, we anticipate the trialists will prefer to power a SMART based on the IPWE to avoid making the additional model assumptions. 
We advocate this approach in practice
as it can assuage concerns about misspecified models while still benefiting from the efficiency gains of the IAIPWE.
If a trial does reach the final analysis, using the AIPWE offers efficiency gains by effectively performing covariate adjustment.
Here, the covariates to be used in the $Q$-functions should be specified before the trial begins.

The framework presented here forms the basis for additional methodology for interim monitoring for SMARTs with random times between stages and specialized endpoints. 
The IAIPWE has potential use in adaptive trials in which randomization probabilities, or even the set of treatments, varies with accumulating information (\citep[][Chapter 17]{jennison1999}; \citealp{wang2019_ispy}).
We will report on these developments in future work.

\section*{Acknowledgements}

The authors thank Dr. Anastasios Tsiatis for helpful remarks and
insights into the demonstration of the independent increments
property.

\section*{Supporting Information}

R code for implementation is available from the authors. 

\section*{Appendix A: Discussion on Coarsening}

\subsection*{Some Examples of Coarsening}

Consider the SMART 
from our data application
and a subject who does
      not respond to PCST-Full and is subsequently given PCST-Plus,
      but has not yet reached their final follow up by interim
      analysis time $t_1$.  When estimating the value of regime $1$,
      the patient is coarsened at level $\Rc^1(t_1) = 4$ because all
      treatments received are consistent with regime 1, but they have
      not completed the trial.  However, when estimating the value of
      regime $3$, the subject is coarsened at level $\Rc^3(t_1) = 3$
      because their first treatment is consistent with regime $3$,
      they made it to the second treatment assignment by the time of
      the interim analysis, but their second treatment is inconsistent
      with regime $3$.  At the final analysis, after the subject
      completes their final follow up, the subject will have
      coarsening levels $\Rc^1(t_{S}) = \infty$ and
      $\Rc^3(t_{S}) = 3$.  The coarsening level when estimating the
      value of regime $1$ is now infinite because the subject finished
      the trial with treatments consistent with regime $1$; however,
      their coarsening level for regime $3$ remains the same because
      they were coarsened due to inconsistent treatments.

\subsection*{Coarsening Under Fixed Arrival Times}

When arrival times for stages and treatment assignment are independent, we can simplify the notation for the IAIPW estimator to get the form for $K=2$ given in the paper. 
For ease of notation, let $A_k \in \{ 0, 1\}$.
Let $\pi_k^d(A_k, \bH_k) = P(A_k = 1 \vert \bH_k)$, 
$\nu_k(t) = P(\kappa(t) \geq k \vert \Gamma(t) = 1)$, and 
$C_k^\ell = \prod_{v=1}^k I \lbrace A_v = d^\ell(\bH_v) \rbrace $. 
We focus on the portion of the augmentation term without the arbitrary function, and suppress dependence on $t$ and $\ell$. 
Write 
\begin{equation*}
    \frac{ I\lbrace \Rc=r \rbrace - \lambda_r I\lbrace \Rc \geq r \rbrace}
         {K_r}
    = I\lbrace \Rc \geq r \rbrace \frac{ I\lbrace \Rc=r \rbrace - \lambda_r}
         {K_r}.
\end{equation*}

Consider cases $r$ odd and even separately. Then

\begin{equation*}
\begin{split}
    I\lbrace \Rc \geq r \rbrace \frac{ I\lbrace \Rc=r \rbrace - \lambda_r}
         {K_r} 
         &= I\lbrace \Rc \geq r \rbrace \left[ \frac{ I\lbrace \Rc=r \rbrace - P(\Rc = r \vert \Rc \geq r)}
         { P(\Rc > r)} \right]
         \\
    &= I\lbrace \Rc \geq r \rbrace \left[ \frac{ I\lbrace \Rc=r \rbrace - \frac{P( \Rc = r)}{P( \Rc \geq r)} }
         { P(\Rc > r)} \right]
         \\
    &= \begin{cases}
    I\lbrace \Rc \geq r \rbrace \left[ \frac{ I\lbrace \Rc=r \rbrace - 
    P \left\lbrace d_{k(r)}(\bH_{k(r)}) \ne A_{k(r)} \right\rbrace }
         { P(\Rc > r)} \right] & \text{, r odd} \\
    I\lbrace \Rc \geq r \rbrace \left[ \frac{ I\lbrace \Rc=r \rbrace - 
        P \left\lbrace \kappa = k(r) \vert \Gamma = 1 \right\rbrace }
         { P(\Rc > r)} \right] & \text{, r even.} \\
        \end{cases} \\
\end{split}
\end{equation*}
Simplify the probability statements using our propensities and enrollment notation, and the fact that the term is $0$ for all individuals coarsened before $r$. 
We can express the denominator as individuals who have been consistent with the regime through $k(r)-1$. Let $\pi_0 = 1$. 
\begin{equation*}
\begin{split}
    &= \begin{cases}
    I\lbrace \Rc \geq r \rbrace \left[ \frac{ I\lbrace \Rc=r \rbrace - 
    \left\lbrace C_{k(r)} (1-\pi_{k(r)}^d) + (1-C_{k(r)}) \pi_{k(r)}^d \right\rbrace }
         { \nu_{k(r)} \left\lbrace \prod_{k=0}^{k(r)-1} \pi_k^d \right\rbrace 
         \left\lbrace C_{k(r)}\pi_{k(r)}^d + (1-C_{k(r)})(1- \pi_{k(r)}^d) \right\rbrace } \right] & \text{, r odd} \\
    I\lbrace \Rc \geq r \rbrace \left[ \frac{ I\lbrace \Rc=r \rbrace - 
        \left\lbrace \nu_{k(r)} - \nu_{k(r)+1} \right\rbrace }
         { \nu_{k(r)+1} \left\lbrace \prod_{k=1}^{k(r)} \pi_k^d \right\rbrace } \right] & \text{, r even} \\
        \end{cases} \\
\end{split}
\end{equation*}

Further algebra and simplifications from $K=2$ yield the estimator in the paper.
If the time to next treatment varies by treatment assignment, then the probability of coarsening due to time conditioned on the treatment assignment is no longer expressed with $\nu$ as defined and proper adjustments can be made.

\section*{Appendix B: Double Robustness Property of IAIPW Estimator}

We show the estimator is consistent if either the propensity and proportion models are correctly specified, or the regression models are correctly specified. 
Let $\btheta_{p}$ characterize the parameters for the propensity and proportion models, and let $\btheta_L$ characterize those of the arbitrary function $L_r$. 
In both cases, we will assume for models chosen, 
$\widehat{\btheta}_p \overset{p}{\to} \btheta_p^*$ 
and 
$\widehat{\btheta}_L \overset{p}{\to} \btheta_L^*$,
where $\btheta_p^*$ and $\btheta_L^*$ constant. 
If the models are correctly specified, then $\btheta_p^* = \btheta_p^0$ and $\btheta_L^* = \btheta_L^0$, respectively,
superscript $0$ denoting true parameters. 
We assume that the enrollment process and treatment assignment and outcome are independent. 
For fixed $t$ and arbitrary $l$,
the estimator converges to 
\begin{equation*}
\begin{split}
&\mathbb{E}
\bigg[
\frac{
I\left\lbrace 
\Rc_i^{\ell}(t) = \infty
\right\rbrace 
}{
\widehat{K}^\ell_{2K, i, n(t)}(t; \btheta^*_p) 
}Y_i
\\ 
&\quad + 
\sum_{r=1}^{2K} \frac{
    I\left\lbrace
        \Rc^{\ell}_i(t) = r
     \right\rbrace - 
    \widehat{\lambda}^{\ell}_{r,i}(t; \btheta^*_p) 
    I\left\lbrace 
        \Rc^{\ell}_{i}(t) \geq r
    \right\rbrace
    }{
    \widehat{K}^\ell_{r,i,n(t)}(t; \btheta^*_p)
    }
L^{\ell}_{k(r)}(
\overline{\bX}_{k(r), i}; 
\btheta^*_{L,k(r)}) 
\bigg] \\
&=
\mathbb{E}
\bigg[
\frac{
I\left\lbrace 
\Rc_i^{\ell}(t) = \infty
\right\rbrace 
}{
\widehat{K}^\ell_{2K, i, n(t)}(t; \btheta^*_p) 
}Y^*(\bd^{\ell}) \bigg] + 
\\
&\quad \mathbb{E} \bigg[
\sum_{r=1}^{2K} \frac{
    I\left\lbrace
        \Rc^{\ell}_i(t) = r
     \right\rbrace - 
    \widehat{\lambda}^{\ell}_{r,i}(t; \btheta^*_p) 
    I\left\lbrace 
        \Rc^{\ell}_{i}(t) \geq r
    \right\rbrace
    }{
    \widehat{K}^\ell_{r,i,n(t)}(t; \btheta^*_p)
    }
L^{\ell}_{k(r)}(
\overline{\bX}_{k(r), i}; 
\btheta^*_{L,k(r)}) .
\bigg]
\end{split}    
\end{equation*}
By definition $Y_i I\{ \Rc_i^\ell(t) = \infty \} = Y^*(\bd^\ell)$. 
By Lemma 10.4 of \cite{tsiatis2006book},
\begin{equation*}
\begin{split}
&\mathbb{E}
\Bigg( Y^*(\bd^l) + 
\bigg[
\frac{
I\left\lbrace 
\Rc_i^{\ell}(t) = \infty
\right\rbrace 
}{
\widehat{K}^\ell_{2K, i, n(t)}(t; \btheta^*_p) 
}
-1 \bigg] Y^*(\bd^{\ell}) \Bigg) + 
\\
&\quad \mathbb{E}\bigg[
\sum_{r=1}^{2K} \frac{
    I\left\lbrace
        \Rc^{\ell}_i(t) = r
     \right\rbrace - 
    \widehat{\lambda}^{\ell}_{r,i}(t; \btheta^*_p) 
    I\left\lbrace 
        \Rc^{\ell}_{i}(t) \geq r
    \right\rbrace
    }{
    \widehat{K}^\ell_{r,i,n(t)}(t; \btheta^*_p)
    }
L^{\ell}_{k(r)}(
\overline{\bX}_{k(r), i}; 
\btheta^*_{L,k(r)}) 
\bigg] 
\\
&= 
\mathbb{E} \{ Y^*(\bd) \} -
\mathbb{E}
\bigg[
1-
\frac{
I\left\lbrace 
\Rc_i^{\ell}(t) = \infty
\right\rbrace 
}{
\widehat{K}^\ell_{2K, i, n(t)}(t; \btheta^*_p) 
}
\bigg] Y^*(\bd^{\ell}) + 
\\
&\quad \mathbb{E}\bigg[
\sum_{r=1}^{2K} \frac{
    I\left\lbrace
        \Rc^{\ell}_i(t) = r
     \right\rbrace - 
    \widehat{\lambda}^{\ell}_{r,i}(t; \btheta^*_p) 
    I\left\lbrace 
        \Rc^{\ell}_{i}(t) \geq r
    \right\rbrace
    }{
    \widehat{K}^\ell_{r,i,n(t)}(t; \btheta^*_p)
    }
L^{\ell}_{k(r)}(
\overline{\bX}_{k(r), i}; 
\btheta^*_{L,k(r)}) 
\bigg] 
\\
&= 
\mathbb{E} \{ Y^*(\bd) \} -
\mathbb{E}\bigg[
\sum_{r=1}^{2K} \frac{
    I\left\lbrace
        \Rc^{\ell}_i(t) = r
     \right\rbrace - 
    \widehat{\lambda}^{\ell}_{r,i}(t; \btheta^*_p) 
    I\left\lbrace 
        \Rc^{\ell}_{i}(t) \geq r
    \right\rbrace
    }{
    \widehat{K}^\ell_{r,i,n(t)}(t; \btheta^*_p)
    }
[Y^*(\bd) - 
L^{\ell}_{k(r)}(
\overline{\bX}_{k(r), i}; 
\btheta^*_{L,k(r)}) ]
\bigg] 
\end{split}
\end{equation*}
Therefore, the IAIPW estimator is consistent if the second term is 0.
First, consider the case the propensity and proportion models are correctly specified.
Then for $r=1,\dots,2K$, the hazard functions are correctly specified, i.e. $\lambda^{\ell}_{r,i}(t; \btheta^*_p) = \lambda^{\ell *}_r(t) $ for all $r=1,\dots, 2K$. 
Define $\mathcal{W}_{r,i}$ as the random vector $\lbrace I(\Rc_i^{\ell}(t) =1),\dots,I(\Rc_i^{\ell}(t) =r-1), W_i  \rbrace$. Then by iterated expectations and the definition of the hazard functions, for all $r \ne \infty$, 
\begin{equation*}
\begin{split}
&\mathbb{E}
\bigg[
\frac{
    \mathbb{E} [I\left\lbrace
        \Rc^{\ell}_i(t) = r
     \right\rbrace  \vert \mathcal{W}_r ] - 
    \widehat{\lambda}^{\ell}_{r,i}(t; \btheta^*_p) 
    I\left\lbrace 
        \Rc^{\ell}_{i}(t) \geq r
    \right\rbrace
    }{
    \widehat{K}^\ell_{r,i,n(t)}(t; \btheta^*_p)
    }
[Y^*(\bd) - 
L^{\ell}_{k(r)}(
\overline{\bX}_{k(r), i}; 
\btheta^*_{L,k(r)}) ]
\bigg] 
\\
&=\mathbb{E}
\bigg[
\frac{
    \widehat{\lambda}^{\ell}_{r,i}(t; \btheta^*_p) 
    I\left\lbrace 
        \Rc^{\ell}_{i}(t) \geq r
    \right\rbrace - 
    \widehat{\lambda}^{\ell}_{r,i}(t; \btheta^*_p) 
    I\left\lbrace 
        \Rc^{\ell}_{i}(t) \geq r
    \right\rbrace
    }{
    \widehat{K}^\ell_{r,i,n(t)}(t; \btheta^*_p)
    }
[Y^*(\bd) - 
L^{\ell}_{k(r)}(
\overline{\bX}_{k(r), i}; 
\btheta^*_{L,k(r)}) ]
\bigg] \\
&=0. 
\end{split}
\end{equation*}

Now consider when the arbitrary functions 
$L^{\ell}_{k(r)}(
\overline{\bX}_{k(r), i}; 
\btheta^*_{L,k(r)})$
are correctly specified for 
$\mathbb{E}\lbrace Y^*(\bd) \vert 
\overline{\bX}_k^*(\bd_k^{\ell}) = \bx_k^* \rbrace$. 
Under the assumption of coarsening at random and using iterated expectations on $[I\lbrace \Rc^{\ell}_i(t) \geq r \rbrace, \overline{\bX}_k^*(\bd_k^{\ell}) ]$
\begin{equation*}
\begin{split}
&
\mathbb{E}\bigg[
\frac{
    I\left\lbrace
        \Rc^{\ell}_i(t) = r
     \right\rbrace - 
    \widehat{\lambda}^{\ell}_{r,i}(t; \btheta^*_p) 
    I\left\lbrace 
        \Rc^{\ell}_{i}(t) \geq r
    \right\rbrace
    }{
    \widehat{K}^\ell_{r,i,n(t)}(t; \btheta^*_p)
    }
[Y^*(\bd) - 
\mathbb{E}\lbrace Y^*(\bd) \vert 
\overline{\bX}_k^*(\bd_k^{\ell}) = \bx_k^* \rbrace]
\bigg] 
\\
&\mathbb{E}\bigg[
\frac{
    I\left\lbrace
        \Rc^{\ell}_i(t) = r
     \right\rbrace - 
    \widehat{\lambda}^{\ell}_{r,i}(t; \btheta^*_p) 
    I\left\lbrace 
        \Rc^{\ell}_{i}(t) \geq r
    \right\rbrace
    }{
    \widehat{K}^\ell_{r,i,n(t)}(t; \btheta^*_p)
    }
[\mathbb{E}\lbrace Y^*(\bd) \vert 
\overline{\bX}_k^*(\bd_k^{\ell}) = \bx_k^* \rbrace - 
\mathbb{E}\lbrace Y^*(\bd) \vert 
\overline{\bX}_k^*(\bd_k^{\ell}) = \bx_k^* \rbrace]
\bigg] 
\\
&=0.
\end{split}
\end{equation*}
Therefore the estimator is consistent if either the propensity or regression models are correctly specified. 

\section*{Appendix C: Conditions and Proof of Theorem 1} \label{s:sketch_thm1}

Theorem 1 follows from the asymptotic results in Section 7 of \cite{boos2013stefanski} with an adjustment to account for $n(t)$. 

Let $\bpsi(\Y(t), \bH(t), \btheta )$ be the set of equations such that 
\begin{equation*}
    \bpsi(\Y(t), \bH(t), \btheta) = \sum_{i=1}^{N} \bpsi(Y_i \Delta_i(t), \bH_i(t), \widehat{\btheta}) = \0.
\end{equation*}
And, let $\btheta_0$ be the unique solution to $\E\lbrace \bpsi(\Y(t), \bH(t), \btheta_0 )\rbrace = \0$.
Assume that 
\begin{itemize}
    \item[C.1] $\bpsi(\Y(t), \bH(t), \btheta )$ and its first two partial derivatives with respect to $\btheta(t)$ exist for all $\btheta$ in a neighborhood of $\btheta_0$, and is $o_p(N^{-1/2})$. 
    \item[C.2] The second derivative of $\bpsi(\Y(t), \bH(t), \btheta )$ is bounded.
    \item[C.3] $\E \lbrace - \bpsi'(Y, \bH(t), \btheta_0 )\rbrace$ exists and is nonsingular.
    \item[C.4] $\E\lbrace \bpsi(Y, \bH(t), \btheta_0 ) \bpsi(Y, \bH(t), \btheta_0 )^T \rbrace$ exists and is finite.
    \item[C.5] The proportion of individuals at the interim analysis time $t$ converges to a constant, i.e., $n(t) / N \overset{p}{\to} c > 0$.
\end{itemize}

Then,
$\sqrt{N} \{ \widehat{\mathcal{V}}(t) - \mathcal{V}(t) \} \overset{d}{\to} \mathcal{N}(\0, \bSigma_T)$ as $N \rightarrow \infty$ for 
$$\bSigma_T = \1 \E \lbrace - \bpsi'(Y, \bH(t), \btheta_0 )\rbrace ^{-1} \E\lbrace \bpsi(Y, \bH(t), \btheta_0 ) \bpsi(Y, \bH(t), \btheta_0 )^T \rbrace \E \lbrace - \bpsi'(Y, \bH(t), \btheta_0 )\rbrace^{-1} \1$$
where $\1$ is the matrix $[ \0 \vert I_{L+1, L+1}]$ such that $\1 \btheta = \mathcal{V}$. 

While it may seem reasonable to consider the estimating equations purely as a function of observed individuals at an interim analysis, the estimation of the quantities $\Gamma_i(t)$ must be viewed as draws over $N$ individuals. 

\textit{Proof:}

For ease of notation, write 
\begin{equation*}
    \widehat{V}_{IA}^l (t) = \frac{1}{n(t)} \sum_{i=1}^N \Gamma_i(t) V_i(t). 
\end{equation*}
First, consider that $c$ may be estimated via estimating equation $0 = \sum_{i=1}^N \Gamma_i - (1/c)$ if $n(t)$ is not fixed. 
Then, we can express the estimating equation for our estimator equivalently as 
\begin{equation*}
\begin{split}
    0 &= \frac{1}{N} \sum_{i=1}^N \Gamma_i(t) \left\{ \widehat{V}_{IA}^l (t) -   V_i(t) \right\} c \\
    &= c \frac{1}{N} \sum_{i=1}^N \Gamma_i(t) \left\{ \widehat{V}_{IA}^l (t) -   V_i(t) \right\} \\
    &= \frac{1}{n(t)} \sum_{i=1}^N \Gamma_i(t) \left\{ \widehat{V}_{IA}^l (t) -   V_i(t) \right\} \\
    &= \widehat{V}_{IA}^l (t)  - \frac{1}{n(t)} \sum_{i=1}^N \Gamma_i(t)  V_i(t) \\
\end{split}
\end{equation*}
Because the estimation remains a set of unbiased estimating equations, then by Theorem 7.2 of \citet{boos2013stefanski} 
$\sqrt{N} \{ \widehat{\mathcal{V}}(t) - \mathcal{V}(t) \} \overset{d}{\to} \mathcal{N}(\0, \bSigma_T)$ as $N \rightarrow \infty$.

The result for contrast matrix $\bC$ follows directly as 
$\sqrt{N}\{\bC \widehat{\mathcal{V}}(t) - \bC \mathcal{V}(t) \} \overset{d}{\to} \mathcal{N}(\0, \bC \bSigma_T \bC^\top)$.

\section*{Appendix D: Stopping Boundaries and Power under Arbitrary $S$}

We use the testing procedure outlined in the Section \ref{s:interim}.
To find stopping boundaries $\{ c_{\alpha}(s) \} _{s=1}^S$ that control the family-wise error rate across all planned analyses at level $\alpha$, we use the joint distribution of the $Z$ statistics across all analyses.
For chosen $\alpha$-spending function, the boundaries satisfy
\begin{equation*}
    \pr \lbrace \text{Reject}\  H_{0D} \vert H_{0D} \ \text{is true} \rbrace
    = \pr \left[ \left.\bigcup_{\ell=1}^L \bigcup_{s=1}^S \lbrace Z^\ell(t_s)
      \geq c_{\alpha}(s) \rbrace  \right| H_{0D} \right] \leq \alpha,
\end{equation*}
which can be solved for by integration of the multivariate normal probability distribution function. 
For chosen stopping boundaries and specified alternative where the expectation of $\bZ$ is $\bmu_A  = \bmu_A\{n(t_1),\ldots,n(t_S)\}$, 
the power is approximately
\begin{equation*} 
    1- \int \dots \int_D \frac{1}{(2\pi)^{LS/2} \mathrm{det}(\bSigma_{H})^{-1/2}} \mathrm{exp} \lbrace -\frac{1}{2} (\bZ-\boldsymbol{\mu}_A)^{\top} \bSigma_{H}^{-1} (\bZ-\boldsymbol{\mu}_A) \rbrace 
    dz^1(1)dz^2(1) \cdots dz^L(S)
\end{equation*}
for domain
$D=(-\infty, c_{\alpha}(1)] \times (-\infty, c_{\alpha}(1)]
\times \dots \times (-\infty, c_{\alpha}(S)]$, 
and one can solve for the sample sizes needed to achieve a specified power or the power for selected sample sizes.

\section*{Appendix E: Proof of Theorem 2}

\subsection*{Proof}
We can write the set of all estimating equations used in the IAIPW Estimator over time as $\mathrm{vec} \{ \bpsi(\Y(t_s), \bH(t_s), \btheta ) \}_{s=1}^S$ where $\mathrm{vec}(\cdot)$ is the vectorization operator. 
Then under the conditions from Section B for times $\{ t_s\}_{s=1}^S$, 
\begin{equation}
    \sqrt{N} [ \mathrm{vec} \{ \widehat{\mathcal{V}}(t_s) \}_{s=1}^S - \mathrm{vec} \{ \mathcal{V}(t) \}_{s=1}^S] \overset{d}{\to} \mathcal{N}(\0, \bSigma_{T_s})
\end{equation}
for covariance $\bSigma_{T_s}$ given in (7.10) in \cite{boos2013stefanski} and the discussion below. 
Let $\1_T$ be the $SL \times S(L+1)$ matrix of block diagonal matrices $[-1_{1\times L} \vert I_{L \times L}] $. 
Then by Slutsky's Theorem 
\begin{equation}
    \bZ = \sqrt{N} \mathrm{diag}(\1_T \widehat{\bSigma}_{T_s}\1_T^{\top} 
    )^{-1/2} \1_T [ \mathrm{vec} \{ \widehat{\mathcal{V}}(t_s) \}_{s=1}^S - \mathrm{vec} \{ \mathcal{V}(t) \}_{s=1}^S ] 
    \to \mathcal{N} (\bmu_{H}, \bSigma_{H}) 
\end{equation}
where 
$$\bmu_H = \E \left(  \mathrm{diag}(\1_T \widehat{\bSigma}_{T_s}\1_T^{\top} 
    )^{-1/2} \1_T [ \mathrm{vec} \{ \widehat{\mathcal{V}}(t_s) \}_{s=1}^S - \mathrm{vec} \{ \mathcal{V}(t) \}_{s=1}^S ]  \right)$$
and 
$$\bSigma_H = \mathrm{bdiag}(\1_T \bSigma_{T_s} \1_T^{\top}) ^{-1/2} \1_T \bSigma_{T_s} \1_T^{\top} \mathrm{bdiag}(\1_T \bSigma_{T_s} \1_T^{\top})^{-1/2}. $$

\subsection*{Discussion}

We can determine the value of $\iota$ given in Theorem 2 by finding the value of $\bSigma_{H}$ for a general hypothesis, $H$, since $\iota^{-1}$ are entries $(1,L+1),(2,L+2), \ldots, (L, L+L)$ of $\bSigma_{H}$.
In the case that the information proportion varies between regimes, the $\iota^{-1,\ell}$ will be regime-specific. 

    The matrix $\bSigma_{H}$ can be computed analytically. 
    First, one must find the estimating equations used for all estimated parameter and stack these over all planned analyses. 
    Then, the covariance of the estimating equations can be written as 
    \begin{equation*}
    \begin{split}
    \bSigma_{\bpsi} = 
        &\E [ - \mathrm{vec} \lbrace \bpsi'(Y(t_s), \bH(t_s), \btheta_0 )\rbrace_{s=1}^S ] ^{-1}  \times \\
        &\E [ \mathrm{vec} \lbrace \bpsi(Y(t_s), \bH(t_s), \btheta_0 )\rbrace_{s=1}^S \mathrm{vec} \lbrace \bpsi(Y(t_s), \bH(t_s), \btheta_0 )\rbrace_{s=1}^{S \quad \top} ] \times \\
        &\E [ - \mathrm{vec} \lbrace \bpsi'(Y(t_s), \bH(t_s), \btheta_0 )\rbrace_{s=1}^S ] ^{-1 \top}
    \end{split}
    \end{equation*}
    following from \citet{boos2013stefanski}. 
    However, this gives the covariance of all estimated parameters. 
    We are interested in finding the covariance of the estimated values. 
    Construct block diagonal matrix $\mathrm{bdiag}(\1_1,\ldots,\1_S)$ where $\1_s$ is the matrix $[0 \vert I_{L+1, L+1}]$ such that $\1_s \btheta = \mathcal{V}$, as given in Appendix C. 
    Then $\bSigma_{T_s} =\mathrm{bdiag}(\1_1,\ldots,\1_S) \bSigma_{\bpsi} \mathrm{bdiag}(\1_1,\ldots,\1_S)^{\top} $. 
    Let $\1_T$ be the matrix of $S$ block diagonal matrices $[-1_{1 \times L} \vert I_{L \times L}]$. 
    Then, following the application of Slutsky's Theorem, 
    \begin{equation*}
    \begin{split}
        \bSigma_{H} = &\mathrm{bdiag} \left( \1_T 
        \mathrm{bdiag}(\1_1,\ldots,\1_S) \bSigma_{\bpsi} \mathrm{bdiag}(\1_1,\ldots,\1_S)^{\top}
        \1_T^\top \right)^{-1/2} \times \\
        & \1_T \bSigma_{\bpsi} \1_T^\top
         \mathrm{bdiag} \left( \1_T 
        \mathrm{bdiag}(\1_1,\ldots,\1_S) \bSigma_{\bpsi} \mathrm{bdiag}(\1_1,\ldots,\1_S)^{\top}
        \1_T^\top \right)^{-1/2 \top}. \\
    \end{split}
    \end{equation*}

\section*{Appendix F: Sample Size Calculations for Wald-type Test Statistics }

Below we outline how to determine the total planned sample size required to attain power $1-\beta$ for the Wald-type test statistic. 
\begin{enumerate}
    \item[(1)] Choose $N$ sufficiently small such that the power at $N$ is below $1-\beta$.
    \item[(2)] Increase $N$ by $\lambda$ and update $\bmu_A, \phi_A$.
    \item[(3)] Take $B$ draws from the distribution of $\bZ$ and form the corresponding correlated sequential $\chi^2$ statistics. 
    \item[(4)] Determine number of rejections under previously selected stopping boundaries $\{ c_{\alpha}(s) \}$.
    \item[(5)] If simulated power exceeds tolerance of nominal power,
    and $\lambda \leq 1$, stop.  
    Else, if simulated power exceeds tolerance of nominal power, and $\lambda > 1$,  update $\lambda$ by discount factor $\gamma < 1$, go to step (2).  
    Else, go to step (2)
\end{enumerate}
The authors have initial values of $N=50$, $\lambda=10$, $\gamma=0.1$, and $B=10000$ to perform well in practice. 
We note that the information proportion for the interim AIPW estimator may require numerical simulation to estimate. 
However, the inverse of the information proportion is bounded between the proportion of individuals who have completed the study at interim analysis $s$ relative to the total planned planned sample size and the proportion of individuals who have enrolled in the study at interim analysis $s$ relative to the total planned planned sample size.

\section*{Appendix G: Additional Simulations}

\subsection*{Regime Estimates for Additional Value Patterns}

Table \ref{tab:schema2_pocock_value_ests_vp13} summarizes the estimator used, the Monte Carlo mean value, standard deviation, and mean standard error for each estimator at both the interim analysis and final analysis. 
The mean square error (MSE) ratio is the ratio of the Monte Carlo MSE for the IPWE divided by that of the indicated method. 
A MSE ratio of greater than one indicates the estimator is more efficient than the IPWE. 
For (VP1), both the AIPWE and IAIPWE are more efficient than the IPWE at both analyses, and the IAIPWE is more efficient than the AIPWE. 
At the interim analysis, the standard errors for the IPWE underestimate the sampling variation in most cases, whereas the standard errors for the AIPWE overestimate the sampling variation. 
The IAIPWE consistently estimates the sampling variation with the exception of regime 6 at the interim analysis. 
The MC means demonstrate the consistency of the estimators as expected since the propensities are known in a SMART. 
Clearly, the IAIPWE estimates of the value for each regime more efficiently than the IPWE or AIPWE. 
Under (VP3) which has a smaller sample size than (VP1) to attain the power $80\%$, the IAIPWE and AIPWE are similarly efficient to the IPWE at the final analysis time. 
Therefore the IAIPWE is as efficient as the IPWE with a lower sample size. 
Here, the benefit of the IAIPWE is the lower overall sample size required for equivalent standard errors. 

\begin{center}
{\small\tabcolsep=1pt
\begin{longtable}{lll llll l llll}
      \caption{\label{tab:schema2_pocock_value_ests_vp13}
     For the schema in Figure~\ref{fig:PCST-schema}, interim analysis performance results for testing hypothesis $H_{0D}$ against $H_{AD}$ with a fixed control value under Pocock Boundaries under (VP1) and (VP3) with method-based sample size $N$.
     MC Mean is the Monte Carlo mean of the estimates, 
     MC SD is the Monte Carlo standard deviation of estimates,
     ASE is the Monte Carlo mean of the standard errors, and 
     MSE Ratio is the ratio of the MC mean square error for the IPWE divided by that of the indicated estimator for the three estimates at the interim analysis (a) and final analysis (b) for $B=1000$ simulations.
    } \\
    \toprule
    \toprule
    VP & & \multicolumn{4}{c}{(a) Interim Analysis} & & \multicolumn{4}{c}{(b) Final Analysis} \\ 
    & Method & Regime & MC Mean & MC SD & ASE & MSE Ratio & & MC Mean & MC SD & ASE & MSE Ratio \\
    \hline 
    \endfirsthead
    
    \multicolumn{12}{l}%
    {{\bfseries \tablename\ \thetable{} -- continued from previous page}} \\
    \toprule
    \toprule
    VP & & \multicolumn{4}{c}{(a) Interim Analysis} & & \multicolumn{4}{c}{(b) Final Analysis} \\ 
    & Method & Regime & MC Mean & MC SD & ASE & MSE Ratio & & MC Mean & MC SD & ASE & MSE Ratio \\
    \hline 
    \endhead
    
    \hline \multicolumn{12}{r}{{Continued on next page}} \\ 
    \hline
    \endfoot

    \bottomrule
    \endlastfoot
    
1 & IPWE & 1 & 47.47 & 1.52 & 1.44 & 1.00 & & 47.50 & 0.80 & 0.79 & 1.00 \\ 
1 & IPWE & 2 & 47.52 & 1.51 & 1.44 & 1.00 & & 47.52 & 0.84 & 0.79 & 1.00 \\ 
1 & IPWE & 3 & 47.48 & 1.47 & 1.44 & 1.00 & & 47.48 & 0.78 & 0.79 & 1.00 \\ 
1 & IPWE & 4 & 47.53 & 1.48 & 1.44 & 1.00 & & 47.51 & 0.82 & 0.79 & 1.00 \\ 
1 & IPWE & 5 & 47.51 & 1.47 & 1.43 & 1.00 & & 47.51 & 0.82 & 0.79 & 1.00 \\ 
1 & IPWE & 6 & 47.55 & 1.43 & 1.44 & 1.00 & & 47.55 & 0.78 & 0.79 & 1.00 \\ 
1 & IPWE & 7 & 47.48 & 1.48 & 1.44 & 1.00 & & 47.49 & 0.80 & 0.79 & 1.00 \\ 
1 & IPWE & 8 & 47.52 & 1.47 & 1.44 & 1.00 & & 47.53 & 0.76 & 0.79 & 1.00 \\ 
1 & AIPWE & 1 & 47.47 & 1.43 & 1.47 & 1.14 & & 47.48 & 0.76 & 0.77 & 1.11 \\ 
1 & AIPWE & 2 & 47.50 & 1.42 & 1.46 & 1.13 & & 47.51 & 0.76 & 0.76 & 1.22 \\ 
1 & AIPWE & 3 & 47.49 & 1.39 & 1.46 & 1.13 & & 47.47 & 0.75 & 0.76 & 1.06 \\ 
1 & AIPWE & 4 & 47.52 & 1.40 & 1.47 & 1.12 & & 47.50 & 0.75 & 0.76 & 1.21 \\ 
1 & AIPWE & 5 & 47.50 & 1.39 & 1.46 & 1.12 & & 47.49 & 0.77 & 0.76 & 1.12 \\ 
1 & AIPWE & 6 & 47.53 & 1.33 & 1.45 & 1.16 & & 47.53 & 0.75 & 0.76 & 1.08 \\ 
1 & AIPWE & 7 & 47.42 & 1.45 & 1.45 & 1.04 & & 47.46 & 0.78 & 0.76 & 1.05 \\ 
1 & AIPWE & 8 & 47.44 & 1.41 & 1.46 & 1.07 & & 47.51 & 0.72 & 0.77 & 1.13 \\ 
1 & IAIPWE & 1 & 47.47 & 1.37 & 1.38 & 1.24 & & 47.48 & 0.76 & 0.77 & 1.10 \\ 
1 & IAIPWE & 2 & 47.50 & 1.37 & 1.36 & 1.21 & & 47.51 & 0.76 & 0.76 & 1.21 \\ 
1 & IAIPWE & 3 & 47.50 & 1.33 & 1.36 & 1.23 & & 47.47 & 0.76 & 0.76 & 1.05 \\ 
1 & IAIPWE & 4 & 47.53 & 1.35 & 1.38 & 1.20 & & 47.50 & 0.75 & 0.77 & 1.21 \\ 
1 & IAIPWE & 5 & 47.51 & 1.34 & 1.37 & 1.20 & & 47.49 & 0.77 & 0.77 & 1.13 \\ 
1 & IAIPWE & 6 & 47.54 & 1.27 & 1.35 & 1.26 & & 47.53 & 0.75 & 0.76 & 1.08 \\ 
1 & IAIPWE & 7 & 47.43 & 1.40 & 1.36 & 1.12 & & 47.46 & 0.79 & 0.76 & 1.05 \\ 
1 & IAIPWE & 8 & 47.46 & 1.36 & 1.37 & 1.16 & & 47.51 & 0.72 & 0.77 & 1.12 \\ 
3 & IPWE & 1 & 50.61 & 1.60 & 1.59 & 1.00 & & 50.54 & 0.88 & 0.88 & 1.00 \\ 
3 & IPWE & 2 & 49.04 & 1.61 & 1.60 & 1.00 & & 49.02 & 0.93 & 0.88 & 1.00 \\ 
3 & IPWE & 3 & 49.10 & 1.64 & 1.59 & 1.00 & & 49.07 & 0.89 & 0.88 & 1.00 \\ 
3 & IPWE & 4 & 47.53 & 1.59 & 1.59 & 1.00 & & 47.54 & 0.87 & 0.88 & 1.00 \\ 
3 & IPWE & 5 & 47.47 & 1.60 & 1.59 & 1.00 & & 47.52 & 0.87 & 0.88 & 1.00 \\ 
3 & IPWE & 6 & 47.52 & 1.59 & 1.58 & 1.00 & & 47.51 & 0.87 & 0.87 & 1.00 \\ 
3 & IPWE & 7 & 47.42 & 1.60 & 1.60 & 1.00 & & 47.52 & 0.85 & 0.88 & 1.00 \\ 
3 & IPWE & 8 & 47.47 & 1.54 & 1.59 & 1.00 & & 47.52 & 0.86 & 0.88 & 1.00 \\ 
3 & AIPWE & 1 & 50.61 & 1.64 & 1.72 & 0.95 & & 50.55 & 0.86 & 0.88 & 1.06 \\ 
3 & AIPWE & 2 & 49.10 & 1.60 & 1.69 & 1.00 & & 49.04 & 0.88 & 0.88 & 1.12 \\ 
3 & AIPWE & 3 & 49.06 & 1.61 & 1.7 & 1.03 & & 49.07 & 0.86 & 0.88 & 1.08 \\ 
3 & AIPWE & 4 & 47.56 & 1.54 & 1.7 & 1.07 & & 47.56 & 0.87 & 0.88 & 1.00 \\ 
3 & AIPWE & 5 & 47.49 & 1.58 & 1.71 & 1.02 & & 47.50 & 0.88 & 0.88 & 0.97 \\ 
3 & AIPWE & 6 & 47.51 & 1.55 & 1.69 & 1.04 & & 47.52 & 0.85 & 0.88 & 1.03 \\ 
3 & AIPWE & 7 & 47.44 & 1.58 & 1.71 & 1.03 & & 47.52 & 0.89 & 0.88 & 0.91 \\ 
3 & AIPWE & 8 & 47.46 & 1.58 & 1.72 & 0.95 & & 47.53 & 0.86 & 0.88 & 1.00 \\ 
3 & IAIPWE & 1 & 50.60 & 1.60 & 1.6 & 1.00 & & 50.55 & 0.86 & 0.89 & 1.05 \\ 
3 & IAIPWE & 2 & 49.10 & 1.57 & 1.57 & 1.04 & & 49.04 & 0.88 & 0.89 & 1.12 \\ 
3 & IAIPWE & 3 & 49.04 & 1.55 & 1.58 & 1.12 & & 49.07 & 0.85 & 0.89 & 1.09 \\ 
3 & IAIPWE & 4 & 47.55 & 1.50 & 1.58 & 1.12 & & 47.56 & 0.87 & 0.89 & 1.00 \\ 
3 & IAIPWE & 5 & 47.49 & 1.54 & 1.58 & 1.08 & & 47.49 & 0.89 & 0.89 & 0.95 \\ 
3 & IAIPWE & 6 & 47.51 & 1.51 & 1.55 & 1.11 & & 47.51 & 0.86 & 0.88 & 1.02 \\ 
3 & IAIPWE & 7 & 47.44 & 1.52 & 1.57 & 1.11 & & 47.51 & 0.89 & 0.88 & 0.91 \\ 
3 & IAIPWE & 8 & 47.46 & 1.53 & 1.59 & 1.01 & & 47.53 & 0.86 & 0.88 & 1.00 \\ 

\end{longtable}
}
\end{center}

\subsection*{Responders Receive a Single Treatment Option}

\begin{figure}
    \centering
    \includegraphics[width=\linewidth]{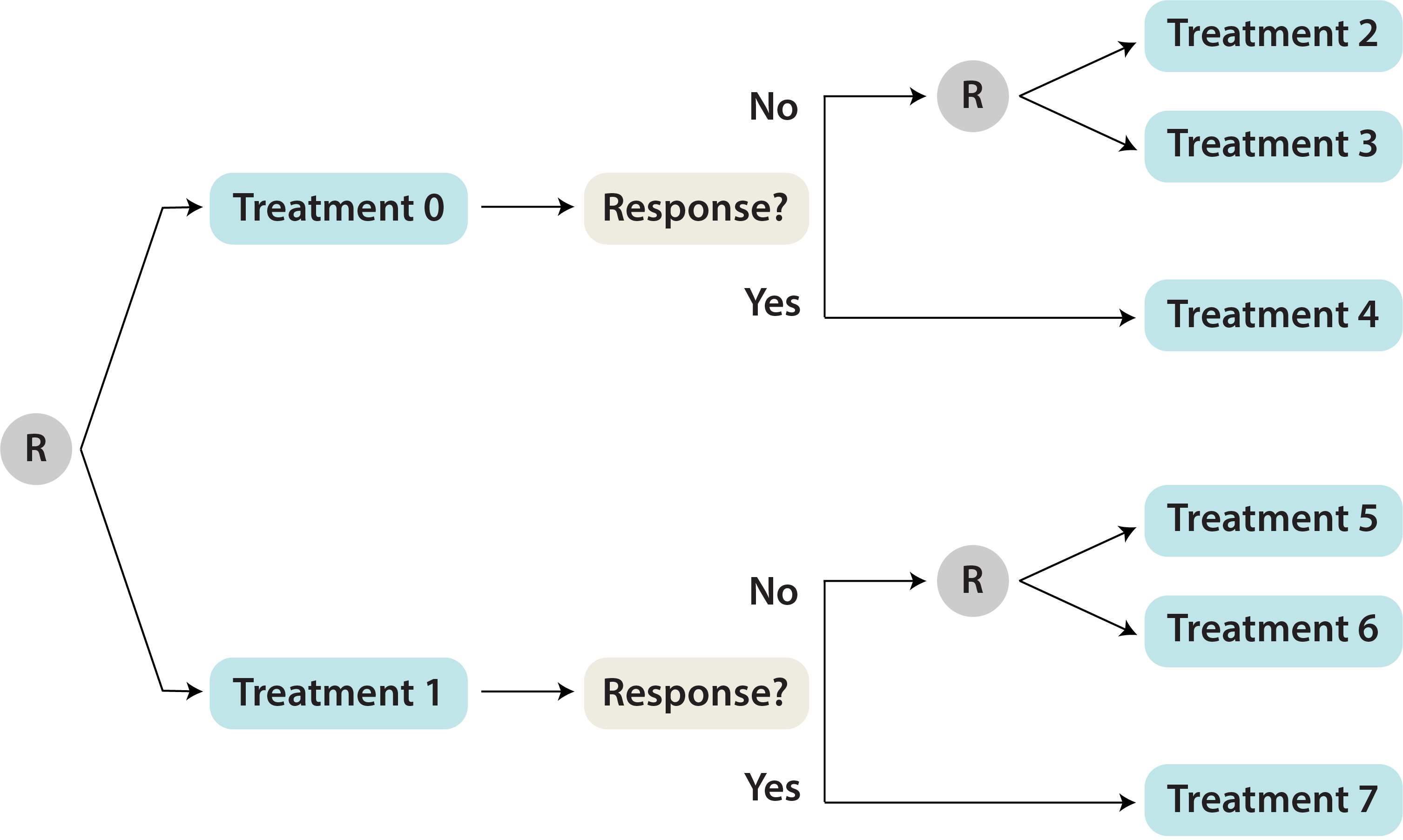}
    \caption{  \label{fig:sim_schema_deterministic}
        Schema in which responders only have one treatment available at stage $K=2$. The design embeds four regimes of the form "Give intervention $a$; if non-response, give $b$; otherwise, if response give $c$."
        Regimes $1,\ldots,4$ take $(a,b,c)$ to be $(0,2,4)$, $(0,3,4)$, $(1,5,7)$, and $(1,6,7)$, respectively. 
    }
\end{figure}

We consider the trial design in Figure~\ref{fig:sim_schema_deterministic} and testing null hypothesis $H_{0D}$ for superiority. 
Enrollment times are drawn uniformly between 0 and 1000 days and follow-up
times occur every 100 days.
The first interim analysis is conducted when 30\% of individuals have
completed the trial which, under the enrollment mechanism, corresponds
to having approximately 50\% enrollment.
We generate two baseline covariates   
$X_{1,1} \sim \mathrm{Uniform}(25, 75)$ and 
$X_{1,2} \sim \mathrm{Bernoulli}(0.5)$ as well as 
an interim outcome 
$X_{2,1} \sim \mathrm{Uniform}(0, 1)$ and response 
status $R_2\sim\mathrm{Bernoulli}(0.4)$; for notational
consistency, $R_2$ is considered part of $\bX_2$. 
The initial treatment is generated
as $A_1 \sim \mathrm{Bernoulli}(0.5)$ and, the second treatment is generated
as $A_2 \vert R_2=0 \sim\mathrm{Bernouilli}(0.5)$  and $A_2 \vert R_2 = 1$ is $0$. 
Outcomes are normally distributed with 
variance $\sigma^2 = 100$ and conditional mean
\begin{multline*}
\mu_{S1}(\overline{\bX}_2, \overline{\bA}_2; \bbeta) =
\beta_0 + \beta_1 X_{1,1} + \beta_{2} X_{1,2} +
A_1\left\lbrace
\beta_3 + \beta_4 X_{1,1} + \beta_5 X_{1,2}
\right\rbrace   + 
\beta_6 R_2 X_{2,1} \\  + \beta_7(1-R_2)X_{2,1} +   
(1-R_2)A_2\left\lbrace
\beta_8 + \beta_9 A_1 + \beta_{10} X_{1,1} + \beta_{11} X_{1,2} + 
\beta_{12} X_{2,1}
\right\rbrace,
\end{multline*}
where $\bbeta = (\beta_0, \beta_1,\ldots, \beta_{12})^{\top}$.  
Values of $\bbeta$ were
chosen to encode three value patterns (VPs): (VP1) all regimes
are equivalent 
$(\bbeta = (10, 0.5, 12.5, 0, 0, 0, 12.5, 12.5, 0, 0, 0, 0, 0)^{\top} )$; 
(VP2) there is a single best embedded regime
$(\bbeta = (10, 0.5, 12.5, 0, 0, 0, 12.5, 12.5, 0, 5, 0, 0, 0)^{\top})$; 
and
(VP3) embedded regimes starting with $A_1=0$ are optimal
$(\bbeta = (12.5, 0.5, 12.5, -2.5, 0, 0,\\ 12.5, 12.5, 0, 0, 0, 0, 0)^{\top})$.  
In (VP2), embedded regime $4$ attains a higher value $50.5$,
and in (VP3), embedded regimes $1$ and $2$ attain the higher value $50.0$.
All other regimes for each VP have value $47.5$.
The clinically
meaningful difference of $\delta=3$ and $2.5$ from the fixed control mean value $V(\bd^0) = 47.5$ and nominal power $80\%$ are used for sample size calculations for (VP2) and (VP3), respectively. 

Table \ref{tab:schema1_main3} summarizes the true value pattern, the estimator used, the total planned sample size to achieve desired power under a specified alternative, the proportion of early rejections of $H_{0D}$, the proportion of total rejections, the expected sample size, and expected stopping time for analyses using the IPWE, AIPWE and IAIPWE.
Results are presented for both when the sample size is determined for each estimator and when the sample for the IPWE is used for all estimators. 
The results may differ for each estimator as the IPWE and AIPWE use only individuals with complete trajectories. 
We calculate the total planned sample size to achieve power $80\%$ under (VP2) as the sample size for investigating the type I error rate under true (VP1), else to achieve power $80\%$ under the respective VPs.
The expected sample size is average number of individuals enrolled in the trial regardless of their contribution to the estimator used when the trial is stopped. 
We test the null $H_{0D}$ against the alternative $H_{AD}$ for $\delta=0$ using the testing procedure outlined in the Section \ref{s:interim}  with $S=2$ planned analyses at day $500$, and, if applicable, trial completion. 
We see that type I error rates are controlled and the nominal power is attained across three value pattern and stopping boundaries. 
The sandwich estimators of the variance of the values overestimates the asymptotic variance of the values for small $N$, which results in deflated early rejections for the AIPWE in these scenarios. 
The use of partial information for individuals by the IAIPWE results in smaller expected sample sizes and earlier expected stopping times for the alternative compared with the IPWE or AIPWE. 
As anticipated by the performance in conventional single-stage clinical trials, OBF boundaries may be too conservative if the analysis is performed when the proportion of information is low.

\begin{table}[H]
\caption{\label{tab:schema1_main3}
    For the schema in Figure~\ref{fig:sim_schema_deterministic}, interim analysis performance results for testing $H_{0D}$ against $H_{AD}$ with a fixed control value using Pocock and OBF boundaries. 
    Value pattern indicates the true value pattern. 
    Method indicates the estimator used. 
    The total planned sample size $N$ is determined by either each method (a) or by the IPWE (b). 
    Total planned sample sizes are determined to maintain a nominal type I error rate of $\alpha=0.05$ and achieve a power of $80\%$ under the respective value patterns, using alternative (VP2) to determine the sample size for the null (VP1).  
    Early Reject and Total Reject are the rejection rates at the first analysis and for the overall procedure, respectively. 
    $\mathbb{E}$(SS) is the expected sample size, i.e, the average number of individuals enrolled in the trial when the trial is completed. 
    $\mathbb{E}$(Stop) is the expected stopping time, i.e., the average number of days that the trial ran. 
    Monte Carlo standard deviations are given in parentheses. 
    }
    \begin{center}
\resizebox{.98\textwidth}{!}{%
    \begin{tabular}{ll lp{3em}p{3em}ll c lp{3em}p{3em}ll}
    \toprule
    \toprule    
    & & \multicolumn{5}{c}{(a) N Based on Method} & & \multicolumn{5}{c}{(b) N Based on IPWE} \\

    VP & Method & N & Early Reject & Total Reject & $\mathbb{E}$(SS) & $\mathbb{E}$(Stop) & & N & Early Reject & Total Reject & $\mathbb{E}$(SS) & $\mathbb{E}$(Stop)   \\
    \hline
    1 & IPWE & 676 & & 0.066 & 676 (0) & 1199 (1) & & 676 & & 0.066 & 676 (0) & 1199 (1) \\
    1 & AIPWE & 458 & & 0.048 & 458 (0) & 1198 (2) & & 676 & & 0.058 & 676 (0) & 1199 (1) \\
    2 & IPWE & 676 & & 0.798 & 676 (0) & 1199 (1) & & 676 & & 0.798 & 676 (0) & 1199 (1) \\
    2 & AIPWE & 458 & & 0.783 & 458 (0) & 1198 (2) & & 676 & & 0.915 & 676 (0) & 1199 (1) \\
    3 & IPWE & 636 & & 0.806 & 636 (0) & 1199 (1) & & 636 & & 0.806 & 636 (0) & 1199 (1) \\
    3 & AIPWE & 463 & & 0.789 & 463 (0) & 1198 (2) & & 636 & & 0.907 & 636 (0) & 1199 (1) \\
    \multicolumn{3}{l}{Pocock}&  \\
    1 & IPWE & 766 & 0.040 & 0.067 & 751 (75) & 1171 (137) & & 766 &  0.040 & 0.067 & 751 (75) & 1171 (137) \\
    1 & AIPWE & 521 & 0.024 & 0.045 & 515 (39) & 1181 (106) & & 766 &  0.027 & 0.050 & 756 (62) & 1180 (113) \\
    1 & IAIPWE & 517 & 0.047 & 0.065 & 505 (54) & 1165 (148) & & 766 &  0.037 & 0.059 & 752 (72) & 1173 (132) \\
    2 & IPWE & 766 & 0.291 & 0.796 & 655 (174) &  995 (318) & & 766 &  0.291 & 0.796 & 655 (174) & 995 (318) \\
    2 & AIPWE & 521 & 0.223 & 0.797 & 463 (108) & 1042 (291) & & 766 &  0.357 & 0.923 & 630 (183) & 949 (335) \\
    2 & IAIPWE & 517 & 0.305 & 0.797 & 438 (119) & 985 (322) & & 766 &  0.462 & 0.925 & 590 (191) & 876 (349) \\
    3 & IPWE & 730 & 0.308 & 0.809 & 617 (169) & 984 (323) & & 730 &  0.308 & 0.809 & 617 (169) & 984 (323) \\
    3 & AIPWE & 532 & 0.265 & 0.810 & 461 (117) & 1013 (308) & & 730 &  0.370 & 0.921 & 595 (177) & 940 (338) \\
    3 & IAIPWE & 524 & 0.355 & 0.804 & 431 (126) & 950 (334) & & 730 &  0.486 & 0.926 & 553 (183) & 859 (349) \\
    \multicolumn{3}{l}{O'Brien Fleming} & \\
    1 & IPWE & 676 & 0.001 & 0.066 & 676 (11) & 1198 (22) & & 676 &  0.001 & 0.066 & 676 (11) & 1198 (22) \\
    1 & AIPWE & 458 & 0.001 & 0.048 & 458 (7) & 1197 (22) & & 676 &  0.001 & 0.058 & 676 (11) & 1198 (22) \\
    1 & IAIPWE & 459 & 0.002 & 0.048 & 459 (10) & 1196 (31) & & 676 &  0.004 & 0.059 & 675 (21) & 1196 (44) \\
    2 & IPWE & 676 & 0.009 & 0.797 & 673 (32) & 1192 (66) & & 676 &  0.009 & 0.797 & 673 (32) & 1192 (66) \\
    2 & AIPWE & 458 & 0.005 & 0.783 & 457 (15) & 1194 (49) & & 676 &  0.015 & 0.915 & 671 (41) & 1188 (85) \\
    2 & IAIPWE & 459 & 0.033 & 0.784 & 451 (41) & 1175 (124) & & 676 &  0.068 & 0.915 & 653 (85) & 1151 (176) \\
    3 & IPWE & 636 & 0.014 & 0.806 & 632 (37) & 1189 (82) & & 636 &  0.014 & 0.806 & 632 (37) & 1189 (82) \\
    3 & AIPWE & 463 & 0.005 & 0.789 & 462 (49) & 1194 (49) & & 636 &  0.010 & 0.906 & 633 (31) & 1192 (70) \\
    3 & IAIPWE & 463 & 0.051 & 0.789 & 451 (51) & 1162 (154) & & 636 &  0.079 & 0.908 & 611 (86) & 1143 (189) \\
    \bottomrule
    \end{tabular}
}%
    \end{center}
\end{table}

Table \ref{tab:schema1_qmis} contains the same information as Table~\ref{tab:schema1_main3} when the $Q$-functions are misspecified by not including terms for $X_{1,1}$. 
The sample size to attain the desired power increases for the IAIPWE and AIPWE, but still remains smaller than the IPWE. 
The type I error rates are still controlled and the nominal power attained if the stopping boundaries are chosen under the misspecification. 

\begin{table}[H]
\centering
\caption{\label{tab:schema1_qmis}
For the schema in Figure~\ref{fig:sim_schema_deterministic}, interim analysis performance results for testing $H_{0D}$ against $H_{AD}$ with a fixed control value using Pocock and OBF boundaries. 
    Summary of results as described in Table~\ref{tab:schema1_main3} when the $Q$-functions are misspecified.
}
\begin{footnotesize}
\begin{tabular}{lllp{3em}p{3em}ll}
    \toprule
    \toprule
    VP & Method  & $N$ & Early Reject & Total Reject & $\mathbb{E}$(SS) & $\mathbb{E}$(Stop) \\
    \hline
    \multicolumn{3}{l}{Single Analysis} & \\
    1 & IPWE & 676 & & 0.066 & 676 (0) & 1199 (1) \\
    1 & AIPWE & 579 & & 0.060 & 579 (0) & 1198 (2) \\
    2 & IPWE & 676 & & 0.798 & 676 (0) & 1199 (1) \\
    2 & AIPWE & 579 & & 0.796 & 676 (0) & 1198 (2) \\
    3 & IPWE & 636 & & 0.806 & 636 (0) & 1198 (2) \\
    3 & AIPWE & 562 & & 0.811 & 562 (0) & 1198 (2) \\
    \multicolumn{3}{l}{Pocock}&  \\
    1 & IPWE & 766 & 0.040 & 0.067 & 751 (75) & 1178 (137) \\
    1 & AIPWE & 659 & 0.034 & 0.059 & 648 (60) & 1175 (127) \\
    1 & IAIPWE & 654 & 0.048 & 0.068 & 638 (70) & 1165 (149) \\
    2 & IPWE & 766 & 0.291 & 0.796 & 655 (174) & 995 (318) \\
    2 & AIPWE & 659 & 0.231 & 0.791 & 583 (139) & 1037 (295) \\
    2 & IAIPWE & 654 & 0.294 & 0.803 & 558 (149) & 993 (318) \\
    3 & IPWE & 730 & 0.308 & 0.809 & 617 (169) & 983 (333) \\
    3 & AIPWE & 646 & 0.263 & 0.803 & 561 (142) & 1015 (308) \\
    3 & IAIPWE & 637 & 0.340 & 0.811 & 529 (151) & 961 (331) \\
    \multicolumn{3}{l}{O'Brien Fleming} & \\
    1 & IPWE & 676 & 0.001 & 0.066 & 676 (11) & 1198 (22) \\
    1 & AIPWE & 580 & 0.001 & 0.060 & 580 (9) & 1198 (22) \\
    1 & IAIPWE & 580 & 0.002 & 0.061 & 579 (12) & 1196 (31) \\
    2 & IPWE & 676 & 0.009 & 0.797 & 673 (32) & 1192 (66) \\
    2 & AIPWE & 580 & 0.009 & 0.798 & 577 (27) & 1192 (66) \\
    2 & IAIPWE & 580 & 0.036 & 0.798 & 570 (54) & 1173 (130) \\
    3 & IPWE & 636 & 0.014 & 0.806 & 632 (37) & 1189 (82) \\
    3 & AIPWE & 562 & 0.006 & 0.810 & 560 (21) & 1194 (54) \\
    3 & IAIPWE & 562 & 0.037 & 0.810 & 552 (53) & 1172 (132) \\
    \bottomrule
\end{tabular}
\end{footnotesize}
\end{table}

Table \ref{tab:schema1_chi2} contains the same information as in Table~\ref{tab:schema1_main3} when testing the null hypothesis for homogeneity $H_{0H}$ using the $\chi^2$ testing procedure. 
We again calculate the sample size determined to achieve power $80\%$ under (VP2) as the sample size for investigating the type I error rate under true (VP1). 
We consider $S=2$ planned analyses at day $500$, and if applicable, trial completion. 
Table \ref{tab:schema1_chi2}
shows that there is a slight increase in the total planned sample size required to achieve power equivalent to that for testing $H_{0D}$. 
The true OBF boundaries for a $\chi^2$ test make early stopping statistically improbable for interim analyses, which is reflected in the low early rejection rates and the difference between the expected sample size using OBF boundaries and the expected sample size performing a single analysis. 
The procedure achieves nominal power with a slightly inflated type I error rate. 

\begin{table}[H]
\centering
\caption{\label{tab:schema1_chi2}
For the schema in Figure~\ref{fig:sim_schema_deterministic}, interim analysis performance results for testing $H_{0H}$ against $H_{AH}$ with a fixed control value using Pocock and OBF boundaries under the $\chi^2$ testing procedure. 
    Summary of results as described in Table~\ref{tab:schema1_main3}. 
}
\begin{footnotesize}
\begin{tabular}{lllp{3em}p{3em}ll}
    \toprule
    \toprule
    VP & Method  & $N$ & Early Reject & Total Reject & $\mathbb{E}$(SS) & $\mathbb{E}$(Stop) \\
    \hline
    \multicolumn{3}{l}{Single Analysis} & \\
    1 & IPWE & 892 & & 0.060 & 892 (0) & 1199 (1) \\
    1 & AIPWE & 473 & & 0.049 & 473 (0) & 1198 (2) \\
    2 & IPWE & 892 & & 0.772 & 892 (0)& 1199 (1) \\
    2 & AIPWE & 473 & & 0.793 & 473 (0)& 1198 (2) \\
    3 & IPWE & 1377 & & 0.783 & 1377 (0)& 1198 (1) \\
    3 & AIPWE & 773 & & 0.766 & 773 (0) & 1199 (1) \\
    \multicolumn{3}{l}{Pocock} &  \\
    1 & IPWE & 1045 & 0.035 & 0.055 & 1027 (95) & 1175 (129) \\
    1 & AIPWE & 559 & 0.044 & 0.065 & 547 (57) & 1167 (143) \\
    1 & IAIPWE & 537 & 0.066 & 0.092 & 519 (66) & 1152 (173) \\
    2 & IPWE & 1045 & 0.237 & 0.775 & 922 (222) & 1033 (297) \\
    2 & AIPWE & 559 & 0.271 & 0.792 & 484 (124) & 1009 (311) \\
    2 & IAIPWE & 537 & 0.324 & 0.803 & 450 (126) & 972 (327) \\
    3 & IPWE & 1581 & 0.239 & 0.777 & 1393 (336)& 1032 (298) \\
    3 & AIPWE & 903 & 0.225 & 0.774 & 802 (188) & 1042 (291) \\
    3 & IAIPWE & 867 & 0.320 & 0.777 & 729 (201) & 975 (326) \\
    \multicolumn{3}{l}{O'Brien Fleming} & \\
    1 & IPWE & 926 & 0.000 & 0.049 & 926 (0)  & 1199 (1) \\
    1 & AIPWE & 497 & 0.000 & 0.043 & 497 (0)  & 1198 (2) \\
    1 & IAIPWE & 484 & 0.001 & 0.047 & 484 (7) & 1197 (22) \\
    2 & IPWE & 926 & 0.003 & 0.803 & 925 (25) & 1197 (22) \\
    2 & AIPWE & 497 & 0.001 & 0.773 & 497 (8) & 1197 (22) \\
    2 & IAIPWE & 484 & 0.020 & 0.800 & 479 (34) & 1184 (98) \\
    3 & IPWE & 1395 & 0.003 & 0.773 & 1393 (37) & 1197 (38) \\
    3 & AIPWE & 796 & 0.002 & 0.760 & 795 (17) & 1197 (31) \\
    3 & IAIPWE & 778 & 0.019 & 0.765 & 771 (53) & 1185 (95) \\
    \bottomrule
\end{tabular}
\end{footnotesize}
\end{table}

Finally, we consider $S=3$ with interim analyses at days $500$ and $700$.
We test $H_{0D}$ against $H_{AD}$ with $\delta=0$. 
Table \ref{tab:schema1_s3} contains the same entries as in Table~\ref{tab:schema1_main3} and the proportion of rejections that occur at the first analysis $s=1$, at the second analysis $s=2$ if the trial continued, and the total rejections if the $H_{0D}$ was rejected at any analysis. 
The IAIPWE again has the lowest expected sample size and earliest expected stopping times. 
For OBF boundaries, the total planned sample size is marginally higher for the IAIPWE than the AIPWE due to Monte Carlo error. 
Both nominal type I error rates and power are achieved.

\begin{table}[H]
\centering
\caption{\label{tab:schema1_s3}
For the schema in Figure~\ref{fig:sim_schema_deterministic}, interim analysis performance results for testing $H_{0D}$ against $H_{AD}$ with a fixed control value using Pocock and OBF boundaries. 
    Summary of results as described in Table~\ref{tab:schema1_main3} with $S=3$ planned analyses on days $500, 700,$ and trial end.
    Rejections for $s=1$ and $s=2$ are given as the proportion of rejections that occur at that analysis without a prior rejection. 
}
\begin{footnotesize}
\begin{tabular}{lll p{3em} p{3em} p{3em} ll}
    \toprule
    \toprule
    VP & Method  & $N$ & Early Reject $s=1$ & Early Reject $s=2$ & Total Reject & $\mathbb{E}$(SS) & $\mathbb{E}$(Stop) \\
    \hline
    \multicolumn{3}{l}{Single Analysis} & \\
    1 & IPWE & 676 & & & 0.048 & 676 (0) & 1198.6 (1) \\
    1 & AIPWE & 458 & & & 0.046 & 458 (0) & 1197.9 (2) \\
    2 & IPWE & 676 & & & 0.784 & 676 (0) & 1198.6 (1) \\
    2 & AIPWE & 458 & & & 0.780 & 458 (0) & 1197.9 (2) \\
    3 & IPWE & 636 & & & 0.809 & 636 (0) & 1198.5 (2) \\
    3 & AIPWE & 462 & & & 0.807 & 462 (0) & 1197.9 (2) \\
    \multicolumn{3}{l}{Pocock} &  \\
    1 & IPWE & 792 & 0.026 & 0.018 & 0.061 & 777 (70) & 1172 (128) \\
    1 & AIPWE & 540 & 0.023 & 0.013 & 0.050 & 532 (44) & 1176 (118) \\
    1 & IAIPWE & 532 & 0.039 & 0.017 & 0.070 & 519 (55) & 1162 (148) \\
    2 & IPWE & 792 & 0.264 & 0.200 & 0.788 & 641 (172) & 915 (313) \\
    2 & AIPWE & 540 & 0.218 & 0.200 & 0.789 & 449 (113) & 946 (304) \\
    2 & IAIPWE & 532 & 0.280 & 0.215 & 0.794 & 423 (116) & 896 (314) \\
    3 & IPWE & 759 & 0.313 & 0.205 & 0.820 & 594 (169) & 878 (318) \\
    3 & AIPWE & 552 & 0.274 & 0.213 & 0.804 & 441 (121) & 901 (313) \\
    3 & IAIPWE & 540 & 0.372 & 0.200 & 0.806 & 407 (122) & 839 (319) \\
    \multicolumn{3}{l}{O'Brien Fleming} & \\
    1 & IPWE & 679 & 0.000 & 0.006 & 0.051 & 678 (16) & 1196 (16) \\
    1 & AIPWE & 460 & 0.000 & 0.004 & 0.045 & 459 (9) & 1196 (32) \\
    1 & IAIPWE & 462 & 0.002 & 0.009 & 0.047 & 460 (16) & 1192 (56) \\
    2 & IPWE & 679 & 0.015 & 0.157 & 0.781 & 642 (83) & 1110 (196) \\
    2 & AIPWE & 460 & 0.007 & 0.122 & 0.780 & 442 (49) & 1132 (171) \\
    2 & IAIPWE & 462 & 0.039 & 0.207 & 0.773 & 424 (68) & 1068 (231) \\
    3 & IPWE & 639 & 0.017 & 0.174 & 0.808 & 600 (81) & 1100 (204) \\
    3 & AIPWE & 464 & 0.010 & 0.161 & 0.802 & 439 (55) & 1111 (193) \\
    3 & IAIPWE & 466 & 0.050 & 0.268 & 0.807 & 417 (75) & 1030 (250) \\
    \bottomrule
\end{tabular}
\end{footnotesize}
\end{table}

\subsection*{Variable Time Between Analyses}

For illustrative purposes, we consider the case when time between stages is not fixed. 
The outcomes are distributed as given in the previous setting where responders receive a single treatment. 
Enrollment times are drawn uniformly between $0$ and $1000$. 
Treatment two is assigned at a follow-up which occurs uniformly between $90$ and $110$ days after enrollment. 
The final outcome is observed uniformly between $90$ and $110$ days after treatment two is assigned. 
\begin{table}[H]
 \caption{\label{tab:schema1_vartime}
For the schema in Figure~\ref{fig:sim_schema_deterministic}, interim analysis performance results for testing $H_{0D}$ against $H_{AD}$ with a fixed control value using Pocock and OBF boundaries. 
    Summary of results as described in Table~\ref{tab:schema1_main3} when time between follow-ups is not fixed.
}
    \begin{center}
    \begin{footnotesize}
    \begin{tabular}{ll lp{3em}p{3em}ll }
    \toprule
    \toprule    
    VP & Method & N & Early Reject & Total Reject & $\mathbb{E}$(SS) & $\mathbb{E}$(Stop)  \\
    \hline
    1 & IPWE & 676 & &  0.066 & 676 (0) & 1206 (1)   \\
    1 & AIPWE & 458 &  & 0.048 & 458 (0) & 1205 (2)  \\
    2 & IPWE & 676 &  & 0.798 & 676 (0) & 1206 (1)   \\
    2 & AIPWE & 458 &  & 0.783 & 458 (0) & 1205 (2)  \\
    3 & IPWE & 636 &  & 0.806 & 636 (0) & 1206 (1)   \\
    3 & AIPWE & 463 & &  0.789 & 463 (0) & 1205 (2)  \\
    \multicolumn{3}{l}{Pocock}&  \\
    1 & IPWE & 766 & 0.043 & 0.068 & 750 (78) & 1177 (144) \\
    1 & AIPWE & 521 & 0.025 & 0.046 & 515 (40) & 1188 (110) \\
    1 & IAIPWE & 517 & 0.047 & 0.066 & 505 (54) & 1172 (149) \\
    2 & IPWE & 766 & 0.282 & 0.794 & 658 (172) &  1008 (318)  \\
    2 & AIPWE & 521 & 0.225 & 0.798 & 463 (109) & 1046 (295)  \\
    2 & IAIPWE & 517 & 0.312 & 0.797 & 436 (120) & 985 (327) \\
    3 & IPWE & 730 & 0.305 & 0.811 & 618 (169) & 991 (326) \\
    3 & AIPWE & 532 & 0.268 & 0.810 & 461 (118) & 1016 (313) \\
    3 & IAIPWE & 524 & 0.362 & 0.805 & 429 (126) & 950 (339)  \\
    \multicolumn{3}{l}{O'Brien Fleming} & \\
    1 & IPWE & 676 & 0.001 & 0.067 & 676 (10) & 1206 (23)  \\
    1 & AIPWE & 458 & 0.001 & 0.048 & 458 (7) & 1204 (23)  \\
    1 & IAIPWE & 459 & 0.002 & 0.048 & 459 (10) & 1203 (32)  \\
    2 & IPWE & 676 & 0.010 & 0.797 & 673 (34) & 1199 (70) \\
    2 & AIPWE & 458 & 0.005 & 0.783 & 457 (15) & 1201 (50)  \\
    2 & IAIPWE & 459 & 0.034 & 0.784 & 451 (42) & 1181 (128) \\
    3 & IPWE & 636 & 0.017 & 0.806 & 631 (41) & 1194 (91)  \\
    3 & AIPWE & 463 & 0.003 & 0.789 & 462 (13) & 1203 (39)  \\
    3 & IAIPWE & 463 & 0.046 & 0.789 & 452 (49) & 1172 (148) \\
    \bottomrule
    \end{tabular}
    \end{footnotesize}
    \end{center}
\end{table}

Table~\ref{tab:schema1_vartime} contains the same entries as in Table~\ref{tab:schema1_main3} when the time between follow-ups is no longer fixed. 
We see that there is little difference in the performance of the estimators compared to when the time between stages is fixed. 
This suggests the IAIPWE is robust to variability in the time between follow-ups if this remains independent of the treatments.

\subsection*{Motivating Design with an Estimated Control Arm}

\begin{figure}[H] 
\begin{center}
\includegraphics[width=1\textwidth]{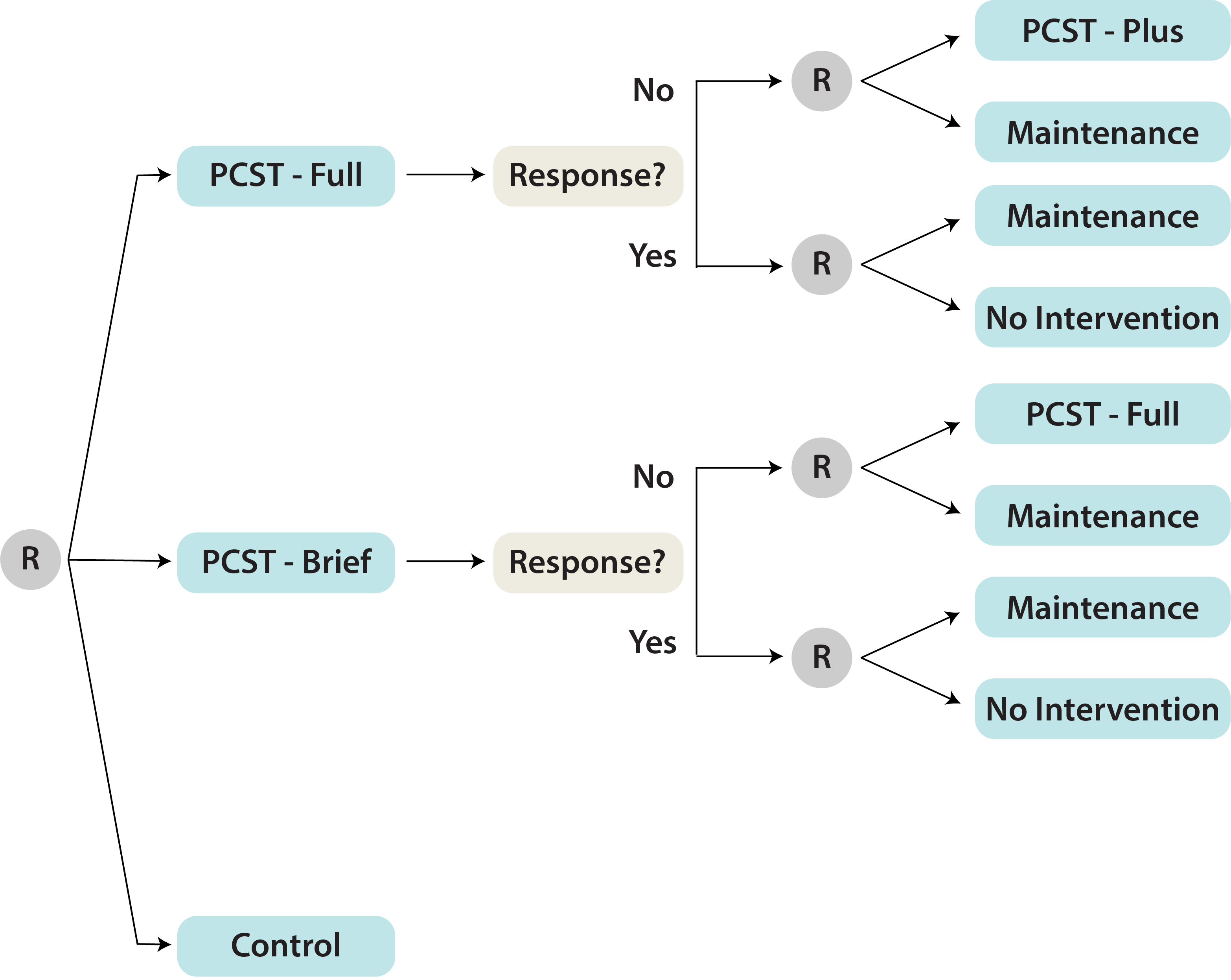}
\caption{
\label{fig:sim_schema_control_arm}
Schema for the SMART evaluating regimes involving behavioral interventions for pain management in breast cancer patients with an additional arm for standard of care. 
}
\end{center}
\end{figure}

We consider an additional schema with inclusion of a control arm shown in Figure~\ref{fig:sim_schema_control_arm}.
Individuals are randomly assigned to treatments labeled PCST-Full ($0$), PCST-Brief ($1$), or Control ($2$) with equal probability and enrolled in the trial as stated in trial design 1. 
We encode the control arm as treatment $A_1=2$ to align our notation with the mean model from the simulations in the Section \ref{s:sim}.
Interim analyses are conducted at day $700$ and trial end to mitigate over-estimation of the variance at day $500$ for small $n(t_1)$. 
The outcome has variance $100$ and mean
\begin{multline*}
\mu_{S3}(\overline{\bX}_2, \overline{\bA}_2) = 
\mu_{S2}(\overline{\bX}_2, \overline{\bA}_2) + 
I\left\lbrace
A_1 = 2
\right\rbrace \left(
\beta_{26} + \beta_{27} X_{1,1} + \beta_{28}X_{1,2} + \beta_{29} X_{2,1}
\right). \\ 
\end{multline*}
Let (VP2) indicate embedded regimes $\ell=1,\ldots,4$ attain a higher mean outcome than the standard of care by $\delta=3$. 
Let (VP3) indicate embedded regimes $\ell=1,\ldots,4$ attain a higher mean outcome than the standard of care by $\delta=5$. 

Table \ref{tab:schema3_main3} summarizes the results with entries as in Table~\ref{tab:schema1_main3}.
We use the sample size determined to achieve power $80\%$ under (VP2) as the sample size for investigating the type I error rate under (VP1), else the sample size is determined under the respective VP. 
We test the $H_{0D}$ against the alternative $H_{AD}$ for $\delta=0$ with $S=2$ planned analyses.
As expected, estimation of a control arm increases the required sample size to achieve the same power when the control is fixed. 
Therefore, a more extreme treatment difference is required to have a similar overall sample size when estimating the mean under a control arm in a SMART. 
All methods attain the desired power, and the IAIPWE consistently has lower expected sample sizes than the IPWE and AIPWE. 
In some cases the AIPWE and IPWE have similar expected stopping times dues to the minimal difference in estimated information proportion or the over-estimated variance at the first analysis by the AIPWE. 
However, when the sample size of for all estimators is determined using the IPWE, the superiority of the AIPWE is seen in uniformly earlier stopping times and lower expected sample sizes. 
Differences in the total planned sample sizes for the IPWE under (a) and (b) are due to Monte Carlo error. 

\begin{table}[H]
   \caption{\label{tab:schema3_main3}
    For the schema in Figure~\ref{fig:sim_schema_control_arm}, interim analysis performance results for testing $H_{0D}$ against $H_{AD}$ with a fixed control value using Pocock and OBF boundaries. 
    Summary of results as described in Table~\ref{tab:schema1_main3}.
    }
    \begin{center}
\resizebox{.98\textwidth}{!}{%
    \begin{tabular}{ll lp{3em}p{3em}ll c lp{3em}p{3em}ll}
    \toprule
    \toprule    
    & & \multicolumn{5}{c}{(a) N Based on Method} & & \multicolumn{5}{c}{(b) N Based on IPWE} \\
    VP & Method & N & Early Reject & Total Reject & $\mathbb{E}$(SS) & $\mathbb{E}$(Stop) & & N & Early Reject & Total Reject & $\mathbb{E}$(SS) & $\mathbb{E}$(Stop)   \\
    \hline
    1 & IPWE & 1203 & & 0.050 & 1203 (0) & 1199 (1) & & 1203 & & 0.050 & 1203 (0)  & 1199 (1) \\
    1 & AIPWE & 1002 & & 0.045 & 1002 (0) & 1199 (1) & & 1203 & & 0.048 & 1203 (0) & 1199 (1) \\
    2 & IPWE & 1203 & & 0.791 & 1203 (0) & 1199 (1) & & 1203 & & 0.815 & 1203 (0) & 1199 (1)  \\
    2 & AIPWE & 1002 & & 0.794 & 1002 (0) & 1199 (1) & & 1203 & & 0.881 & 1203 (0) & 1199 (1) \\
    3 & IPWE & 433 & & 0.817 & 433 (0) & 1198 (2) & & 434 & & 0.816 & 434 (0) & 1198 (2) \\
    3 & AIPWE & 361 & & 0.790 & 361 (0) & 1197 (3) & & 434 & & 0.866 & 434 (0) & 1198 (2) \\
    \multicolumn{3}{l}{Pocock}&  \\
    1 & IPWE & 1341 & 0.037 & 0.056 & 1326 (77) & 1181 (94) & & 1342 & 0.033 & 0.058 & 1329 (72) & 1183 (89) \\
    1 & AIPWE & 1125 & 0.036 & 0.053 & 1113 (63) & 1181 (93) & & 1342 & 0.027 & 0.045 & 1331 (65) & 1186 (81) \\
    1 & IAIPWE & 1122 & 0.038 & 0.055 & 1107 (78) & 1180 (95) & & 1342 & 0.029 & 0.046 & 1328 (84) & 1185 (84) \\
    2 & IPWE & 1341 & 0.478 & 0.796 & 1149 (201) & 961 (250) & & 1342 & 0.458 & 0.804 & 1158 (200) & 971 (249) \\
    2 & AIPWE & 1125 & 0.449 & 0.796 & 974 (168) & 975 (248) & & 1342 & 0.543 & 0.871 & 1124 (201) & 928 (249) \\
    2 & IAIPWE & 1122 & 0.470 & 0.794 & 929 (205) & 965 (249) & & 1342 & 0.565 & 0.872 & 1064 (244) & 917 (248) \\
    3 & IPWE & 484 & 0.528 & 0.827 & 407 (73) & 935 (249) & & 483 & 0.499 & 0.821 & 411 (73) & 949 (249) \\
    3 & AIPWE & 406 & 0.438 & 0.811 & 353 (61) & 980 (247) & & 483 & 0.511 & 0.877 & 409 (73) & 943 (249) \\
    3 & IAIPWE & 404 & 0.464 & 0.812 & 335 (74) & 967 (248) & & 483 & 0.540 & 0.878 & 388 (88) & 929 (248) \\
    \multicolumn{3}{l}{O'Brien Fleming} & \\
    1 & IPWE & 1206 & 0.003 & 0.050 & 1205 (19) & 1198 (27) & & 1207 & 0.002 & 0.054 & 1206 (17) & 1198 (22) \\
    1 & AIPWE & 1008 & 0.004 & 0.046 & 1007 (19) & 1197 (32) & & 1207 & 0.000 & 0.052 & 1207 (0) & 1199 (1) \\
    1 & IAIPWE & 1007 & 0.007 & 0.047 & 1004 (31) & 1195 (42) & & 1207 & 0.001 & 0.052 & 1207 (14) & 1199 (16) \\
    2 & IPWE & 1208 & 0.156 & 0.792 & 1150 (131) & 1122 (181) & & 1207 & 0.167 & 0.820 & 1147 (135) & 1116 (186) \\
    2 & AIPWE & 1008 & 0.137 & 0.798 & 967 (104) & 1131 (172) & & 1207 & 0.201 & 0.877 & 1134 (145) & 1099 (200) \\
    2 & IAIPWE & 1007 & 0.175 & 0.797 & 943 (140) & 1112 (190) & & 1207 & 0.250 & 0.876 & 1097 (192) & 1074 (216) \\
    3 & IPWE & 434 & 0.189 & 0.821 & 409 (51) & 1104 (195) & & 435 & 0.185 & 0.814 & 411 (50) & 1106 (193) \\
    3 & AIPWE & 363 & 0.124 & 0.792 & 349 (36) & 1136 (164) & & 435 & 0.164 & 0.866 & 414 (48) & 1116 (184) \\
    3 & IAIPWE & 363 & 0.166 & 0.792 & 341 (50) & 111 (185) & & 435 & 0.222 & 0.866 & 400 (66) & 1087 (207) \\
    \bottomrule
    \end{tabular}
}%
    \end{center}
  
\end{table}

\section*{Appendix H: Generative Models for Data Application}

Using the generative model below, the expected outcome for regimes $\ell=1,\ldots,8$ are $(37.5, 35.0, 35.0, 32.5, 26.5, 23.0, 23.0, 19.5)$. 

\begin{table}[H]
\caption{\label{smartGenModel} Generative model for cancer pain SMART.
$\bbeta = (\beta_0,\beta_1,\ldots,\beta_{12})^{\top} = (1, 0, 0.2, 0, 10, -10, 1, 0, -10,  -5, -2, 10, -2)^{\top}$.}
\begin{tabular}{llp{9cm}}
\toprule
\toprule 
Variable & Description & Distribution \\ \hline 
$X_{1,1}$ & height (cm) & $\mathrm{Normal}(152, 25)$ \\
$X_{1,2}$ & weight (kg) & $\mathrm{Normal}(55, 100)$ \\
$X_{1,3}$ & comorbidities & $\mathrm{Bernoulli}(0.6)$ \\ 
$X_{1,4}$ & use of pain medication & $\mathrm{Bernoulli}(0.4)$ \\ 
$X_{1,5}$ & received chemo & $\mathrm{Bernoulli}(0.6)$ \\
$A_1$     & initial treatment & $\mathrm{Bernoulli}(0.5)$ \\ 
$R_2$     & response          & $\mathrm{Bernoulli}(0.5)$ \\ 
$X_{2,0}$ & reduction in pain & $\mathrm{Uniform}(30, 40)$ if $R_2=1$ and 
$\mathrm{Uniform}(0,20)$ otherwise \\ 
$X_{2,1}$ & adherence & $\mathrm{Uniform}(0.5, 1)$ \\ 
$A_2$ & second treatment & $\mathrm{Bernoulli}(0.5)$ \\ 
$Y$   & response & $\beta_0 + \beta_1 X_{1,1} + \beta_2 X_{1,2} + \beta_3 X_{1,3} + \beta_4 X_{,14}+ \beta_5 X_{1,5} + \beta_6 X_{2,0} + \beta_7 X_{2,1} + \beta_8 A_1 + \beta_9 A_2 + \beta_{10} A_1 A_2 + \beta_{11} R_2 + \beta_{12} A_1 R_2 + \epsilon$ \\ 
$\epsilon$ & error & $\mathrm{Normal}(0, 900)$ \\ \hline
\end{tabular}

\end{table}

The fixed control value of 22.5\% is based on positing that the average responder and non-responder exhibit a 30\% and 15\% reduction in pain, respectively. 
Consistent with the null hypothesis of no treatment difference, the response status at week eight under null hypothesis (2) would be similar between regimes. 
Because the final outcome is measured at six months,
participants who respond to the first stage treatment
may still exhibit different treatment effects across regimes.

\section*{Appendix I: Data Application Tabular Results}

\begin{table}[H]
    \centering
    \caption{\label{tab:pcst-table_ipwe}
Pocock boundaries, OBF boundaries, estimated values $\times 10^{-1}$ (standard errors), and test statistics
at interim and final analysis time for the behavioral pain management case study data set using the IPWE.
Results are presented for the interim analysis (a) then the final analysis (b). 
}
\begin{footnotesize}
 \begin{tabular}{l lllr lllr}
    \toprule
    \toprule    
     & \multicolumn{4}{c}{(a) Interim Analysis} & \multicolumn{4}{c}{(b) Final Analysis} \\
    Regime & Pocock & OBF & Value (SE) & $Z$-score & Pocock & OBF & Value (SE) & $Z$-score \\
    \hline
1 & 2.66 & 4.30 & 5.13 (7.06) & 2.12 & 2.66 & 2.44 & 3.75 (4.57) & 3.28 \\
2 & 2.66 & 4.30 & 3.23 (5.54) & 1.82 & 2.66 & 2.44 & 3.26 (3.93) & 2.56 \\
3 & 2.66 & 4.30 & 4.71 (9.06) & 1.78 & 2.66 & 2.44 & 3.87 (4.48) & 3.61 \\
4 & 2.66 & 4.30 & 2.81 (7.72) & 1.46 & 2.66 & 2.44 & 3.38 (3.84) & 2.93 \\
5 & 2.66 & 4.30 & 2.85 (6.12) & 0.46 & 2.66 & 2.44 & 2.53 (3.64) & 0.78 \\
6 & 2.66 & 4.30 & 1.57 (6.15) & 0.27 & 2.66 & 2.44 & 2.42 (4.08) & 0.41 \\
7 & 2.66 & 4.30 & 2.19 (6.16) & $-$0.61 & 2.66 & 2.44 & 1.87 (3.96) & $-$0.95 \\
8 & 2.66 & 4.30 & 0.90 (5.91) & $-$0.83 & 2.66 & 2.44 & 1.76 (4.35) & $-$1.13 \\
    \bottomrule
\end{tabular}
\end{footnotesize}

\end{table}

\begin{table}[H]
    \centering
    \caption{\label{tab:pcst-table_aipwe}
Pocock boundaries, OBF boundaries, estimated values $\times 10^{-1}$ (standard errors), and test statistics
at interim and final analysis time for the behavioral pain management case study data set using the AIPWE.
Results are presented for the interim analysis (a) then the final analysis (b). 
}
\begin{footnotesize}
 \begin{tabular}{l lllr lllr}
    \toprule
    \toprule    
     & \multicolumn{4}{c}{(a) Interim Analysis} & \multicolumn{4}{c}{(b) Final Analysis} \\
    Regime & Pocock & OBF & Value (SE) & $Z$-score & Pocock & OBF & Value (SE) & $Z$-score \\
    \hline
1 & 2.66 & 4.30 & 4.97 (6.87) & 3.96 & 2.66 & 2.44 & 3.79 (4.45) & 3.46 \\
2 & 2.66 & 4.30 & 3.29 (5.20) & 1.99 & 2.66 & 2.44 & 3.21 (3.81) & 2.53 \\
3 & 2.66 & 4.30 & 4.33 (8.05) & 2.58 & 2.66 & 2.44 & 3.86 (4.3) & 3.74 \\
4 & 2.66 & 4.30 & 2.65 (6.53) & 0.61 & 2.66 & 2.44 & 3.29 (3.79) & 2.73 \\
5 & 2.66 & 4.30 & 2.87 (5.70) & 1.09 & 2.66 & 2.44 & 2.53 (3.61) & 0.78 \\
6 & 2.66 & 4.30 & 1.81 (6.19) & $-$0.70 & 2.66 & 2.44 & 2.37 (3.88) & 0.32 \\
7 & 2.66 & 4.30 & 2.21 (6.09) & $-$0.06 & 2.66 & 2.44 & 1.98 (3.91) & $-$0.69 \\
8 & 2.66 & 4.30 & 1.15 (6.28) & $-$1.75 & 2.66 & 2.44 & 1.82 (4.26) & $-$1.00 \\
    \bottomrule
\end{tabular}
\end{footnotesize}
\end{table}

\begin{table}[H]
    \centering
    \caption{\label{tab:pcst-table}
Interim analysis performance results for testing null hypothesis $H_{0D}$ against $H_{AD}$ with a fixed control.
Results include the
Pocock boundaries, OBF boundaries, value estimates $\times 10^{-1}$ (standard errors), and test statistics
at interim and final analysis time for the behavioral pain management case study data set using the IAIPWE.
Results are presented for the interim analysis (a) then the final analysis (b).
}
\begin{footnotesize}
 \begin{tabular}{l lllr lllr}
    \toprule
    \toprule    
     & \multicolumn{4}{c}{(a) Interim Analysis} & \multicolumn{4}{c}{(b) Final Analysis} \\
    Regime & Pocock & OBF & Value (SE) & $Z$-score & Pocock & OBF & Value (SE) & $Z$-score \\
    \hline
1 & 2.66 & 4.20 & 5.02 (6.49) & 4.27 & 2.66 & 2.43 & 3.79 (4.45) & 3.46 \\
2 & 2.66 & 4.20 & 3.31 (4.94) & 2.14 & 2.66 & 2.43 & 3.21 (3.81) & 2.53 \\
3 & 2.66 & 4.20 & 4.35 (7.58) & 2.77 & 2.66 & 2.43 & 3.86 (4.3) & 3.74 \\
4 & 2.66 & 4.20 & 2.64 (6.47) & 0.60 & 2.66 & 2.43 & 3.29 (3.79) & 2.73 \\
5 & 2.66 & 4.20 & 2.94 (5.51) & 1.26 & 2.66 & 2.43 & 2.53 (3.61) & 0.78 \\
6 & 2.66 & 4.20 & 1.83 (6.14) & $-$0.68 & 2.66 & 2.43 & 2.37 (3.88) & 0.32 \\
7 & 2.66 & 4.20 & 2.34 (5.89) & 0.16 & 2.66 & 2.43 & 1.98 (3.91) & $-$0.69 \\
8 & 2.66 & 4.20 & 1.23 (6.19) & $-$1.65 & 2.66 & 2.43 & 1.82 (4.26) & $-$1.00 \\
    \bottomrule
\end{tabular}
\end{footnotesize}

\end{table}

\section*{Appendix J: Independent Increments for Independent Enrollment and Propensities}

   Without the augmentation terms, the independent increments
      property follows intuitively because there is no correlation
      between outcomes for different individuals.  When considering
      the augmentation terms, it is not immediately obvious that
      independent increments holds.  The property of independent
      increments follows from the form of the estimator and can hold
      when the correlation between sequential observations is
      non-zero.  The influence function of our estimator has a
      martingale structure, which leads to many of the cross-term
      covariances among sequential influence functions to be zero.
      The non-zero covariances exhibit an interesting pattern where
      portions of the cross-term covariances for different terms
      cancel out.  What is left after summing these terms is the
      variance of the influence function at a single analysis time,
      proving the independent increments.  We may consider this as a
      result of the coarsening structure of the estimator
      appropriately accounting for how information will be accumulated
      over time.  Therefore, even though individuals may have
      correlated information between analysis times, the estimator
      possesses the independent increments property.

Estimated values of regimes at the same analysis time may be correlated. 
From the previous results at any interim analysis $s$, 
$\sqrt{N} \{ \widehat{\mathcal{V}}(t_s) - \mathcal{V}(t_s) \} \overset{d}{\to} \mathcal{N}(\0, \bSigma_{T})$ and
$\sqrt{N} \{ \1  \widehat{\mathcal{V}}(t_s) - \1 \mathcal{V}(t_s) \} \overset{d}{\to} \mathcal{N}(\0, \1 \bSigma_{T} \1^{\top})$. 

Denote the information at analysis $s$ as $\mathcal{I}(t_s) =  n(t_s) (\1 \bSigma_{T} \1^{\top} )^g$. Then, let $W(t_s) = \mathcal{I}(t_s) \1 \widehat{\mathcal{V}}(t_s)$. 
It is straightforward to see
\begin{equation*}
\begin{split}
    & \E W(t_s) =  \mathcal{I}(t_s) \1 {\mathcal{V}}(t_s) \\
    & \mathrm{var} W(t_s) =  \mathcal{I}(t_s) \1 \bSigma_T \1^{\top} / n(t_s) \mathcal{I}(t_s)  = \mathcal{I}(t_s) . \\
\end{split}
\end{equation*}

Define the influence functions for the IAIPW estimator with $C_r$ for ease of notation as 
\begin{equation*}
    IF^l(t_s) =  Y C_0(t_s) + \sum_{r=1}^{2K} \left[ C_r(t_s) 
    \E\{Y^*(\bd) \vert \overline{X}_{k(r)-1}, {X}^*_{k(r)}  \} 
    \right] 
    - \E \{ Y^*(\bd) \}
\end{equation*}

It is sufficient to show independent increments by showing that the influence functions of the estimator $V_{IA}^l(t_s)$,
$\mathrm{cov} \{ IF^l(t_s) , IF^l(t_{s'}) \} = \mathrm{var} \{IF^l(t_{s'})\} $ for $t_s < t_{s'}$.

By construction, $\E \{ IF^l(t_s) \} = 0$.
Furthermore, $\sum_{r=0}^{2K} C_r = 1$ for individuals enrolled in the study. 
Thus,
\begin{equation}
    IF^l(t_s) =  Y^*(\bd) - \E \{ Y^*(\bd) \} -
    \sum_{r=1}^{2K} 
    C_r(t_s) 
    \left[   Y^*(\bd)  - 
    \E\{Y^*(\bd) \vert \overline{X}_{k(r)-1}, {X}^*_{k(r)}  \} 
    \right] 
\end{equation}
We will begin by showing that these terms have a martingale structure. 
In particular, we will show that the expectation of the cross terms for $r \ne r' \pm 1$ is $0$.
Let 
$$\xi_{k(r)}(t_s) = \{ X_1, A_1, \kappa_1, \dots, X_{k(r)-1}, A_{k(r)-1}, \kappa_{k(r)-1}, I(\kappa_{k(r)} = k(r)) X_{k(r)}, X_{k(r)}^*(\bd), Y^*(d) \},$$
and suppress $t_s$ when we consider only one analysis. 
\begin{equation*}
\begin{split}
    & \E \Big( 
    C_r(t_s) 
    \big[   Y^*(\bd)  - 
    \E\{Y^*(\bd) \vert \overline{X}_{k(r)-1}, {X}^*_{k(r)}  \} 
    \big] 
    C_{r'}(t_s) 
    \big[  Y^*(\bd)  - 
    \E\{Y^*(\bd) \vert \overline{X}_{k(r')-1}, {X}^*_{k(r')}  \} 
    \big] 
    \Big) \\
    & = 
    \E \bigg\{
    C_r(t_s) 
    \big[  Y^*(\bd)  - 
    \E\{Y^*(\bd) \vert \overline{X}_{k(r)-1}, {X}^*_{k(r)}  \} 
    \big] \times \\
    & \hspace{5mm}
    \E \Big( C_{r'}(t_s) 
    \big[   Y^*(\bd)  - 
    \E\{Y^*(\bd) \vert \overline{X}_{k(r')-1}, {X}^*_{k(r')}  \} 
    \big] 
    \vert \xi_{k(r')-1} 
    \Big)
    \bigg\} \\
    & = 
    \E \bigg\{
    C_r(t_s) 
    \big[   Y^*(\bd)  - 
    \E\{Y^*(\bd) \vert \overline{X}_{k(r)-1}, {X}^*_{k(r)}  \} 
    \big] \times 
    0
    \bigg\} \\
    &= 0.
    \\
\end{split}
\end{equation*}
Similarly, 
\begin{equation*}
    \E \Big( 
    [Y^*(\bd) - \E \{ Y^*(\bd) \} ]
    C_{r'}(t_s) 
    \big[   Y^*(\bd)  - 
    \E\{Y^*(\bd) \vert \overline{X}_{k(r')-1}, {X}^*_{k(r')}  \} 
    \big] 
    \Big) =0. \\
\end{equation*}
Therefore 
\begin{equation} \label{eq:}
    \mathrm{var} \{ IF^l(t_s) \} = \mathrm{var} \{ Y^*(\bd) \} 
    +
    \sum_{k=1}^{K} 
    \E \bigg( 
    \{ C_{2k-1}(t_s) + C_{2k}(t_s) \}
    \left[   Y^*(\bd)  - 
    \E\{Y^*(\bd) \vert \overline{X}_{k-1}, {X}^*_{k}  \} 
    \right] 
    \bigg)^2
\end{equation}

Now, we consider the covariance between influence functions at two time points, $t_s$ and $t_{s'}$ for $s < s'$. By the martingale structure, 
\begin{equation*}
\begin{split}
    & \mathrm{cov} \Big(
    [Y^*(\bd) - \E \{ Y^*(\bd) \} ], 
    C_r(t_s) 
    \left[   Y^*(\bd)  - 
    \E\{Y^*(\bd) \vert \overline{X}_{k(r)-1}, {X}^*_{k(r)}  \} 
    \right] 
    \Big) = 0, 
    \\
    & \mathrm{cov} \Big(
    [Y^*(\bd) - \E \{ Y^*(\bd) \} ], 
    C_r(t_{s'}) 
    \left[   Y^*(\bd)  - 
    \E\{Y^*(\bd) \vert \overline{X}_{k(r)-1}, {X}^*_{k(r)}  \} 
    \right] 
    \Big) = 0, \mathrm{ and } \\
    & \mathrm{cov} \Big(
    C_{r'}(t_{s}) 
    \left[   Y^*(\bd)  - 
    \E\{Y^*(\bd) \vert \overline{X}_{k(r')-1}, {X}^*_{k(r')}  \} 
    \right] 
    , 
    C_{r}(t_{s'}) 
    \left[  Y^*(\bd)  - 
    \E\{Y^*(\bd) \vert \overline{X}_{k(r)-1}, {X}^*_{k(r)}  \} 
    \right] 
    \Big) = 0 
\end{split}
\end{equation*}
for $r < r' + 1$ and  $s < s'$.
We will show that 
\begin{equation*}
\begin{split}
    &\E \Big( \sum_{k=1}^{k'}
    \{ C_{2k-1}(t_{s}) + C_{2k}(t_{s}) \} 
    \left[  Y^*(\bd)  - 
    \E\{Y^*(\bd) \vert \overline{X}_{k-1}, {X}^*_{k}  \} 
    \right] \times \\
    & \hspace{10mm}
     \{ C_{2k'-1}(t_{s'}) + C_{2k'}(t_{s'}) \} 
    \left[  Y^*(\bd)  - 
    \E\{Y^*(\bd) \vert \overline{X}_{k'-1}, {X}^*_{k'}  \} 
    \right] \Big) \\
    &=   \E \Big(  \{ C_{2k'-1}(t_{s'}) + C_{2k'}(t_{s'}) \} 
    \left[  Y^*(\bd)  - 
    \E\{Y^*(\bd) \vert \overline{X}_{k'-1}, {X}^*_{k'}  \} 
    \right] \Big)^2. \\
\end{split}
\end{equation*}
for the case $K=2$ to show that independent increments holds. 
\begin{equation*}
\begin{split}
    &\E \Big( \sum_{k=1}^{k'}
    \{ C_{2k-1}(t_{s}) + C_{2k}(t_{s}) \} 
    \left[  Y^*(\bd)  - 
    \mu_k
    \right]  
     \{ C_{2k'-1}(t_{s'}) + C_{2k'}(t_{s'}) \} 
    \left[  Y^*(\bd)  - 
    \mu_{k'}
    \right] \Big) \\
    &=   \E \Big(  \{ C_{2k'-1}(t_{s}) + C_{2k'}(t_{s}) \} 
    \left[  Y^*(\bd)  - 
    \mu_{k'}
    \right] \Big)^2. \\
\end{split}
\end{equation*}

Let $r$ be odd and let $\nu_k(t_s)= pr\{\kappa(t_s) \geq k \vert \Gamma(t_s) = 1\}$ and $\nu_{K+1}(t_s) = pr\{\Delta(t_s) =1 \vert \Gamma(t_s) = 1\}$. We make use of the abused notation that $k(r) = k(r+1) = k$ and begin by simplifying part of the expressions we will need. 
\begin{multline*}
C_{r}(t_s) + C_{r+1}(t_s) =     \frac{ I \{ \Rc(t_s) = r\} - pr\{\Rc(t_s) = r \vert \Rc(t_s) \geq r, W(t_s) \} I \{ \Rc(t_s) \geq r \} }
    {pr\{ \Rc(t_s) > r \vert W(t_s)  \} }
    + \\
    \frac{ I \{ \Rc(t_s) = r+1\} - pr\{\Rc(t_s) = r+1 \vert \Rc(t_s) \geq r+1, W(t_s) \} I \{ \Rc(t_s) \geq r+1 \} }
    {pr\{ \Rc(t_s) > r+1 \vert W(t_s) \} } \\ 
    =\frac{ I \{ \Rc(t_s) = r\} - \{1 - \pi_k(\bd) \} I \{ \Rc(t_s) \geq r \} }
    {\nu_k(t_s) \prod_{\tilde{k}=1}^{k} \pi_{\tilde{k}}(\bd)  }
    + \\
    \frac{ I \{ \Rc(t_s) = r+1\} - \frac{\nu_k(t_s) - \nu_{k+1}(t_s)}{\nu_k(t_s)}  I \{ \Rc(t_s) \geq r+1 \} }
    {\nu_{k+1}(t_s) \prod_{\tilde{k}=1}^{k} \pi_{\tilde{k}}(\bd)  } \\ 
    =\frac{ \prod_{\tilde{k}=1}^{k-1} I \{ A_{\tilde{k}} = \bd_{\tilde{k}} \}
    I \{ A_k \ne \bd_k, \kappa(t_s) \geq k \}  
    - \{1 - \pi_k(\bd) \}
    \prod_{\tilde{k}=1}^{k-1} I \{ A_{\tilde{k}} = \bd_{\tilde{k}} \}
    I \{ \kappa(t_s) \geq k \}  
    }
    {\nu_k(t_s) \prod_{\tilde{k}=1}^{k} \pi_{\tilde{k}}(\bd)  }
    + \\
    \frac{ \prod_{\tilde{k}=1}^{k} I \{ A_{\tilde{k}} = \bd_{\tilde{k}} \}
    I \{ \kappa(t_s) = k \} \{1-\Delta(t_s) \}
    - \frac{\nu_k(t_s) - \nu_{k+1}(t_s)}{\nu_k(t_s)}  
    \prod_{\tilde{k}=1}^{k} I \{ A_{\tilde{k}} = \bd_{\tilde{k}} \}
    I \{ \kappa(t_s) \geq k \} }
    {\nu_{k+1}(t_s) \prod_{\tilde{k}=1}^{k} \pi_{\tilde{k}}(\bd)  } \\ 
    = \frac{
    I \{ \kappa(t_s) \geq k \} \prod_{\tilde{k}=1}^{k-1} I \{ A_{\tilde{k}} = \bd_{\tilde{k}} \}
    }{
    \prod_{\tilde{k}=1}^{k-1} \pi_{\tilde{k}}(\bd)
    } \times \bigg(
    \frac{ 
    I \{ A_k \ne \bd_k \}  
    - \{1 - \pi_k(\bd) \}
    }
    {\nu_k(t_s) \pi_k(\bd) }
    + \\
    \frac{ I \{ A_k = \bd_k \} 
    I \{ \kappa(t_s) = k \} \{1-\Delta(t_s) \}
    - \frac{\nu_k(t_s) - \nu_{k+1}(t_s)}{\nu_k(t_s)}  
    I \{ A_k = \bd_k \}
    }
    {\nu_{k+1}(t_s) \pi_{k}(\bd)  } \bigg) \\ 
    = \frac{
    I \{ \kappa(t_s) \geq k \} \prod_{\tilde{k}=1}^{k-1} I \{ A_{\tilde{k}} = \bd_{\tilde{k}} \}
    }{
    \nu_k(t_s)
    \prod_{\tilde{k}=1}^{k-1} \pi_{\tilde{k}}(\bd)
    } \times \bigg(
    \frac{ 
     - I \{ A_k = \bd_k \}  
    + \pi_k(\bd) 
    }
    {\pi_k(\bd) }
    + \\
    \frac{ \nu_k(t_s) I \{ A_k = \bd_k \} 
    I \{ \kappa(t_s) = k \} \{1-\Delta(t_s) \}
    - \{ \nu_k(t_s) - \nu_{k+1}(t_s) \} 
    I \{ A_k = \bd_k \}
    }
    { \nu_{k+1}(t_s) \pi_{k}(\bd)  } \bigg) \\ 
    = \frac{
    I \{ \kappa(t_s) \geq k \} \prod_{\tilde{k}=1}^{k-1} I \{ A_{\tilde{k}} = \bd_{\tilde{k}} \}
    }{
    \nu_k(t_s)
    \prod_{\tilde{k}=1}^{k-1} \pi_{\tilde{k}}(\bd)
    } \times \bigg(
    \frac{ 
     - I \{ A_k = \bd_k \} \nu_{k+1}(t_s) 
    + \pi_k(\bd) \nu_{k+1}(t_s)
    }
    {\pi_k(\bd) \nu_{k+1}(t_s) }
    + \\
    \frac{ \nu_k(t_s) I \{ A_k = \bd_k \} 
    I \{ \kappa(t_s) = k \} \{1-\Delta(t_s) \}
    - \nu_k(t_s) I \{ A_k = \bd_k \} 
    + \nu_{k+1}(t_s) 
    I \{ A_k = \bd_k \}
    }
    { \nu_{k+1}(t_s) \pi_{k}(\bd)  } \bigg) \\ 
        = \frac{
    I \{ \kappa(t_s) \geq k \} \prod_{\tilde{k}=1}^{k-1} I \{ A_{\tilde{k}} = \bd_{\tilde{k}} \}
    }{
    \nu_k(t_s)
    \prod_{\tilde{k}=1}^{k-1} \pi_{\tilde{k}}(\bd)
    } \bigg(
    1
    + 
    \frac{ \nu_k(t_s) I \{ A_k = \bd_k \} 
    [I \{ \kappa(t_s) = k \}\{1-\Delta(t_s) \} - 1 ]
    }
    { \nu_{k+1}(t_s) \pi_{k}(\bd)  } \bigg) \\ 
\end{multline*}
We note that $1-\Delta(t_s)$ is only zero when $\kappa(t_s) = K$, so for all other $\kappa(t_s) < K, [I \{ \kappa(t_s) = k \}\{1-\Delta(t_s) \} - 1 ] = [I \{ \kappa(t_s) = k \} - 1 ]$.
In our notation, we define $\prod_{\tilde{k} = 1}^{k-1}(\cdot) = 1$ for $k=1$. 
We will use the below expressions for $\{C_{2k-1}(t_s) + C_{2k}(t_s) \} \{ C_{2k'-1}(t_s') + C_{2k'}(t_s') \}$.

\begin{align}
    & \frac{      
    I \{ \kappa(t_{s'}) \geq k' \} \prod_{\tilde{k}=1}^{k'-1} I \{ A_{\tilde{k}} = \bd_{\tilde{k}} \}
    }{
    \nu_{k'}(t_{s'})
    \prod_{\tilde{k}=1}^{k'-1} \pi_{\tilde{k}}(\bd)
    } \bigg(
    1
    + 
    \frac{ \nu_{k'}(t_{s'}) I \{ A_{k'} = \bd_{k'} \} 
    [I \{ \kappa(t_{s'}) = k' \}\{1-\Delta(t_{s'}) \} - 1 ]
    }
    { \nu_{k'+1}(t_{s'}) \pi_{k'}(\bd)  } \bigg) \times \notag \\  
    & \frac{
    I \{ \kappa(t_s) \geq k \} \prod_{\tilde{k}=1}^{k-1} I \{ A_{\tilde{k}} = \bd_{\tilde{k}} \}
    }{
    \nu_k(t_s)
    \prod_{\tilde{k}=1}^{k-1} \pi_{\tilde{k}}(\bd)
    } \bigg(
    1
    + 
    \frac{ \nu_k(t_s) I \{ A_k = \bd_k \} 
    [I \{ \kappa(t_s) = k \}\{1-\Delta(t_s) \} - 1 ]
    }
    { \nu_{k+1}(t_s) \pi_{k}(\bd)  } \bigg) \notag \\ 
    =
    &\frac{      
    I \{ \kappa(t_{s'}) \geq k' \} I \{ \kappa(t_s) \geq k \} \prod_{\tilde{k}=1}^{k'-1} I \{ A_{\tilde{k}} = \bd_{\tilde{k}} \}
    }{
    \nu_{k'}(t_{s'}) \nu_k(t_s) 
    \prod_{\tilde{k}=1}^{k'-1} \pi_{\tilde{k}}(\bd)
    \prod_{\tilde{k}=1}^{k-1} \pi_{\tilde{k}}(\bd)
    } \times \label{eq:lead_term} \\
    &\bigg(
    1 + \label{eq:line_1} \\
    & \frac{ \nu_{k'}(t_{s'}) I \{ A_{k'} = \bd_{k'} \} 
    [I \{ \kappa(t_{s'}) = k' \}\{1-\Delta(t_{s'}) \} - 1 ]
    }
    { \nu_{k'+1}(t_{s'}) \pi_{k'}(\bd)  } + \label{eq:line_2} \\
    & \frac{ \nu_k(t_s) I \{ A_k = \bd_k \} 
    [I \{ \kappa(t_s) = k \}\{1-\Delta(t_s) \} - 1 ]
    }
    { \nu_{k+1}(t_s) \pi_{k}(\bd)  } + \label{eq:line_3} \\
    & \frac{ \nu_{k'}(t_{s'}) I \{ A_{k'} = \bd_{k'} \} 
    [I \{ \kappa(t_{s'}) = k' \}\{1-\Delta(t_{s'}) \} - 1 ]
    }
    { \nu_{k'+1}(t_{s'}) \pi_{k'}(\bd)  } 
    \frac{ \nu_k(t_s) I \{ A_k = \bd_k \} 
    [I \{ \kappa(t_s) = k \}\{1-\Delta(t_s) \} - 1 ]
    }
    { \nu_{k+1}(t_s) \pi_{k}(\bd)  } \label{eq:line_4} 
\end{align}

\subsection*{Terms for $k'=1$}

For $k' = 1$, we need only consider $k=1$.  Note that \eqref{eq:lead_term} is zero for $k'=1$. We suppress dependence in $\E\{Y^*(\bd) \vert X_1, X^*(\bd) \}$ for ease of notation. 

\begin{multline*}
    \E \bigg\{ [Y^*(\bd) - \E\{ Y^*(\bd) \} ]^2 \bigg( 
    1 + 
    \frac{ I \{ A_{1} = \bd_{1} \} 
    [I\{ \kappa(t_s) = 1 \} - 1 ]
    }
    { \nu_{2}(t_{s}) \pi_{1}(\bd)  }
    +
    \frac{ I \{ A_{1} = \bd_{1} \} 
    [I\{ \kappa(t_{s'}) = 1 \} - 1 ]
    }
    { \nu_{2}(t_{s'}) \pi_{1}(\bd)  } + \\
    \frac{ I \{ A_{1} = \bd_{1} \} 
    [I\{ \kappa(t_s) = 1 \} - 1 ]
    }
    { \nu_{2}(t_{s}) \pi_{1}(\bd)  }
    \frac{ I \{ A_{1} = \bd_{1} \} 
    [I\{ \kappa(t_{s'}) = 1 \} - 1 ]
    }
    { \nu_{2}(t_{s'}) \pi_{1}(\bd)  } 
    \bigg) \bigg\} \\
    = \E \bigg\{ [Y^*(\bd) - \E\{ Y^*(\bd) \} ]^2 \bigg( 
    1 -
    \frac{ I \{ A_{1} = \bd_{1} \} 
    I\{ \kappa(t_s) = 2 \}
    }
    { \nu_{2}(t_{s}) \pi_{1}(\bd)  }
    -
    \frac{ I \{ A_{1} = \bd_{1} \} 
    I\{ \kappa(t_{s'}) = 2 \}
    }
    { \nu_{2}(t_{s'}) \pi_{1}(\bd)  } + \\
    \frac{ I \{ A_{1} = \bd_{1} \} 
    I\{ \kappa(t_s) = 2 \}
    }
    { \nu_{2}(t_{s}) \pi_{1}(\bd)  }
    \frac{ I \{ A_{1} = \bd_{1} \} 
    I\{ \kappa(t_{s'}) = 2 \}
    }
    { \nu_{2}(t_{s'}) \pi_{1}(\bd)  } 
    \bigg) \bigg\} \\
    = \E \bigg\{ [Y^*(\bd) - \E\{ Y^*(\bd) \} ]^2 \bigg( 
    1 -
    \frac{ I \{ A_{1} = \bd_{1} \} 
    I\{ \kappa(t_s) = 2 \}
    }
    { \nu_{2}(t_{s}) \pi_{1}(\bd)  }
    -
    \frac{ I \{ A_{1} = \bd_{1} \} 
    I\{ \kappa(t_{s'}) = 2 \}
    }
    { \nu_{2}(t_{s'}) \pi_{1}(\bd)  } + \\
    \frac{ I \{ A_{1} = \bd_{1} \} 
    I\{ \kappa(t_{s'}) = 2 \} I\{ \kappa(t_s) = 2 \} 
    }
    { \nu_{2}(t_{s'}) \nu_{2}(t_{s}) \pi_{1}(\bd)^2  } 
    \bigg) \bigg\} \\
    = \E \bigg\{ [Y^*(\bd) - \E\{ Y^*(\bd) \} ]^2 \bigg( 
    \frac{ I \{ A_{1} = \bd_{1} \} 
    I\{ \kappa(t_s) = 2 \} 
    }
    { \nu_{2}(t_{s'}) \nu_{2}(t_{s}) \pi_{1}(\bd)^2  } -1
    \bigg) \bigg\} \\
    = \E \bigg[  [Y^*(\bd) - \E\{ Y^*(\bd) \} ]^2 \E \bigg\{ \bigg( 
    \frac{ I \{ A_{1} = \bd_{1} \} 
    I\{ \kappa(t_s) = 2 \} 
    }
    { \nu_{2}(t_{s'}) \nu_{2}(t_{s}) \pi_{1}(\bd)^2  } -1
    \bigg) \vert \xi_1(t_s) \bigg\} \bigg] \\
    = \E \bigg[  [Y^*(\bd) - \E\{ Y^*(\bd) \} ]^2  \bigg( 
    \frac{ 1
    }
    { \nu_{2}(t_{s'})  \pi_{1}(\bd)  } -1
    \bigg) \bigg] 
    = \E \bigg( [Y^*(\bd) - \E\{ Y^*(\bd) \} ] \{ C_1(t_{s'}) + C_2(t_{s'}) \} \bigg)^2 \\
\end{multline*}

\subsection*{Terms for $k'=2$}

We begin with the term $k=1, s$ and $k'=2, s'$. 
\begin{multline*}
    \E \bigg\{ [Y^*(\bd) - \E\{ Y^*(\bd) \vert X_{1} \} ]
    [Y^*(\bd) - \E\{ Y^*(\bd) \vert X_{1}, X_2^*(\bd) \} ] 
    \frac{      
    I \{ \kappa(t_{s'}) \geq 2 \} 
    I \{ A_1 = \bd_1 \} 
    }{
    \nu_{2}(t_{s'})  
    \pi_{1}(\bd)
    } \times \\
    \bigg(
    1 + 
     \frac{ \nu_{2}(t_{s'}) I \{ A_{2} = \bd_{2} \} 
    \{-\Delta(t_{s'})\}
    }
    { \nu_{3}(t_{s'}) \pi_{2}(\bd)  } + 
     \frac{ I \{ A_1 = \bd_1 \} 
    [ - I \{ \kappa(t_s) = 2 \} ]
    }
    { \nu_{2}(t_s) \pi_{1}(\bd)  } + \\
    \frac{ \nu_{2}(t_{s'}) I \{ A_{2} = \bd_{2} \} 
    \{-\Delta(t_{s'})\}
    }
    { \nu_{3}(t_{s'}) \pi_{2}(\bd)  } 
     \frac{ I \{ A_1 = \bd_1 \} 
    [-I \{ \kappa(t_s) = 2 \}]
    }
    { \nu_{2}(t_s) \pi_{1}(\bd)  }
    \bigg)  \bigg\} \\
    = \E \bigg\{ [Y^*(\bd) - \E\{ Y^*(\bd) \vert X_{1} \} ]
    [Y^*(\bd) - \E\{ Y^*(\bd) \vert X_{1}, X_2^*(\bd) \} ] 
    \frac{      
    I \{ \kappa(t_{s'}) \geq 2 \} 
    I \{ A_1 = \bd_1 \} 
    }{
    \nu_{2}(t_{s'})  
    \pi_{1}(\bd)
    } \times \\
     \frac{ I \{ A_1 = \bd_1 \} 
    I \{ \kappa(t_s) = 2 \}
    }
    { \nu_{2}(t_s) \pi_{1}(\bd)  }
    \bigg(  
    \frac{ \nu_{2}(t_{s'}) I \{ A_{2} = \bd_{2} \} 
    \{\Delta(t_{s'})\}
    }
    { \nu_{3}(t_{s'}) \pi_{2}(\bd)  } 
    - 1
    \bigg)  \bigg\} \\
    = - \E \bigg\{ [Y^*(\bd) - \E\{ Y^*(\bd) \vert X_{1} \} ]
    [Y^*(\bd) - \E\{ Y^*(\bd) \vert X_{1}, X_2^*(\bd) \} ] 
    \frac{      
    I \{ \kappa(t_s) = 2 \}
    I \{ A_1 = \bd_1 \} 
    }{
    \nu_{2}(t_{s'}) 
    \nu_{2}(t_{s}) 
    \pi_{1}(\bd)^2
    } \bigg\} + \\
    \E \bigg\{ [Y^*(\bd) - \E\{ Y^*(\bd) \vert X_{1} \} ]
    [Y^*(\bd) - \E\{ Y^*(\bd) \vert X_{1}, X_2^*(\bd) \} ] \times \\
    \frac{      
    I \{ \kappa(t_s) = 2 \}
    I \{ A_1 = \bd_1 \} 
    }{
    \nu_{2}(t_{s'})  
    \nu_{2}(t_{s})  
    \pi_{1}(\bd)^2
    } 
    \frac{ \nu_{2}(t_{s'}) I \{ A_{2} = \bd_{2} \} 
    \{\Delta(t_{s'})\}
    }
    { \nu_{3}(t_{s'}) \pi_{2}(\bd)  } 
     \bigg\} \\
    = - \E \bigg\{ [Y^*(\bd) - \E\{ Y^*(\bd) \vert X_{1} \} ]
    [Y^*(\bd) - \E\{ Y^*(\bd) \vert X_{1}, X_2^*(\bd) \} ] 
    \frac{      
    1
    }{
    \nu_{2}(t_{s'}) 
    \pi_{1}(\bd)
    } \bigg\} + \\
    \E \bigg\{ [Y^*(\bd) - \E\{ Y^*(\bd) \vert X_{1} \} ]
    [Y^*(\bd) - \E\{ Y^*(\bd) \vert X_{1}, X_2^*(\bd) \} ] \times \\
    \frac{      
    I \{ \kappa(t_s) = 2 \}
    I \{ A_1 = \bd_1 \} 
    }{
    \nu_{2}(t_{s'})  
    \nu_{2}(t_{s})  
    \pi_{1}(\bd)^2
    } 
    \frac{ \nu_{2}(t_{s'}) I \{ A_{2} = \bd_{2} \} 
    \{\Delta(t_{s'})\}
    }
    { \nu_{3}(t_{s'}) \pi_{2}(\bd)  } 
     \bigg\} \\
    = -  \E \bigg\{ [Y^*(\bd) - \E\{ Y^*(\bd) \vert X_{1} \} ]
    [Y^*(\bd) - \E\{ Y^*(\bd) \vert X_{1}, X_2^*(\bd) \} ] 
    \frac{      
    1
    }{
    \nu_{2}(t_{s'}) 
    \pi_{1}(\bd)
    } \bigg\} + \\
    \E \bigg\{ [Y^*(\bd) - \E\{ Y^*(\bd) \vert X_{1} \} ]
    [Y^*(\bd) - \E\{ Y^*(\bd) \vert X_{1}, X_2^*(\bd) \} ] 
    \frac{ \nu_{2}(t_{s'}) 
    P\{\Delta(t_{s'}) = 1 ; \kappa(t_{s}) = 2 \}
    }
    {
    \pi_{1}(\bd)
    \nu_{2}(t_{s})  
    \nu_{2}(t_{s'})  
    \nu_{3}(t_{s'})   
    } 
     \bigg\} \\
     = - \E \bigg\{ [Y^*(\bd) - \E\{ Y^*(\bd) \vert X_{1} \} ]
    [Y^*(\bd) - \E\{ Y^*(\bd) \vert X_{1}, X_2^*(\bd) \} ] 
    \frac{      
    1
    }{
    \nu_{2}(t_{s'}) 
    \pi_{1}(\bd)
    } \bigg\} + \\
    \E \bigg\{ [Y^*(\bd) - \E\{ Y^*(\bd) \vert X_{1}, X_2^*(\bd) \} ]^2
    \frac{ \nu_{2}(t_{s'}) 
    P\{\Delta(t_{s'}) = 1 ; \kappa(t_{s}) = 2 \}
    }
    {
    \pi_{1}(\bd)
    \nu_{2}(t_{s})  
    \nu_{2}(t_{s'})  
    \nu_{3}(t_{s'})   
    } 
     \bigg\} + \\
     \E \bigg\{ [\E\{ Y^*(\bd) \vert X_{1}, X_2^*(\bd) \} - \E\{ Y^*(\bd) \vert X_{1} \} ]
    [Y^*(\bd) - \E\{ Y^*(\bd) \vert X_{1}, X_2^*(\bd) \} ] \times \\
    \frac{ \nu_{2}(t_{s'}) 
    P\{\Delta(t_{s'}) = 1 ; \kappa(t_{s}) = 2 \}
    }
    {
    \pi_{1}(\bd)
    \nu_{2}(t_{s})  
    \nu_{2}(t_{s'})  
    \nu_{3}(t_{s'})   
    } 
     \bigg\} \\
     = - \E \bigg\{ 
    [Y^*(\bd) - \E\{ Y^*(\bd) \vert X_{1}, X_2^*(\bd) \} ] ^2
    \frac{      
    1
    }{
    \nu_{2}(t_{s'}) 
    \pi_{1}(\bd)
    } \bigg\} + \\
    \E \bigg\{ [Y^*(\bd) - \E\{ Y^*(\bd) \vert X_{1}, X_2^*(\bd) \} ]^2
    \frac{ \nu_{2}(t_{s'}) 
    P\{\Delta(t_{s'}) = 1 ; \kappa(t_{s}) = 2 \}
    }
    {
    \pi_{1}(\bd)
    \nu_{2}(t_{s})  
    \nu_{2}(t_{s'})  
    \nu_{3}(t_{s'})   
    } 
     \bigg\}
\end{multline*}

Now, we consider $k=k'=2$. 
\begin{multline*}
    \E \bigg\{ 
    [Y^*(\bd) - \E\{ Y^*(\bd) \vert X_{1}, X_2^*(\bd) \} ]^2
    \frac{      
    I \{ \kappa(t_{s'}) \geq 2 \} I \{ \kappa(t_{s}) \geq 2 \} 
    I \{ A_1 = \bd_1 \} 
    }{
    \nu_{2}(t_{s'}) \nu_{2}(t_{s})  
    \pi_{1}(\bd) \pi_1(\bd)
    } \times \\
    \bigg(
    1 -  
     \frac{ \nu_{2}(t_{s'}) I \{ A_{2} = \bd_{2} \} 
    \{\Delta(t_{s'})\}
    }
    { \nu_{3}(t_{s'}) \pi_{2}(\bd)  } -
     \frac{ \nu_{2}(t_{s}) I \{ A_{2} = \bd_{2} \} 
    \{\Delta(t_{s})\}
    }
    { \nu_{3}(t_{s}) \pi_{2}(\bd)  }  + \\
    \frac{ \nu_{2}(t_{s'}) I \{ A_{2} = \bd_{2} \} 
    \{\Delta(t_{s'})\}
    }
    { \nu_{3}(t_{s'}) \pi_{2}(\bd)  } 
         \frac{ \nu_{2}(t_{s}) I \{ A_{2} = \bd_{2} \} 
    \{\Delta(t_{s})\}
    }
    { \nu_{3}(t_{s}) \pi_{2}(\bd)  }
    \bigg)  \bigg\} \\
   =\E \bigg\{ 
    [Y^*(\bd) - \E\{ Y^*(\bd) \vert X_{1}, X_2^*(\bd) \} ]^2
    \frac{      
    I \{ \kappa(t_{s}) \geq 2 \} 
    I \{ A_1 = \bd_1 \} 
    }{
    \nu_{2}(t_{s'}) \nu_{2}(t_{s})  
    \pi_{1}(\bd)^2
    } \times \\
    \bigg(
    1 -  
     \frac{ \nu_{2}(t_{s'}) I \{ A_{2} = \bd_{2} \} 
    \{\Delta(t_{s'})\}
    }
    { \nu_{3}(t_{s'}) \pi_{2}(\bd)  } -
     \frac{ \nu_{2}(t_{s}) I \{ A_{2} = \bd_{2} \} 
    \{\Delta(t_{s})\}
    }
    { \nu_{3}(t_{s}) \pi_{2}(\bd)  }  + \\
     \frac{ \nu_{2}(t_{s'})  \nu_{2}(t_{s}) I \{ A_{2} = \bd_{2} \} 
    \{\Delta(t_{s})\}
    }
    { \nu_{3}(t_{s'}) \nu_{3}(t_{s}) \pi_{2}(\bd)^2  }
    \bigg)  \bigg\} \\ 
   =\E \bigg\{ 
    [Y^*(\bd) - \E\{ Y^*(\bd) \vert X_{1}, X_2^*(\bd) \} ]^2
    \frac{      
    I \{ \kappa(t_{s}) \geq 2 \} 
    I \{ A_1 = \bd_1 \} 
    }{
    \nu_{2}(t_{s'}) \nu_{2}(t_{s})  
    \pi_{1}(\bd)^2
    } \times \\
    \bigg(
     -  
     \frac{ \nu_{2}(t_{s'}) I \{ A_{2} = \bd_{2} \} 
    \{\Delta(t_{s'})\}
    }
    { \nu_{3}(t_{s'}) \pi_{2}(\bd)  } +
     \frac{ \nu_{2}(t_{s'})  \nu_{2}(t_{s}) I \{ A_{2} = \bd_{2} \} 
    \{\Delta(t_{s})\}
    }
    { \nu_{3}(t_{s'}) \nu_{3}(t_{s}) \pi_{2}(\bd)^2  }
    \bigg)  \bigg\} \\
   = - \E \bigg\{ 
    [Y^*(\bd) - \E\{ Y^*(\bd) \vert X_{1}, X_2^*(\bd) \} ]^2
    \bigg(
    \frac{      
    I \{ \kappa(t_{s}) \geq 2 \} 
    }{
    \nu_{2}(t_{s'}) \nu_{2}(t_{s})  
    \pi_{1}(\bd)
    }
     \frac{ \nu_{2}(t_{s'}) 
    \{\Delta(t_{s'})\}
    }
    { \nu_{3}(t_{s'})  } \bigg) \bigg\} + \\
    \E \bigg\{ 
    [Y^*(\bd) - \E\{ Y^*(\bd) \vert X_{1}, X_2^*(\bd) \} ]^2
    \bigg(
    \frac{ I \{ \kappa(t_{s}) \geq 2 \} 
    \Delta(t_s) 
    }
    {\pi_{1}(\bd) \pi_{2}(\bd) \nu_3(t_{s}) \nu_3(t_{s'}) }
    \bigg)  \bigg\} \\
   = - \E \bigg\{ 
    [Y^*(\bd) - \E\{ Y^*(\bd) \vert X_{1}, X_2^*(\bd) \} ]^2
    \bigg(
    \frac{      
     \nu_{2}(t_{s'}) 
     P \{ \Delta(t_{s'}) =1; \kappa(t_s) = 2  \}
    }{
    \nu_{2}(t_{s}) \nu_{2}(t_{s'})  \nu_{3}(t_{s'})
    \pi_{1}(\bd)
    }
     \bigg) \bigg\} + \\
    \E \bigg\{ 
    [Y^*(\bd) - \E\{ Y^*(\bd) \vert X_{1}, X_2^*(\bd) \} ]^2
    \bigg(
    \frac{ I \{ \kappa(t_{s}) \geq 2 \} 
    \Delta(t_s) 
    }
    {\pi_{1}(\bd) \pi_{2}(\bd) \nu_3(t_{s}) \nu_3(t_{s'}) }
    \bigg)  \bigg\} \\
    = - \E \bigg\{ 
    [Y^*(\bd) - \E\{ Y^*(\bd) \vert X_{1}, X_2^*(\bd) \} ]^2
    \bigg(
    \frac{      
     \nu_{2}(t_{s'}) 
     P \{ \Delta(t_{s'}) =1; \kappa(t_s) = 2  \}
    }{
    \nu_{2}(t_{s}) \nu_{2}(t_{s'})  \nu_{3}(t_{s'})
    \pi_{1}(\bd)
    }
     \bigg) \bigg\} + \\
    \E \bigg\{ 
    [Y^*(\bd) - \E\{ Y^*(\bd) \vert X_{1}, X_2^*(\bd) \} ]^2
    \bigg(
    \frac{ 1
    }
    {\pi_{1}(\bd) \pi_{2}(\bd)  \nu_3(t_{s'}) }
    \bigg)  \bigg\} \\
\end{multline*}

Putting together the expressions for $k'=2$, we see 
\begin{multline*}
    - \E \bigg\{ 
    [Y^*(\bd) - \E\{ Y^*(\bd) \vert X_{1}, X_2^*(\bd) \} ] ^2
    \frac{      
    1
    }{
    \nu_{2}(t_{s'}) 
    \pi_{1}(\bd)
    } \bigg\} + \\
    \E \bigg\{ [Y^*(\bd) - \E\{ Y^*(\bd) \vert X_{1}, X_2^*(\bd) \} ]^2
    \frac{ \nu_{2}(t_{s'}) 
    P\{\Delta(t_{s'}) = 1 ; \kappa(t_{s}) = 2 \}
    }
    {
    \pi_{1}(\bd)
    \nu_{2}(t_{s})  
    \nu_{2}(t_{s'})  
    \nu_{3}(t_{s'})   
    } 
     \bigg\} - \\
     \E \bigg\{ 
    [Y^*(\bd) - \E\{ Y^*(\bd) \vert X_{1}, X_2^*(\bd) \} ]^2
    \bigg(
    \frac{      
     \nu_{2}(t_{s'}) 
     P \{ \Delta(t_{s'}) =1; \kappa(t_s) = 2  \}
    }{
    \nu_{2}(t_{s}) \nu_{2}(t_{s'})  \nu_{3}(t_{s'})
    \pi_{1}(\bd)
    }
     \bigg) \bigg\} + \\
    \E \bigg\{ 
    [Y^*(\bd) - \E\{ Y^*(\bd) \vert X_{1}, X_2^*(\bd) \} ]^2
    \bigg(
    \frac{ 1
    }
    {\pi_{1}(\bd) \pi_{2}(\bd)  \nu_3(t_{s'}) }
    \bigg)  \bigg\} \\
    = - \E \bigg\{ 
    [Y^*(\bd) - \E\{ Y^*(\bd) \vert X_{1}, X_2^*(\bd) \} ] ^2
    \frac{      
    1
    }{
    \nu_{2}(t_{s'}) 
    \pi_{1}(\bd)
    } \bigg\} + \\
    \E \bigg\{ 
    [Y^*(\bd) - \E\{ Y^*(\bd) \vert X_{1}, X_2^*(\bd) \} ]^2
    \bigg(
    \frac{ 1
    }
    {\pi_{1}(\bd) \pi_{2}(\bd)  \nu_3(t_{s'}) }
    \bigg)  \bigg\} \\
    = \E \bigg\{ [Y^*(\bd) - \E\{ Y^*(\bd) \vert X_{1}, X_2^*(\bd) \} ]^2
   \frac{ 1 }{
    \nu_2(t_{s'}) \pi_1(\bd)
    } \bigg(
    \frac{ \nu_2(t_{s'})
    }
    { \nu_{3}(t_{s'}) \pi_{2}(\bd)  }
    - 1 \bigg) \bigg\} \\ 
\end{multline*}

Therefore, for both $k'=1, 2$, we have shown 
\begin{equation*}
\begin{split}
    &\E \Big( \sum_{k=1}^{k'}
    \{ C_{2k-1}(t_{s}) + C_{2k}(t_{s}) \} 
    \left[  Y^*(\bd)  - 
    \E\{Y^*(\bd) \vert \overline{X}_{k-1}, {X}^*_{k}  \} 
    \right] \times \\
    & \hspace{10mm}
     \{ C_{2k'-1}(t_{s'}) + C_{2k'}(t_{s'}) \} 
    \left[  Y^*(\bd)  - 
    \E\{Y^*(\bd) \vert \overline{X}_{k'-1}, {X}^*_{k'}  \} 
    \right] \Big) \\
    &=   \E \Big(  \{ C_{2k'-1}(t_{s'}) + C_{2k'}(t_{s'}) \} 
    \left[  Y^*(\bd)  - 
    \E\{Y^*(\bd) \vert \overline{X}_{k'-1}, {X}^*_{k'}  \} 
    \right] \Big)^2. \\
\end{split}
\end{equation*}
So independent increments holds.

\section*{Appendix K: Implementation}

We provide an outline of the steps required to implement an interim analysis for a SMART. 
We assume that the trial design is already determined and the primary aim is a testable hypothesis using the IAIPWE. 

\begin{itemize}
    \item State the regimes of interest. 
    \item Posit models for $\lambda_r^\ell(t), K_r^\ell(t), $ and $L_{k(r)}^\ell(\overline{\bx}_{k(r)})$. 
    \item Choose the number of interim analyses and when they will occur in the trial duration. 
    \item Either analytically or empirically, estimate the covariance
      matrix $\bSigma_{H0}$. 
      See Appendix E for additional remarks. 
    \item Select the type of stopping boundaries desired and the type I error rate. 
    \item Solve (6) from Section 5.2 to determine the stopping boundaries. 
    \item If the goal is to achieve a nominal power, find the sample
      size to achieve the power following the method given in Section
      5.3. If the plan is to use a pre-determined sample size, use the methods in
      Section 5.3 to determine the power for the chosen sample size.
    \item Begin trial using the chosen enrollment procedures.
    \item At the interim analysis time, carry out inference as outlined in Section 5.1. Determine whether to end or continue the trial following stopping rules. 
    \item  If desired, check the information proportion at the interim analysis to verify your initial posited models.
\end{itemize}

\bibliographystyle{apalike}
\bibliography{biomsample}

\end{document}